  \documentclass{aa}
   \usepackage{graphicx}
  \begin{document}


  \title{Doppler boosting effect and flux evolution of superluminal components 
   in QSO 3C345}
     
  \author{S.J.~Qian\inst{1}}
   \institute{
    National Astronomical Observatories, Chinese Academy of Sciences,
    Beijing 100012, China}
 \date{Compiled by using A\&A latex}
 \abstract{The precessing jet-nozzle scenario previously proposed was 
    applied to interpret the VLBI-measured kinematics of five superluminal 
    components (C4, C5, C9, C10 and C22) and their flux density evolution in blazar 3C345.}
    {It is shown that in the inner-trajectory sections 
   their kinematic properties, including trajectory, coordinates, core 
   separation and apparent velocity can be well model-simulated by using 
   the scenario with a precession 
   period of 7.30\,yr (4.58\,yr in the source frame) and a precessing common 
   trajectory, which produces the individual knot-trajectories at their 
   corresponding precession phases.}{Through the model-simulation of their 
   kinematic behavior  their bulk Lorentz factor, viewing angle and
    Doppler factor
   were derived as functions of time. These anticipatively determined Lorentz
   /Doppler factors were used to investigate the knots' Doppler-boosting effect
   and interpret their flux evolution.} 
   {It was found that the light-curves of the five superluminal components
     observed at 15, 22 and 43\,GHz were extraordinarily 
   well coincident with their Doppler
    boosting profiles ($[\delta(t)/\delta_{max}]^{3+\alpha}$,
     $\alpha$--spectral index). Additionally, some flux fluctuations on shorter
    time-scales could be due to variations in knots' intrinsic 
     flux and spectral index.} 
     {The close relation between the flux evolution and the 
    Doppler-boosting effect not only firmly validates the precessing 
    jet-nozzle scenario being fully appropriate to explain the kinematic 
    and emission properties of superluminal components in QSO 3C345,
    but also strongly supports the traditional
   common point-view: superluminal components are physical entities 
   (shocks or plasmoids) participating relativistic motion toward us with 
   acceleration/deceleration along helical trajectories.}
  \maketitle
  \section{Introduction}
   3C345 (z=0.595) is a prototypical quasar emanating emission over the entire
   electromagnetic spectrum from radio, infrared, optical, UV and  X-rays 
   to high-energy $\gamma$ rays. It is also  one of the best-studied 
  blazars (e.g., Biretta et al. \cite{Bi86}, Hardee et al. \cite{Ha87},
  Steffen et al. \cite{St96}, Unwin et al. \cite{Un97}, Zensus et al.
   \cite{Ze97}, Klare \cite{Kl03}, Klare et al. \cite{Kl05}, Lobanov \& Roland
   \cite{Lo05}, Jorstad et al. \cite{Jo05}, \cite{Jo13},
    \cite{Jo17}, Qian et al. \cite{Qi91a}, \cite{Qi96}, \cite{Qi09},
    Schinzel et al. \cite{Sc10}, Schinzel \cite{Sc11a}, Schinzel et al.
    \cite{Sc11b}, Homan et al. \cite{Ho14}). 
   Its remarkable and violent variations in all these wavebands and 
   the spectral energy distribution have been extensively monitored and
    studied, leading to many important results on the emission properties of 
   the source.  Studies of the correlation between the variabilities at
    multi-frequencies (from radio to $\gamma$ rays) play an important role. \\
    3C345 is a remarkable compact flat-spectrum radio source which was 
    one of the firstly discovered quasars to have a relativistic jet, emanating
    superluminal components steadily.\\
    VLBI-observations have revealed the parsec structure of its jet and 
    monitored the kinematic behavior of its superluminal components ejected
    from the radio core over a quite long time-interval since $\sim$1979.\\
    It has been shown that flaring activities of the source in multi-frequencies
    (from radio to $\gamma$-rays) are closely connected with the jet-activity
    and ejection of superluminal components. In addition, VLBI-monitoring 
    observations have shown that the swing of the knot's ejection position 
    angle and the apparent direction of its relativistic jet could be produced
    by jet-precession (or jet-nozzle precession) with a precession period of
    $\sim$5-10 years. The determination of its jet-nozzle precession may be
     very important for understanding the kinematics of superluminal components
     and the properties of the central energy
     engine in its nucleus.\\
      Since 1991 (Qian et al. \cite{Qi91a}, \cite{Qi09}), we have tried to 
    interpret the VLBI-kinematics of superluminal components in 3C345 in terms
     of a precessing jet-nozzle scenario. In the recent work (Qian \cite{Qi22})
    the kinematics of twenty-seven superluminal components observed during a
    38\,yr period was explained in detail. It was found that these superluminal
    knots could be divided into two groups, which were suggestively ejected by
    their respective precessing jet nozzle. Based on this division the 
    precession period of both nozzles was derived to be 7.3$\pm$0.36\,yr in the
    same direction.\\
    Our precessing jet-nozzle scenario proposed to interpret the VLBI-measured
    kinematics in 3C345 and other blazars has been based on two assumptions:
    \begin{itemize}
   \item (1) jet-nozzle precesses with a certain period and ejects superluminal
   components and magnetized plasmas at different precession phases, which form
   the entire jet (or jet-body); thus each component has its corresponding 
   ejection precession phase.
   \item (2) Inner sections of knots' trajectories along which the superluminal 
   components  move are assumed to follow a common helical pattern
    which precesses to produce the individual knot trajectories at their
   corresponding precession phases.    
    \end{itemize}
    By applying the precessing nozzle scenario,  the swing of the ejection
    position angle of the superluminal components could be 
    investigated and  the VLBI-kinematics  of  the superluminal components
    could be fitted by using model-simulation methods, including the observed
    features: their  trajectory, core separation, coordinates and 
    apparent velocity versus time.\\
    Most importantly, the bulk Lorentz factor, viewing angle and Doppler 
    factor of superluminal components as continuous functions of time could 
    be derived. Thus the acceleration/deceleration in the motion of
    superluminal components and the relation between their flux variations
    and Doppler boosting effect  could be investigated. This may be regarded
    as the distinct advantage of our scenario and method with respect to other
    scenarios. For example, in the recent work (Qian \cite{Qi22}),
    the flux evolution of the superluminal knot C9 in 3C345 was investigated
    and for the first time found that its 15GHz and 43 GHz light-curves were 
    very well coincident with its Doppler boosting profile, suggesting the 
    radio flux variations (flaring events) observed in knot C9 being fully
     due to its Doppler boosting effect. This strongly supports the traditional
     or common viewpoint: superluminal components are physical entities
    (shocks or plasmoids) moving with relativistic speeds toward us  along
    helical trajectories at small viewing angles. At 
    the same time this finding also strongly supports the electromagnetic
    mechanisms for the acceleration of superluminal components. It seems that
    the observed trajectories of superluminal components 
    could not result from an underlying jet-pattern 
    (induced by hydrodynamic/magneto-hydrodynamic instabilities)
     lit-up by plasmoids ejected from the nuclear activities.\\
    The precessing nozzle scenario has been applied to investigate the 
    VLBI-kinematic behavior of superluminal knots in several QSOs, e.g., 
    3C279, B1308+326, PG1302+202, NRAO150, 3C454.3 and 
    OJ287 (Qian et al. \cite{Qi14}, \cite{Qi17}, \cite{Qi18a}, \cite{Qi19a}, 
    Qian \cite{Qi13}, \cite{Qi16}, \cite{Qi18b}, Qian et al. \cite{Qi21}).
    Through model-simulation of the VLBI-kinematics of their superluminal
    components, we have shown that all these quasars could have precessing 
    jet-nozzles with certain precession periods. For the four blazars (3C279,
    OJ287, 3C454.3 and 3C345) we have tentatively shown that they could have
    double precessing-jet structures and some precessing common trajectory
    patterns could exist, along which the superluminal components move 
    according to their precession phases. However, we found that only within
    the inner sections of observed trajectories the superluminal components
    followed the precessing common trajectory pattern, while in the outer
    sections they followed their own individual trajectory patterns. Thus as 
   for  their whole trajectories concerned different trajectory patterns 
   should be chosen to make model-fits to the entire kinematics of 
   individual knots.\\
   Interestingly, both jets (designated as jet-A and jet-B) were tentatively 
   found to precess with the same precession period in the same 
   direction. Thus the precession of both jet-nozzles might be due to the 
    keplerian  motion of the putative binary black holes in their nuclei 
   (e.g., see 
    Begelman et al. \cite{Be80}).\\
   Especially in the case of blazar OJ287, its quasi-periodic optical 
   variability has been suggested
    to be connected with its double-jet activity (Villata et al.\cite{Vi98}).
    Qian (\cite{Qi18b}) has summarized both the theoretical and observational
    evidence for the possible existence of a double-jet structure in its 
    nucleus, and interpreted the VLBI-kinematics of its superluminal
    components in detail in terms of precessing nozzle scenario within the
    framework of double-jet structure (Qian \cite{Qi19b}, \cite{Qi19c},
    \cite{Qi20}; also Qian et al. \cite{Qi07}).\\
     Search for periodicities in optical and radio light-curves
   (e.g. Sillanp\"a\"a et al. \cite{Si88}, Babadzhanyants et al.\cite{Ba95},
    Kudryavtseva et al.\cite{Ku06}, Qian et al. \cite{Qi07})  are important
    and could provide key information on the nature of the central engine
     in blazars.\\
     In the case of blazar 3C345, in our earlier studies
     we  found that the observed tracks of knots C4 and C5 could be 
    reproduced by the rotation of a common helical trajectory (Qian et al. 
    \cite{Qi91a}, \cite{Qi91b}, Qian \& Zhang \cite{Qi99}).
    Qian et al. (\cite{Qi09}) analyzed the  distribution of the position 
    angles for seven superluminal components (C4 to C10) at different core 
    separations of 0.15\,mas, 0.20\,mas and 0.25\,mas, and found that their 
     inner trajectories (within core separation 
    $r_n{\stackrel{<}{_\sim}}$0.4\,mas) could be explained in terms of the
    precession of a common trajectory and  a precession period of its
    jet-nozzle of $\sim$7.36\,yr was derived. 
    Based on the position angle swing of its superluminal
     components, some authors argued for the existence of a jet precession
    period: e.g., $\sim$8--10\,yr (Lobanov \& Zensus \cite{Lo99}, Klare et al. \cite{Kl05}, Klare \cite{Kl03}, Lobanov \& Roland \cite{Lo05}).\\
    We would like to point out that our precessing jet-nozzle scenario is well 
    consistent with the magneto-hydrodynamic theories for the 
    formation/collimation/acceleration of relativistic jets in 
    active galactic nuclei (e.g., Artymoviz \cite{Ar98}, Blandford \& Payne 
    \cite{Bl82}, Blandford \& Znajek \cite{Bl77}, Camenzind \cite{Ca86},
   \cite{Ca87}, \cite{Ca90}, Li et al. \cite{Li92},
    Lovelace et al. \cite{Lo86},Meier \& Nakamura \cite{Me06}, 
    Nakamura \& Asada \cite{Na13}, Shi et al. \cite{Sh12}, \cite{Sh15},
     Valhakis \& K\"onigl \cite{Vl03}, \cite{Vl04}).\\
    In this paper we  investigate the relation between the Doppler boosting
    effect and the flux  density evolution for five superluminal components 
    (C4, C5, C9, C10 and C22 of jet-A) in blazar 3C345 within the framework 
    of our precessing jet-nozzle scenario. We confirm for the first time 
    that the radio flux density  light-curves 
    of these knots were closely coincident with their Doppler boosting profiles
    which had been anticipatively derived through the model-fitting of their
    VLBI-kinematics.
   \footnote{Interpretation of the flux evolution in terms of Doppler-boosting
    effect for more superluminal components in blazar 3C345 will be
     presented elsewhere (Qian, in preparation).}
    These results may strongly justify the validity of our 
    precessing jet-nozzle scenario and the correctness or appropriateness
    of the chosen model-parameters and the modeled functions.\\
    As in the previous work (Qian \cite{Qi22}) the observational data for
    knots C4, C5, C9, C10 and C22 we adopted from 
   Schinzel (\cite{Sc11a}) and some data were re-calculated to make
   the compact core as the unified origin of coordinates. Due to core-shift
    effects only 43GHz and 22GHz observations were used (except for knot C5,
   for which only 15GHz data are available). 
   Generally, we would not mark the errors of measurements
   in the figures obtained from the model-fittings for clarity, but one
   should keep in mind that the errors in measurements of knot's position were 
   in the range of $\sim$0.05--0.1\,mas.\\
    We  apply the concordant cosmological model (Spergel et al.\cite{Sp03},
    Hogg \cite{Hog99}) with ${\Omega}_{\lambda}$=0.73 and ${\Omega}_m$=0.27, 
    and  $H_0$=71\,km${s^{-1}}{{Mpc}^{-1}}$. Thus 
    the luminosity distance of 3C345 is 3.49Gpc, angular-diameter distance
    $D_a$=1.37\,Gpc, 1\,mas=6.65\,pc, 1\,mas/yr=34.6c.\\
    \begin{figure*}
    \centering
    \includegraphics[width=8cm,angle=90]{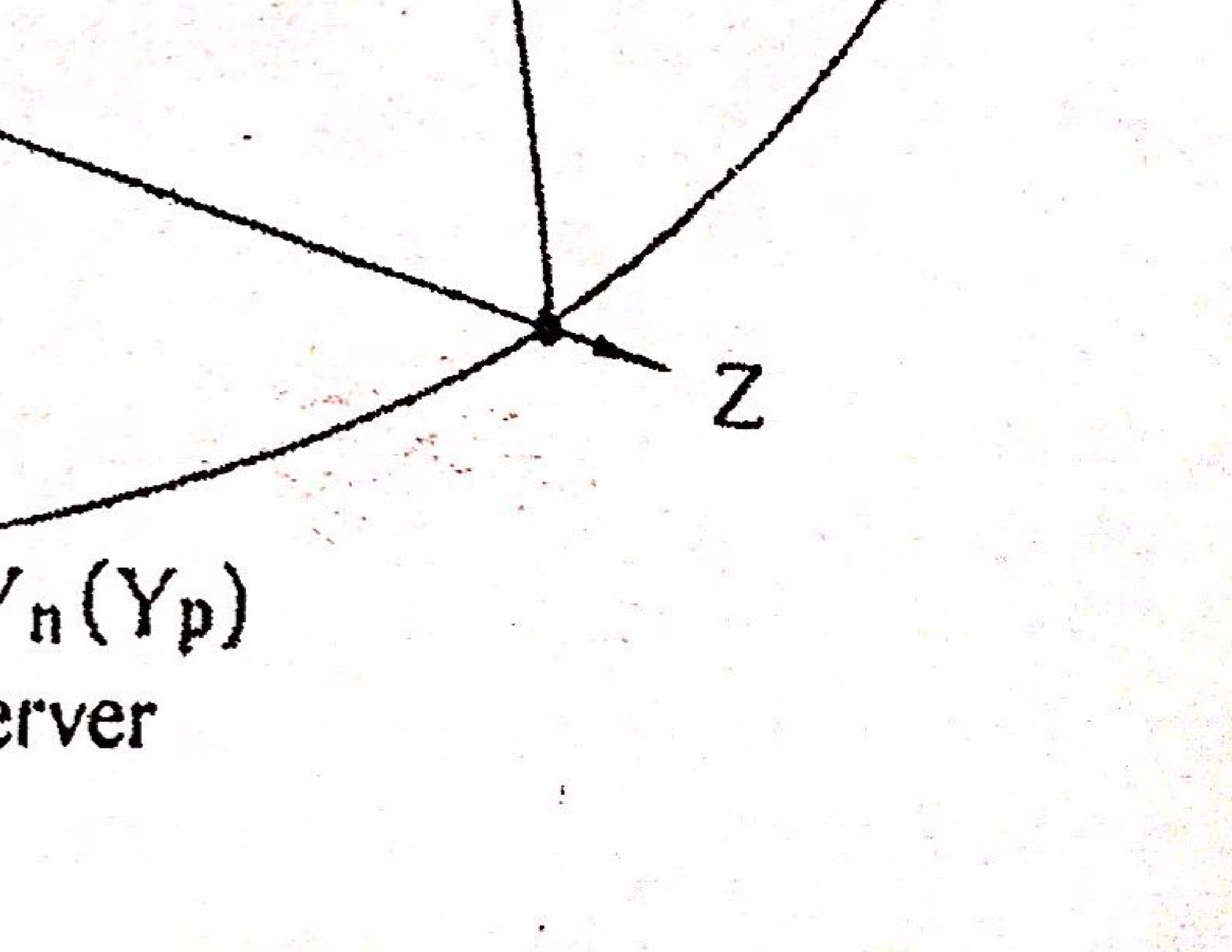}
    \caption{Geometry of the precessing jet-nozzle scenario for 3C345. 
    The jet-axis is defined in the $(X,Z)$-plane by parameters
    ($\epsilon$, $\psi$) and function $\it{x_0(z_0)}$. The common helical 
    trajectory pattern is defined by functions A(Z) and $\phi$(Z) given 
    in section 2 (Figure 2).}
    \end{figure*}
    \begin{figure*}
    \centering
    \includegraphics[width=5cm,angle=-90]{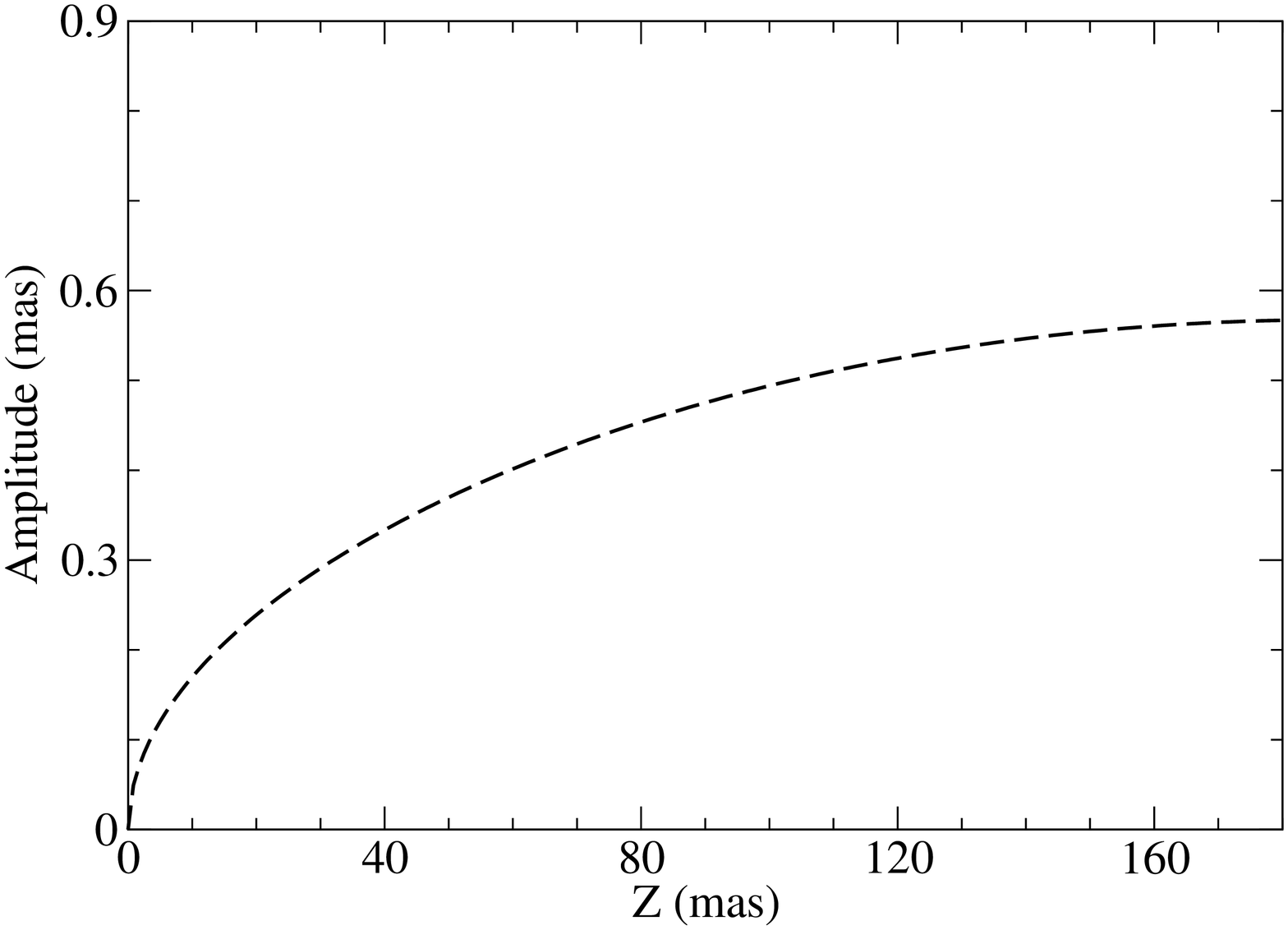}
    \includegraphics[width=5cm,angle=-90]{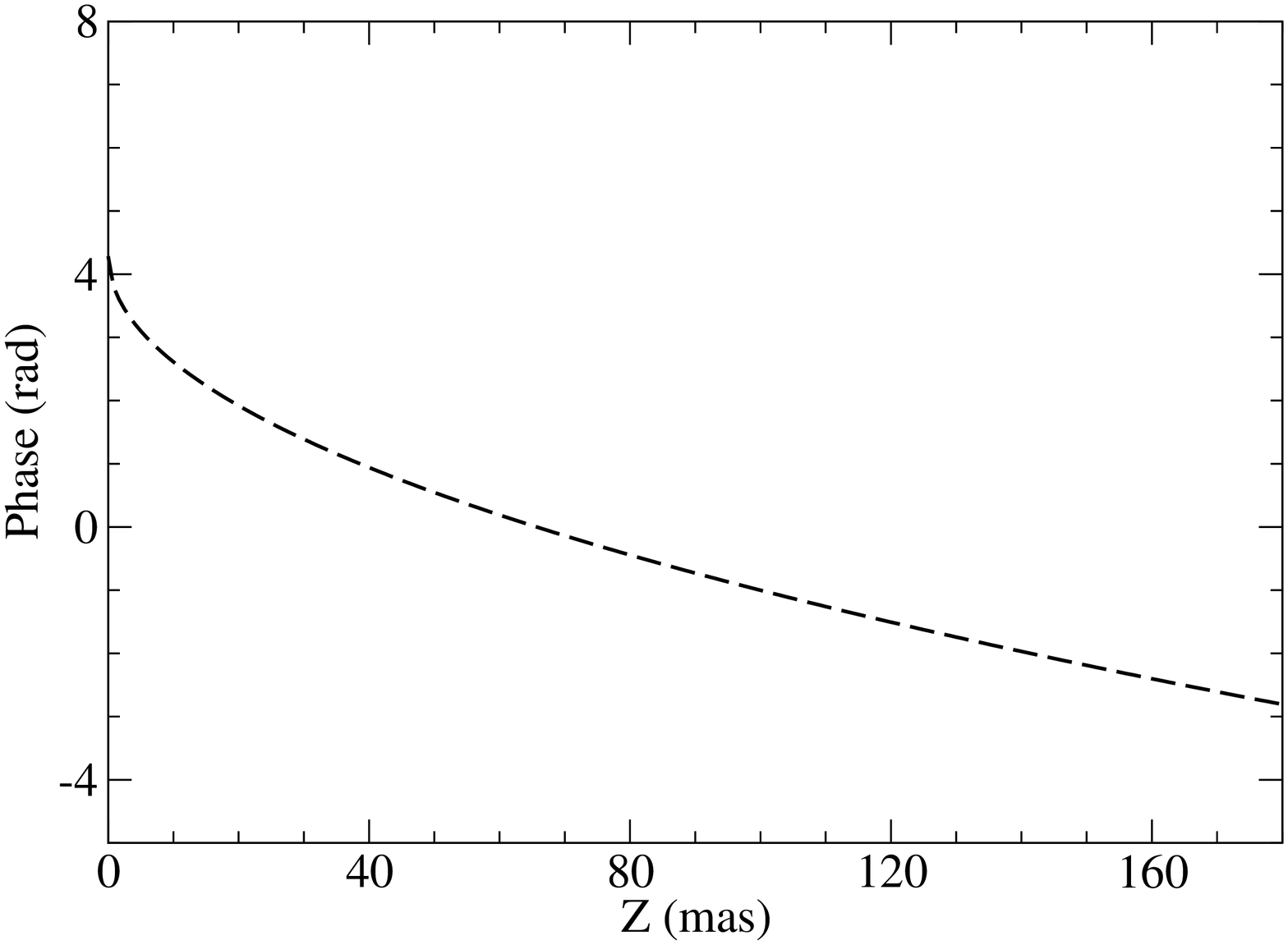}
    \caption{Model parameters of the helical trajectory for knot C4: amplitude 
    A(Z) and phase $\phi$(Z).}
    \end{figure*}
    \section{Geometry of the model}
    The precessing jet-nozzle model has been previously described (Qian et al.
    \cite{Qi19a},\cite{Qi09}, \cite{Qi91a}) for investigating
    the VLBI-kinematics and trajectory-distribution of superluminal components
    on parsec scales in the QSO 3C345. The formalism of the geometry is
    recapitulated as follows (referring to Qian et al. \cite{Qi21}, Qian
     \cite{Qi22}).\\
    A special geometry consisting of four coordinate systems is as
    shown in Figure 1. We assume that the superluminal components move along
    helical trajectories around the curved jet axis (i.e., the axis of
    the helix).\\
    We use coordinate system ($X_n,Y_n,Z_n$) to define the plane of the sky 
    ($X_n,Z_n$) and the direction of observer ($Y_n$), with $X_n$-axis pointing
    toward the negative right ascension and $Z_n$-axis toward the north pole.\\
    The coordinate system ($X,Y,Z$) is used to locate the 
    curved jet-axis in the 
    plane ($X,Z$), where $\epsilon$ represents the angle between $Z$-axis and
    $Y_n$-axis and $\psi$ the angle between $X$-axis and $X_n$-axis. 
    Thus parameters $\epsilon$ and $\psi$ are used to define the plane where 
    the jet-axis locates relative to the coordinate system ($X_n,Y_n,Z_n$).\\
    We use coordinate system (${\it{x}}'$,${\it{y}}'$,${\it{z}}'$) along the
     jet-axis to define the helical trajectory pattern for a knot, introducing
    parameters $A(s_0)$ (amplitude) and $\phi(s_0)$ (phase), where $s_0$
    represents the arc-length along the axis of helix (or curved jet-axis).
    ${\it{z}}'$-axis is along the tangent of the axis of helix.
    ${\it{y}}'$-axis is parallel to the $Y$-axis and $\eta$ is the angle
    between ${\it{x}}'$-axis and $X$-axis (see Figure 1).\\
      In general we assume that the jet-axis can be defined by a function
     $x_0(z_0)$ in the $(X,Z)$-plane as follows.
    \begin{equation}
    {x_0}=p({z_0}){{z_0}^{\zeta}}
    \end{equation}
     where
    \begin{equation}
    p({z_0})={p_1}+{p_2}[1+\exp(\frac{{z_t}-{z_0}}{{z_m}})]^{-1}
    \end{equation}
     $\zeta$, $p_1$, $p_2$, $z_t$ and $z_m$ are constants. The exponential
     term is devised for describing the jet-axis gradually curving toward
     the north, as the trajectory of knot C4 shows on large-scales.
    \begin{equation}
    {s_0}=\int_{0}^{z_0}{\sqrt{1+(\frac{d{x_0}}{d{z_0}})^2}}\,{d{z_0}}
    \end{equation}
    Therefore, the helical trajectory of a knot can be described in the (X,Y,Z)
    system as follows.
    \begin{equation}
    X({s_0})=A({s_0}){\cos{\phi({s_0})}}{\cos{\eta({s_0})}}+{x_0}
    \end{equation}
    \begin{equation}
    Y({s_0})=A({s_0}){\sin{\phi({s_0})}}
    \end{equation}
    \begin{equation}
    Z({s_0})=-A({s_0}){\cos{\phi({s_0})}}{\sin{\eta({s_0})}}+{z_0}
    \end{equation}
    where $\tan{\eta({s_0})}$=$\frac{d{x_0}}{d{z_0}}$. The projection of
   the helical trajectory on the sky-plane (or the apparent trajectory)
     is represented by
   \begin{equation}
    {X_n}={X_p}{\cos{\psi}}-{Z_p}{\sin{\psi}}
    \end{equation}
    \begin{equation}
    {Z_n}={X_p}{\sin{\psi}}+{Z_p}{\cos{\psi}}
    \end{equation}
    where
    \begin{equation}
       {X_p}=X({s_0})
    \end{equation}
    \begin{equation}
     {Z_p}={Z({s_0})}{\sin{\epsilon}}-{Y({s_0})}{\cos{\epsilon}}
    \end{equation}
    (All coordinates and amplitude (A) are measured in units of mas).
    Introducing the functions
    \begin{equation}
    {\Delta}=\arctan[(\frac{dX}{dZ})^2+(\frac{dY}{dZ})^2]^{-\frac{1}{2}}
    \end{equation}
   \begin{equation}
    {{\Delta}_p}=\arctan(\frac{dY}{dZ})
    \end{equation}
    \begin{equation}
    {{\Delta}_s}=\arccos[(\frac{dX}{d{s_0}})^2+(\frac{dY}{d{s_0}})^2+
                        (\frac{dZ}{d{s_0}})^2]^{-\frac{1}{2}}
    \end{equation}
    we can then calculate the viewing angle $\theta$, apparent transverse
    velocity ${\beta}_a$, Doppler factor $\delta$ and the elapsed time T,
    at which the knot reaches distance $z_0$ as follows: 
    \begin{equation}
     {\theta}=\arccos[{\cos{\epsilon}}(\cos{\Delta}+
               \sin{\epsilon}\tan{{\Delta}_p})]
    \end{equation}
    \begin{equation}
     {\Gamma}=(1-{\beta}^2)^{-\frac{1}{2}}
    \end{equation}
    \begin{equation}
    {\delta}=[{\Gamma}(1-{\beta}{\cos{\theta}})]^{-1}
    \end{equation}
    \begin{equation}
     {{\beta}_a}={{\beta}{\sin{\theta}}/(1-{\beta}{\cos{\theta}})}
    \end{equation}
    \begin{equation}
    T=\int^{{s_0}}_{0}{\frac{(1+z)}{{\Gamma\delta}{v}{\cos{{\Delta}_s}}}}
                       {d{s_0}}
    \end{equation}
    The amplitude and phase of the helical trajectory for superluminal  knots 
   are defined as follows.
    \begin{equation}
    {A({Z})} = {{A_0}[\sin({\pi}{Z}/{Z_1})]^{1/2}} 
    \end{equation}
    \begin{equation}
    {\phi}({Z})={{\phi}_0}-{({Z}/{Z_2})^{1/2}}
     \end{equation}
      $A_0$ represents the amplitude coefficient of the common helical
     trajectory pattern and  ${\phi}_0$ is the precession phase of an 
     individual knot.\\
     The aim of our model fitting of the kinematics of the superluminal
     components observed in 3C345 was to show that most components have
    their inner trajectories following the precessing common trajectory 
    and their kinematics can be interpreted in terms of our precessing
    jet nozzle scenario, indicating the possible presence of a supermassive
    black hole binary in its nucleus (for details see Qian \cite{Qi22} where
    a double-jet structure (jet-A and jet-B) in 3C345 was proposed). \\
     The main subject of this paper is to investigate the flux evolution 
    of five superluminal knots (C4, C5, C9 C10 and C22) of jet-A and shows that
    their flux evolution could well be interpreted in terms of the Doppler 
    boosting effect during their accelerated/decelerated  motion along 
   helical trajectories. Thus the kinematic and 
    emission properties of the five components can be unitedly explained 
    within the framework of  our precessing nozzle scenario.\\
      As shown above, the precessing common helical trajectory which 
   the superluminal
   components follow is defined by parameters ($\epsilon$, $\psi$) , and 
   formulas (1)-(2) and (19)-(20) and the corresponding parameters in them.
   For jet-A we assume the following values:\\
      $\epsilon$=0.0349\,rad=$2^{\circ}$; $\psi$=0.125\,rad=$7.16^{\circ}$; 
      $\zeta$=2.0, $p_1$=0; $p_2$=1.34$\times{10^{-4}}$; $z_t$=66\,mas; 
     $z_m$=6\,mas; $A_0$=0.605\,mas, $Z_1$=396\,mas and $Z_2$=3.58\,mas. 
     The precession  phase $\phi_0$ is related to the knot's ejection time:
      \begin{equation}
       {\phi_0}=4.28+{\frac{2\pi}{T_0}}({T_0}-1979.00)
      \end{equation}
      where $T_0$=7.30\,yr--precession period of the jet-nozzle. The functions
     modeled for the amplitude and phase of the precessing helical trajectory
     pattern is shown in Figure 2 (for C4, $\phi_0$=4.28\,rad).\\   
    We would like to indicate here that all the five superluminal components
    move along the precessing common trajectory only in their inner trajectory
    sections. In the outer trajectory sections they move along their own
    individual trajectories. In order to model-fit their outer trajectories
     we shall  introduce changes in parameters $\epsilon$ and $\psi$ 
    (which define the modeled jet-axis direction) only,  without introducing
    any changes in the common helical trajectory pattern.
     \subsection{A note on the model-parameters}
      During the model-simulation of the kinematics of superluminal components 
     in 3C345 the values chosen for the model parameters and the forms for
     the associated functions were not statistical samples
     and not unique. They were only specific and physically applicable sets
     of working ingredients which were obtained through trial and error methods
    over the past few years. But it was shown  that they could be used to
     analyze the  distribution of the observed trajectories and kinematics of 
     superluminal components in blazar 3C345 on VLBI-scales, especially 
     discovering the possible division of its superluminal knots into 
     two groups with different kinematic properties. Our methods
     aimed at: (1) seeking for the possible jet-precession and determining 
     the precession period; (2) searching for a double-jet structure; 
     (3) disentangling the observed superluminal components into two groups 
     attributed to respective jets; (4) investigating the relation between
     knots' flux evolution and Doppler boosting effect; (5) studying the 
     properties of the putative supermassive binary  black holes in its 
     nucleus.  \\
     Similar methods have also been applied to blazar 3C279 (Qian et al. 
     \cite{Qi18a}), OJ287
     (Qian \cite{Qi18b}) and 3C454.3 (Qian et al. \cite{Qi21}). Interestingly,
     we tentatively found that all the four blazars seemed to have 
     double-jet structures.\\ 
     Since in our works model-simulations of the kinematics of superluminal 
     components involved multi-parameters and multi-functions, we introduced 
     a new criterion to judge the validity of the model-fitting
     results, instead of ordinarily used statistical errors estimated for the
     model parameters. That is, a reasonable and effective model-fitting of 
     the kinematics of a superluminal knot was required to satisfy the 
     condition that its observed inner-trajectory had to be fitted to 
     follow the precessing common trajectory predicted by the scenario 
    within $\pm$5$\%$ of the precession period. A good example is presented 
    in Figure 15 for knot C9 below.\\ 
    In addition, since the outer-trajectory of the superluminal components
    deviated from the precessing common trajectory pattern the model-fits to
    their outer-trajectories were performed through changing the parameters
     ($\epsilon$, $\psi$) which defined the modeled jet-axis direction,
     while the  common helical pattern was kept unchanged.\\
    As shown in the previous paper (Qian \cite{Qi22}) the kinematics of
     the superluminal components of both jets in 3C345 could be  well 
    model-fitted and  a precession period of 7.30$\pm$0.36\,yr for both 
    the jet-nozzles were derived.\\
    In Figure 3 are shown the distribution of the precessing common 
    trajectory for jet-A of 3C345 at different precession phases (left panel)
    and the observed trajectories of the superluminal knots C4, C5 and C13
    (right panel) for comparison.
     \section{Basic equations for Doppler boosting effect and flux evolution}
      In order to investigate the relation between  the flux evolution of
      superluminal components and their Doppler boosting effect during their
      accelerated/decelerated motion along helical trajectories,
      the observed flux density $S_{obs}(\nu,t)$ of superluminal components
      can be defined as the Doppler-boosted flux density:
     \begin{equation}
   {S_{obs}}(\nu,t)={S_{int}}(\nu){\times}{\delta(t)}^{3+\alpha}
     \end{equation}
     $S_{int}(\nu)$=$S_0$\,${({\nu/{\nu_0})}^{-\alpha}}$--the 
    intrinsic flux density, $S_0$--the flux density at the fiducial frequency
    $\nu_0$, and $\alpha$--spectral index for the whole range of the observing
   frequencies (10--43\,GHz), $\delta(t)$--Doppler factor.\,
    \footnote{In most general case
    both $S_{int}$ and $\alpha$ could be defined as functions of $\nu$ and t
    , and thus Doppler-boosting profiles varying with time and 
    frequency, and complex flux evolution could be investigated. But 
    here in this paper we
    assume that $S_0(t)$ and $\alpha(\nu,t)$ are constants, not depending on
    time and frequency.}\\
     We also use the normalized flux density ${S_{obs,N}}(\nu,t)$ which is 
    defined as
     \begin{equation}
       {S_{obs,N}}(\nu,t)={S_{obs}}(\nu,t)/{S_{obs,max}(\nu)}
     \end{equation}
      $S_{obs,max}(\nu)$--the observed maximum flux density.\\
      The Doppler-boosting profile (a normalized profile) is defined as:
      \begin{equation}
      {S_D}(t)=
        [\delta(t)/\delta_{max}]^{3+\alpha}
      \end{equation}
     Here $\delta_{max}$ -- the modeled maximum Doppler factor.\\
    We also use the intrinsic flux density, approximately defined as: 
    ${S_{int}}(\nu)$=${S_{obs,max}(\nu)}$/$[\delta_{max}]^{3+\alpha}$.\\
    In the precessing nozzle scenario the Doppler factor $\delta(t)$ of 
    superluminal knots depends on their  motion along helical
    trajectories which are produced by the precessing common trajectory at
    corresponding precession phases (or ejection epochs).\\
    As already shown in the previous
    paper, the kinematics of the superluminal components was well  
    model-simulated, and their bulk Lorentz factor $\Gamma(t)$ and
    Doppler factor $\delta(t)$ as functions of time were predictively derived,
    which  could be used to investigate the relation between the Doppler
    boosting effect and their flux evolution. We found that
    in order to correctly model-fit their flux evolution,
    the bulk Lorentz factor and Doppler factor previously derived should be
    appropriately modified. Moreover, in order to investigate the flux 
    evolution at multi-frequency on the whole trajectory (in both inner
    and outer trajectory sections), variations in parameters $\epsilon$ and 
    $\psi$, intrinsic flux density $S_{int}$ and spectral index
     $\alpha$ should also be taken into consideration.     
     \begin{figure*}
     \centering
     \includegraphics[width=5cm,angle=-90]{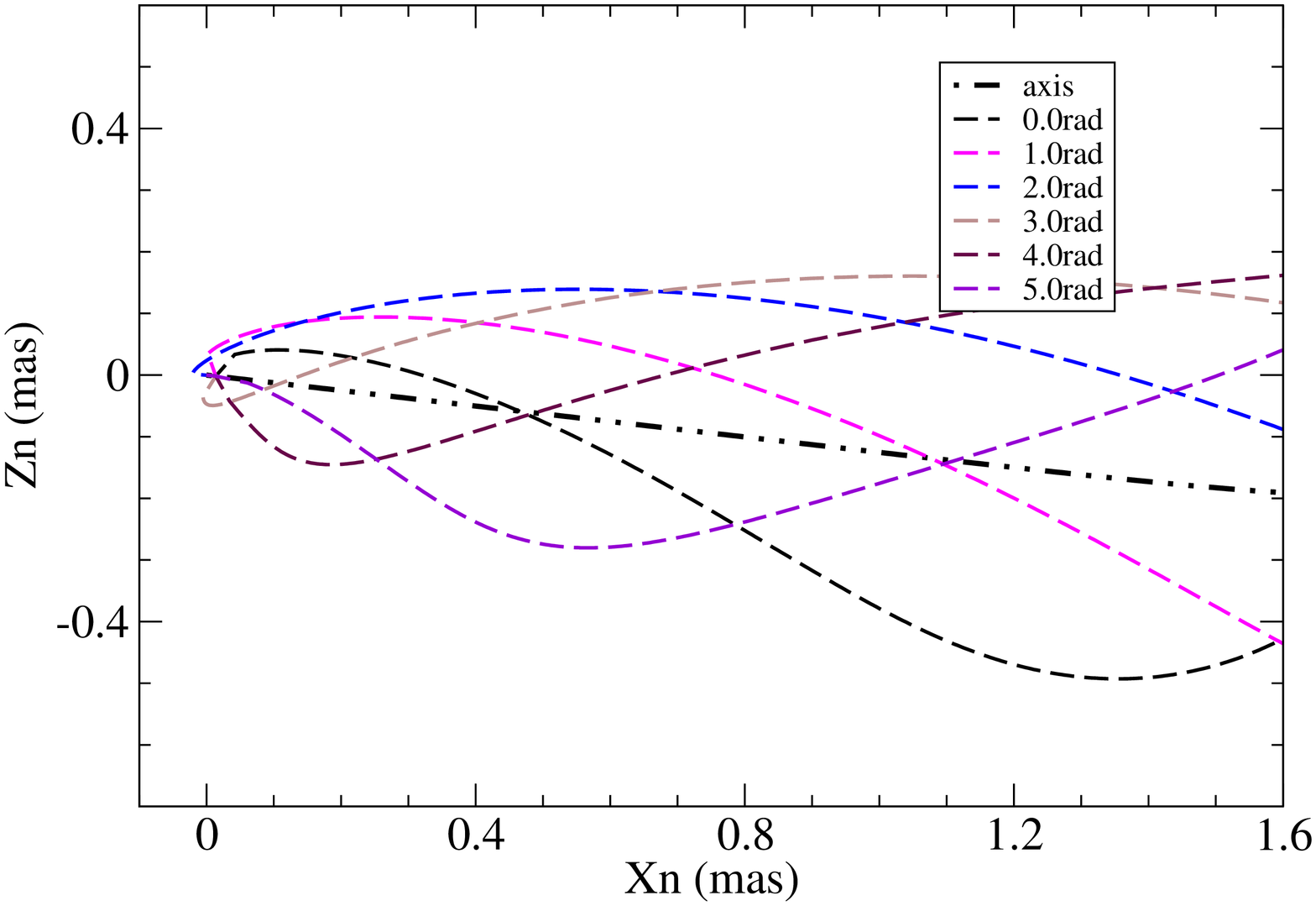}
     \includegraphics[width=5cm,angle=-90]{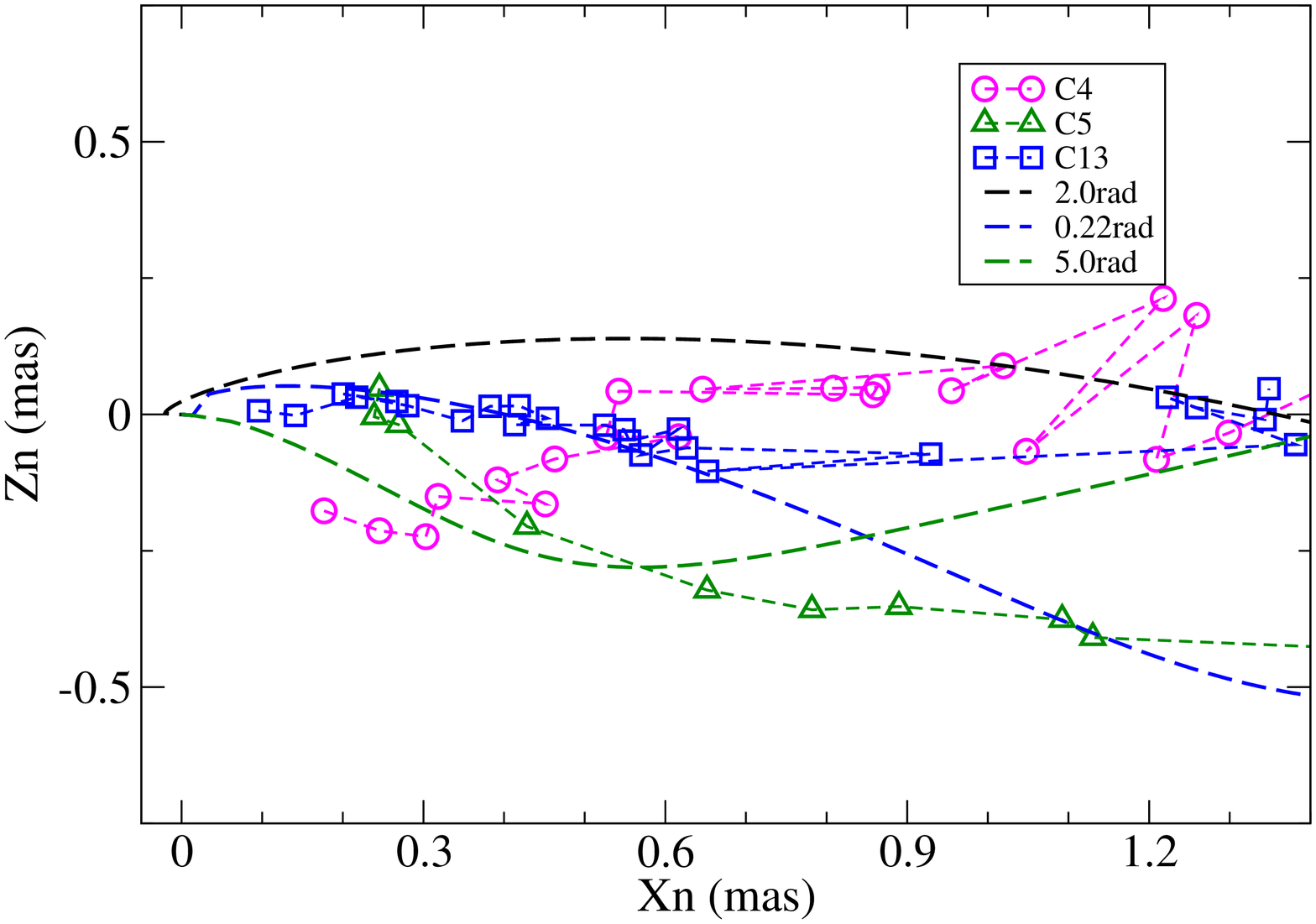}
     \caption{Distribution of the precessing common trajectory for jet-A in 
    3C345 (left panel).
    The jet axis is at position angle $\sim{-97.2^{\circ}}$ with its cone 
    aperture $\sim{42.5^{\circ}}$ (at core separation 0.5\,mas). The opening 
    angle of the jet is $\sim{1.12^{\circ}}$ in space. The observed 
    trajectories of knots C4, C5 and C13 are presented in the right panel for 
    a comparison with the model distribution.}
     \end{figure*}
    \begin{figure*}
    \centering
    \includegraphics[width=5cm,angle=-90]{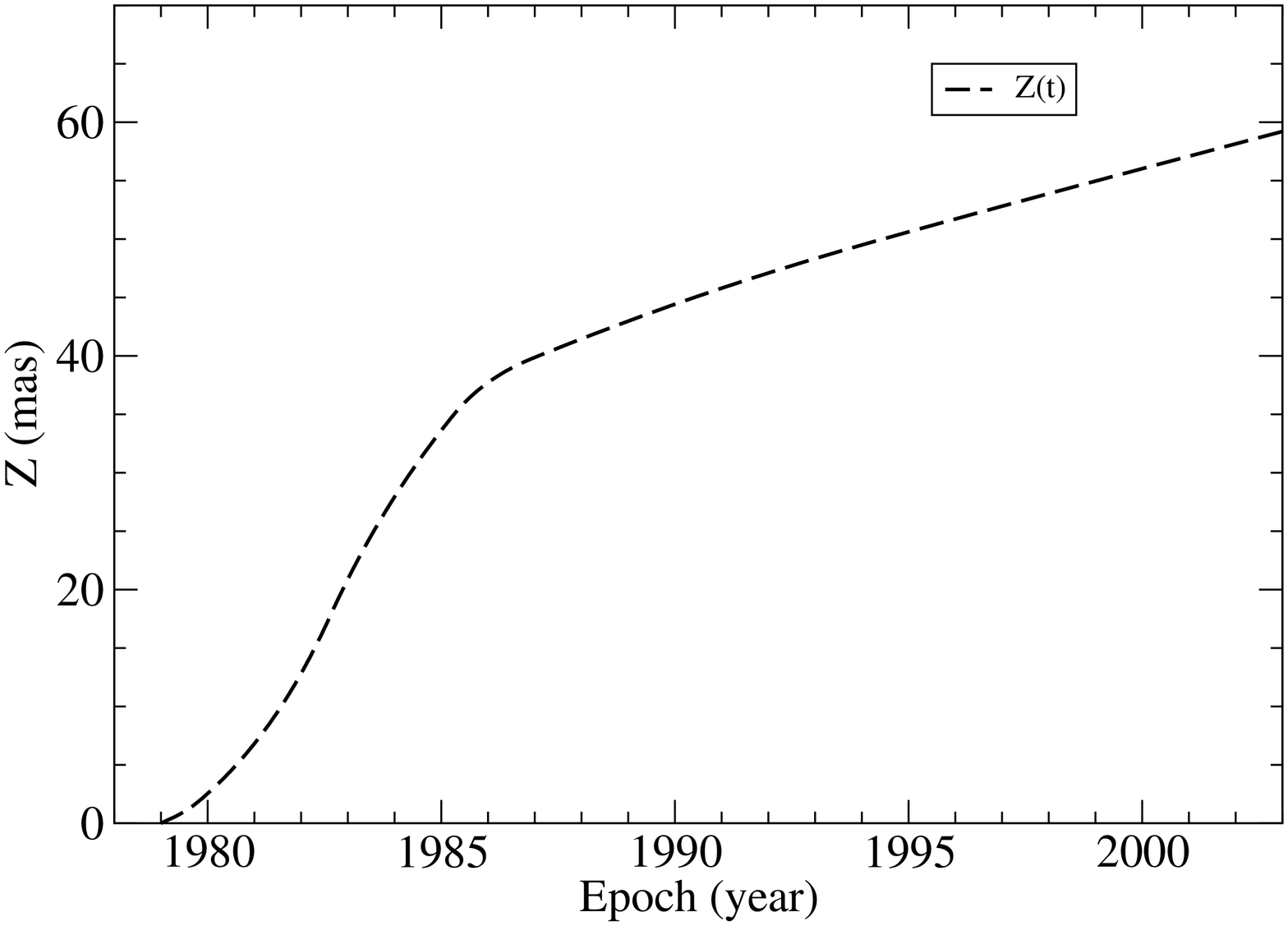}
    \includegraphics[width=5cm,angle=-90]{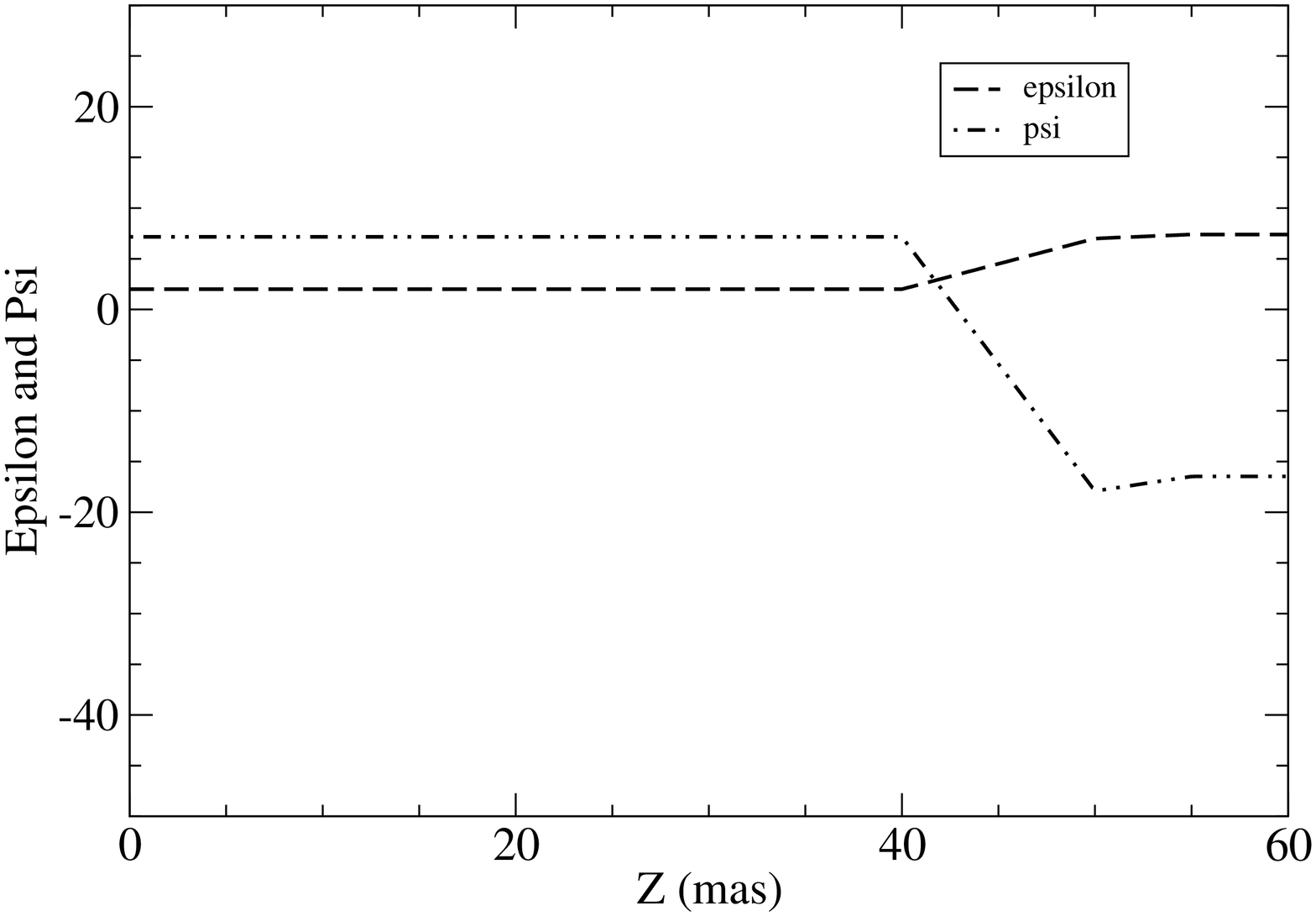}
    \caption{Knot C4. The traveled distance Z(t) along the Z-axis (left panel)
    and the  model parameters $\epsilon$ and $\psi$
    as functions of time. Within Z=40.0\,mas (t$\leq$1987.08) 
    $\epsilon$=$2.0^{\circ}$ and
    $\psi$=$7.16^{\circ}$, and  knot C4 moved along the precessing common
    trajectory, while beyond Z=40.0\,mas $\epsilon$ increased and $\psi$
    decreased and knot C4 moved along its own individual track.}
    \end{figure*}
    \begin{figure*}
    \centering
    \includegraphics[width=6cm,angle=-90]{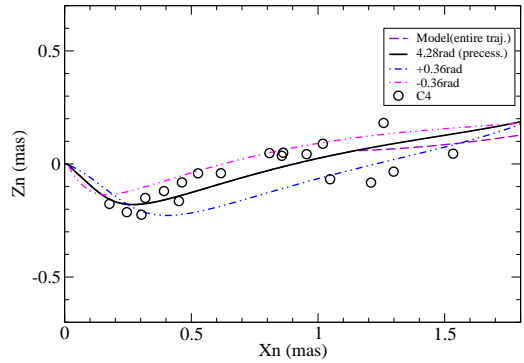}
    \caption{Knot C4: Model-fit to its inner trajectory 
    within $X_n{\simeq}$1.14\,mas. The black line describes the precessing 
    common trajectory-section with the  precession phase $\phi_0$=4.28\,rad and
    the corresponding ejection time $t_0$=1979.0. The lines in magenta and blue
    show the precessing common trajectories for precession phases 
    $\phi_0$+0.31\,rad and $\phi_0$-0.31\,rad (i.e., $\pm$5\% of the precession
     period), respectively. The curve in violet represents the model-fit to the
    whole trajectory.}
    \end{figure*}
    \begin{figure*}
    \centering
    \includegraphics[width=4.5cm,angle=-90]{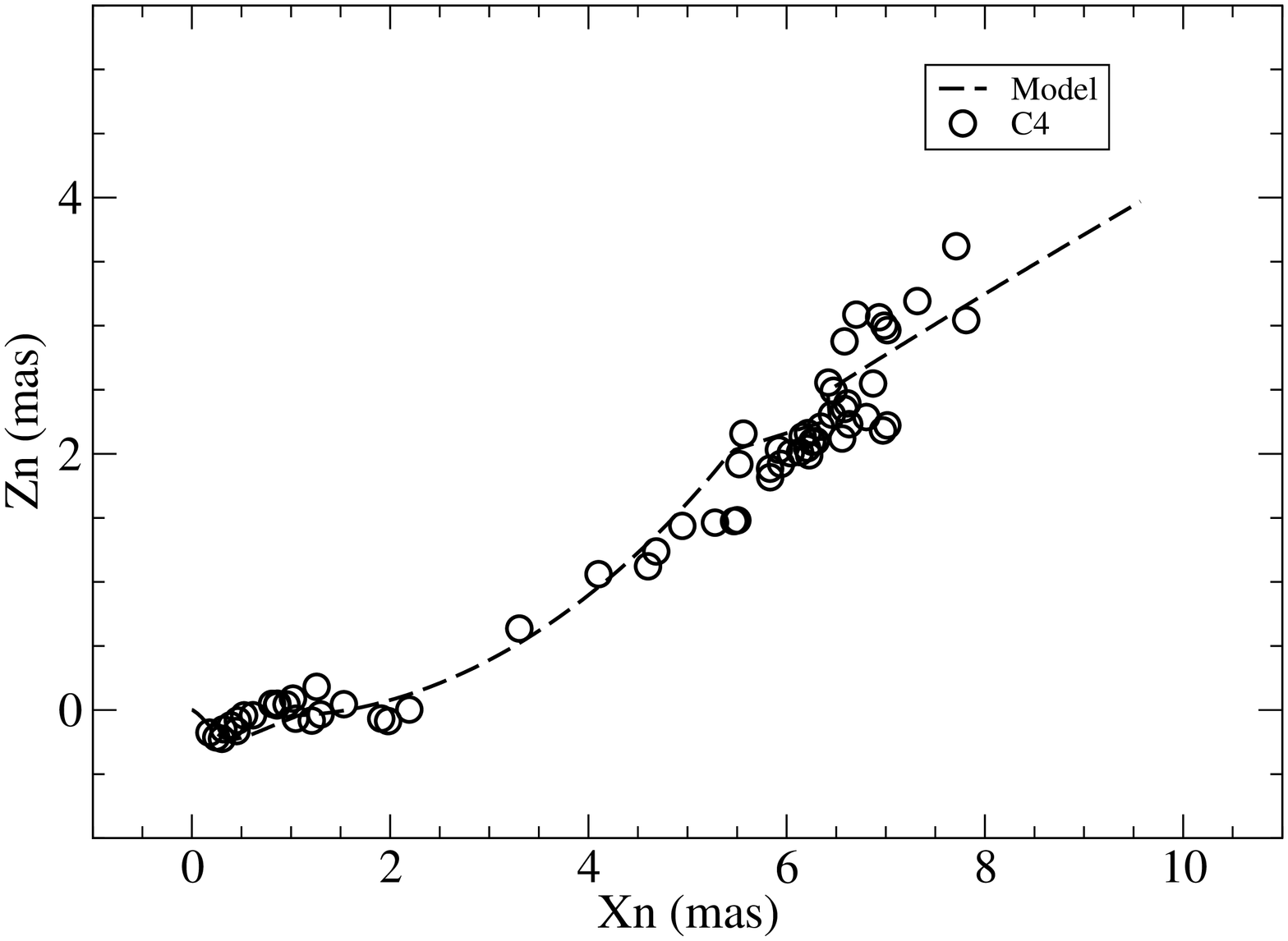}
    \includegraphics[width=4.5cm,angle=-90]{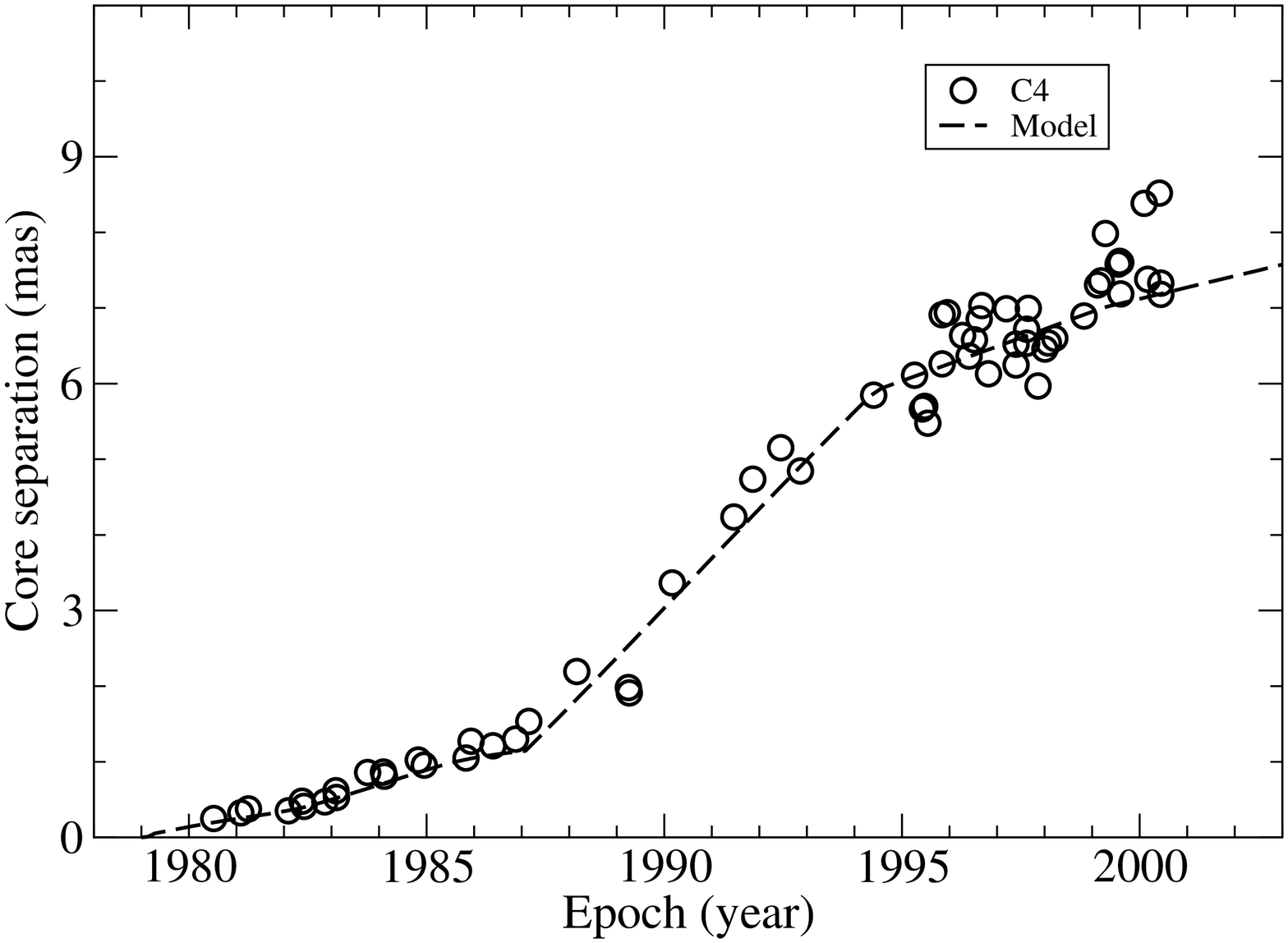}
    \includegraphics[width=4.5cm,angle=-90]{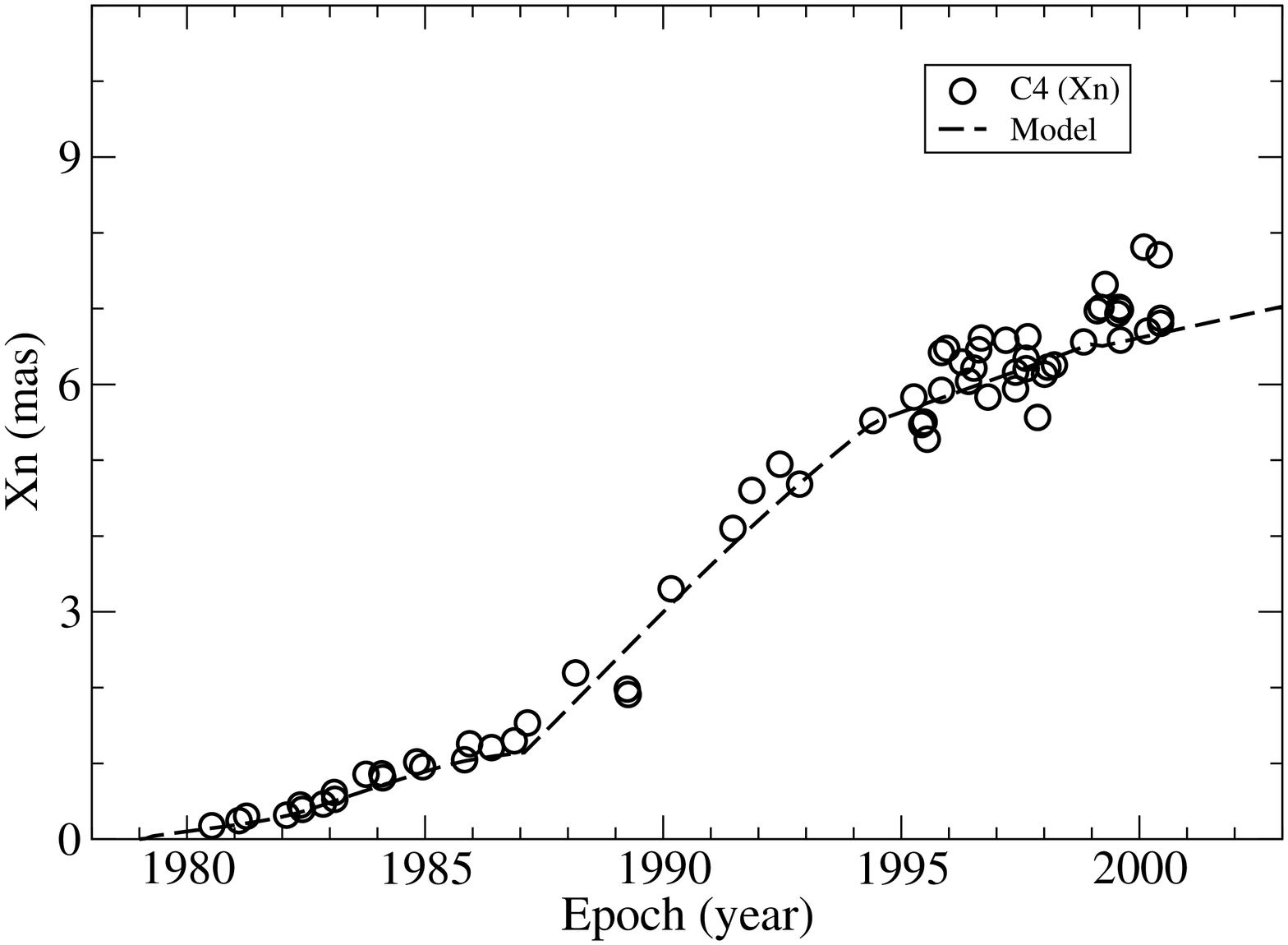}
    \includegraphics[width=4.5cm,angle=-90]{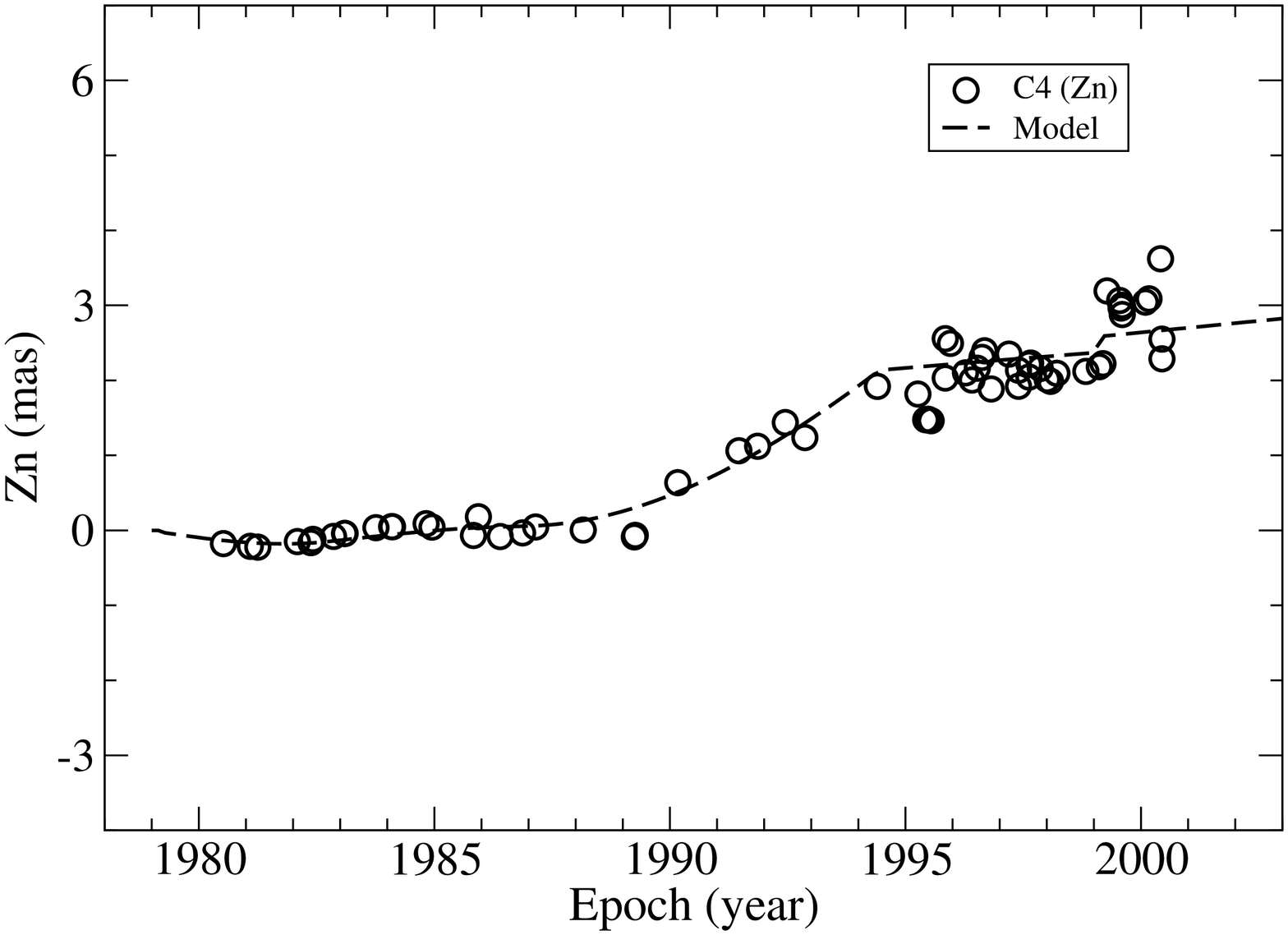}
    \includegraphics[width=4.5cm,angle=-90]{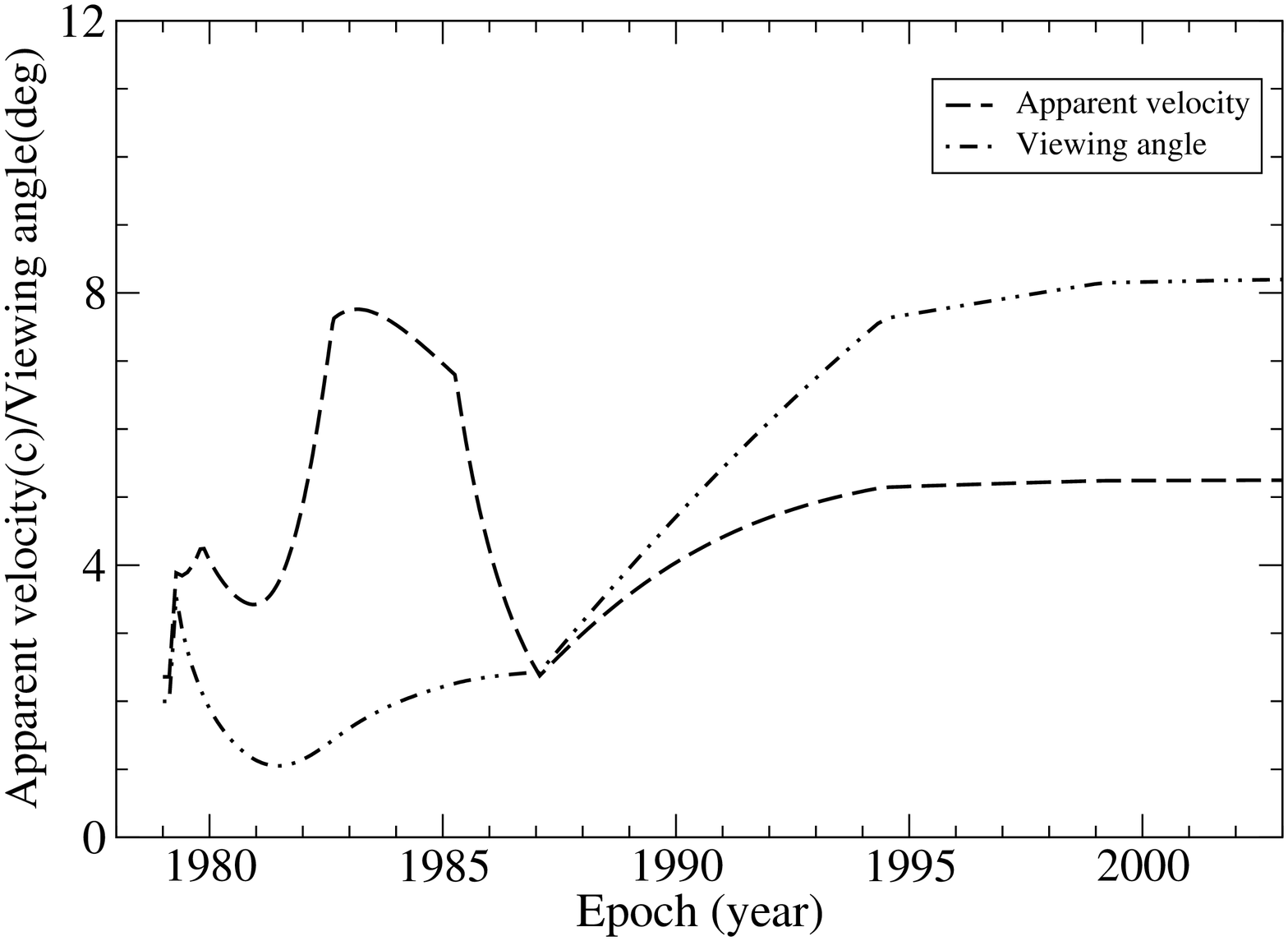}
    \includegraphics[width=4.5cm,angle=-90]{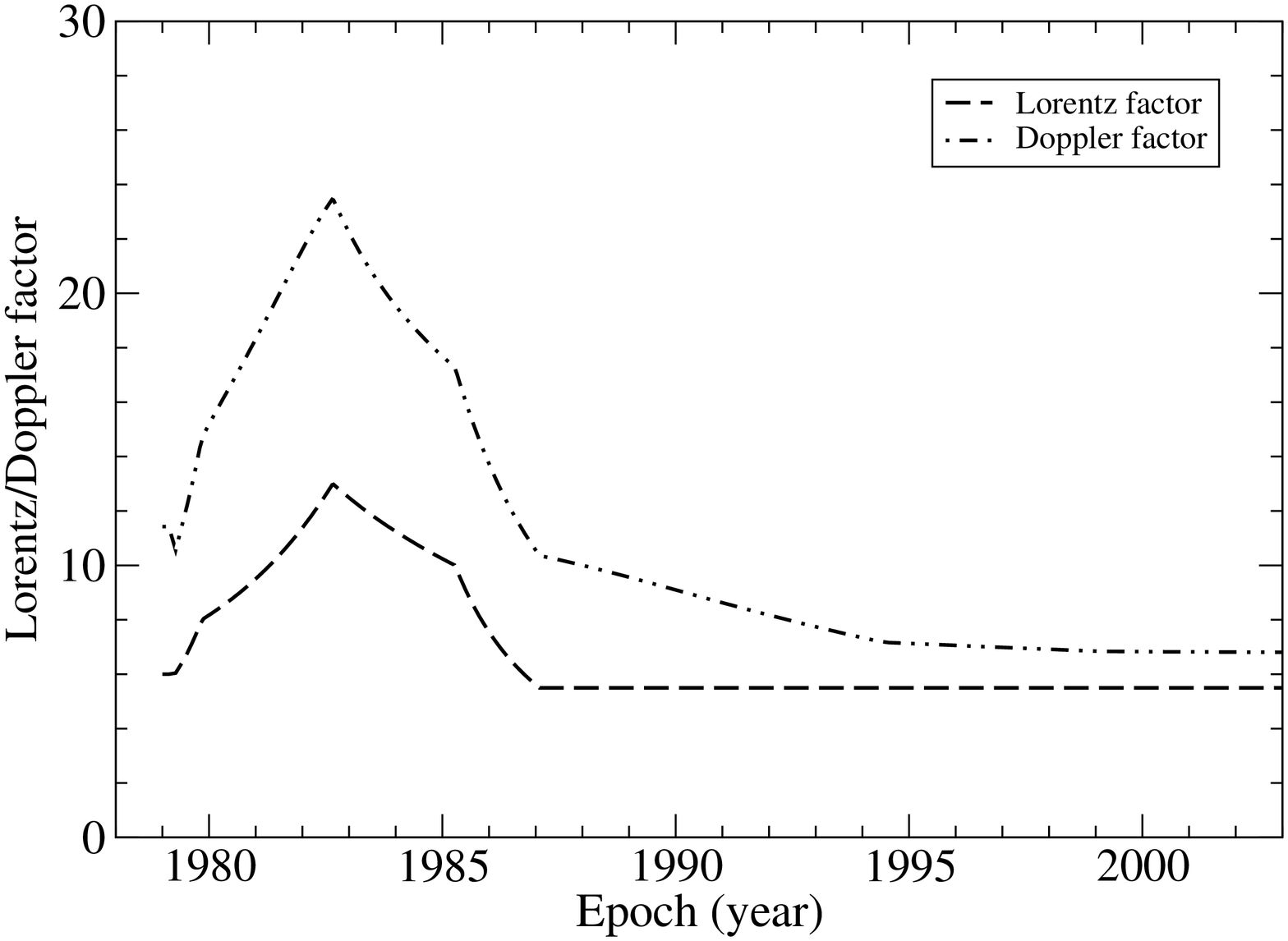}
    \caption{Knot C4: Model fits to the entire kinematics of knot C4  
    (within $r_n{\simeq}$8\,mas), including its whole trajectory, core 
    separation and coordinate $X_n$ (upperthree panels), and coordinate $Z_n$,
    the model-derived apparent speed and viewing angle, and the model-derived
    bulk Lorentz factor and Doppler factor (bottom three panels).}
    \end{figure*}
    \section{Interpretation of kinematics and flux evolution for knot C4} 
    As shown in previous paper (Qian \cite{Qi22}),
     the entire kinematics of knot C4 with its observed core separation
    extended to  
    $\sim$8\,mas could be well model-simulated in terms of our precessing
     nozzle model. However, only its inner trajectory within 
    a core-separation of $\sim$1.8\,mas followed the precessing common 
    trajectory pattern. Thus in order to  model-fit its entire kinematics and
     investigate its flux evolution we had to appropriately
     modify the previously derived bulk Lorentz factor and model-parameters
     $\epsilon$ and $\psi$.\\
    \begin{figure*}
    \centering
    \includegraphics[width=6cm,angle=-90]{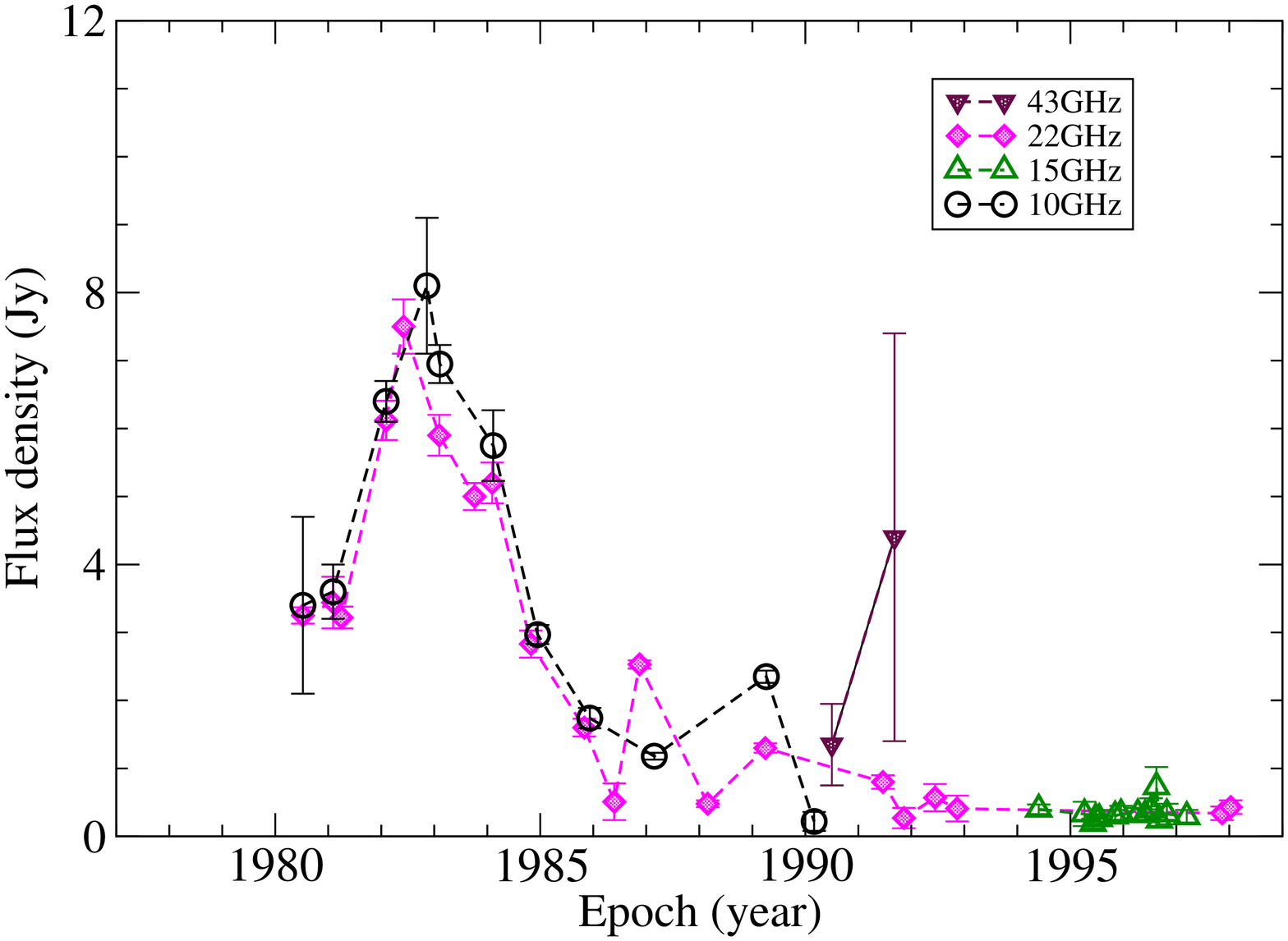}
    \caption{Knot C4: Light curves observed at 10\,GHz and 22\,GHz with 
    a few data-points observed at 15 and 43\,GHz.}
    \end{figure*}
     \begin{figure*}
    \centering
    \includegraphics[width=5cm,angle=-90]{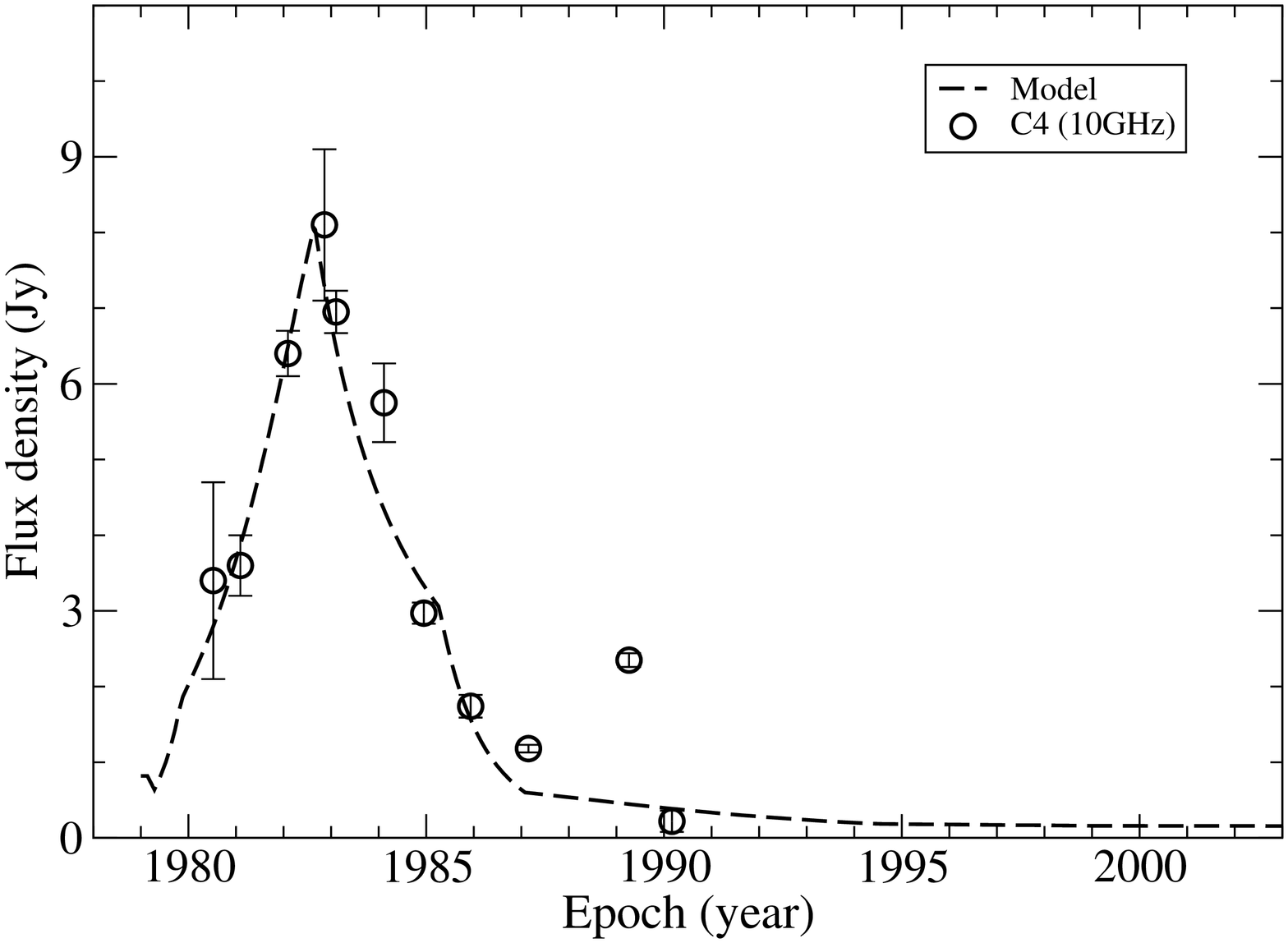}
    \includegraphics[width=5cm,angle=-90]{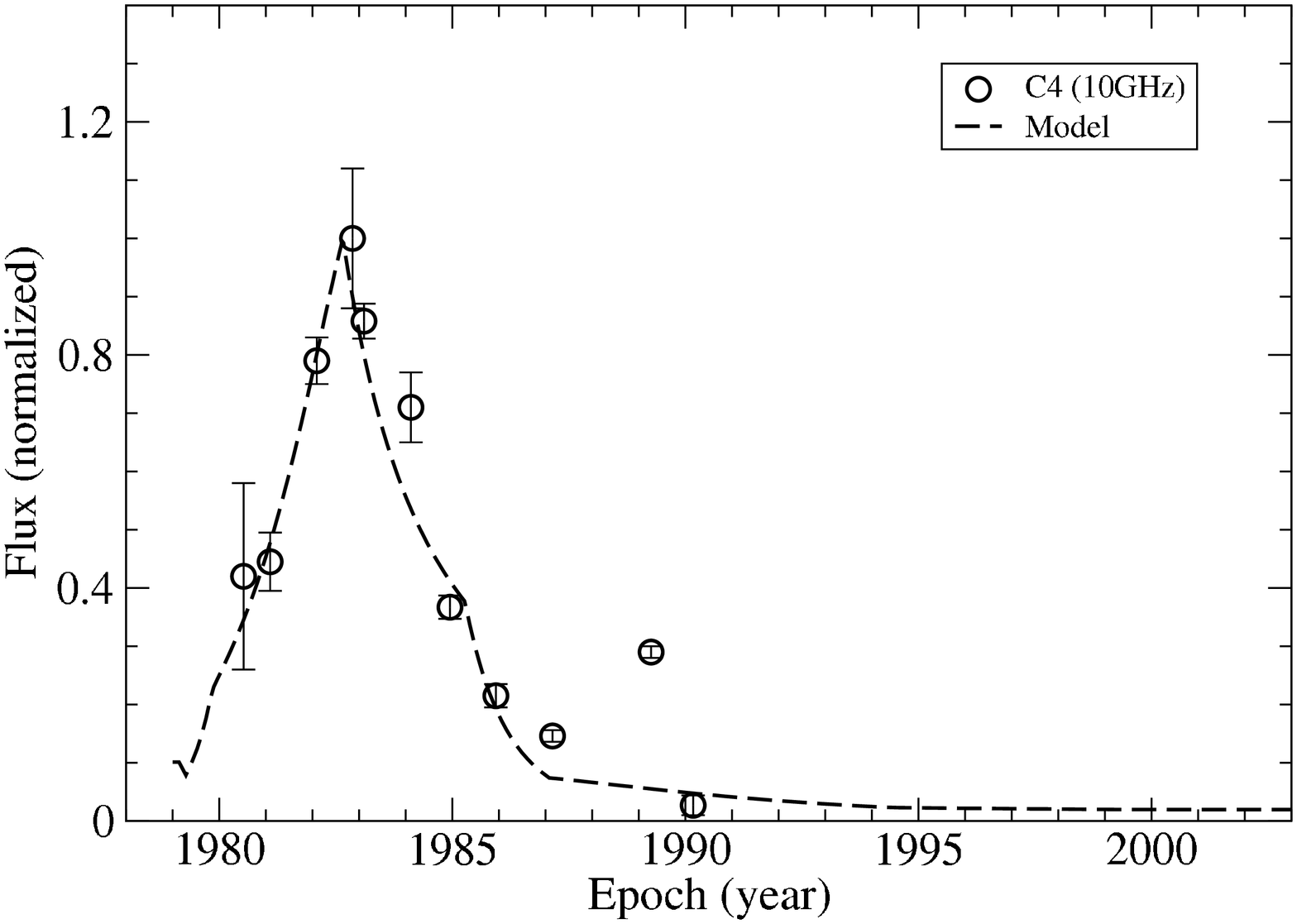}
    \includegraphics[width=5cm,angle=-90]{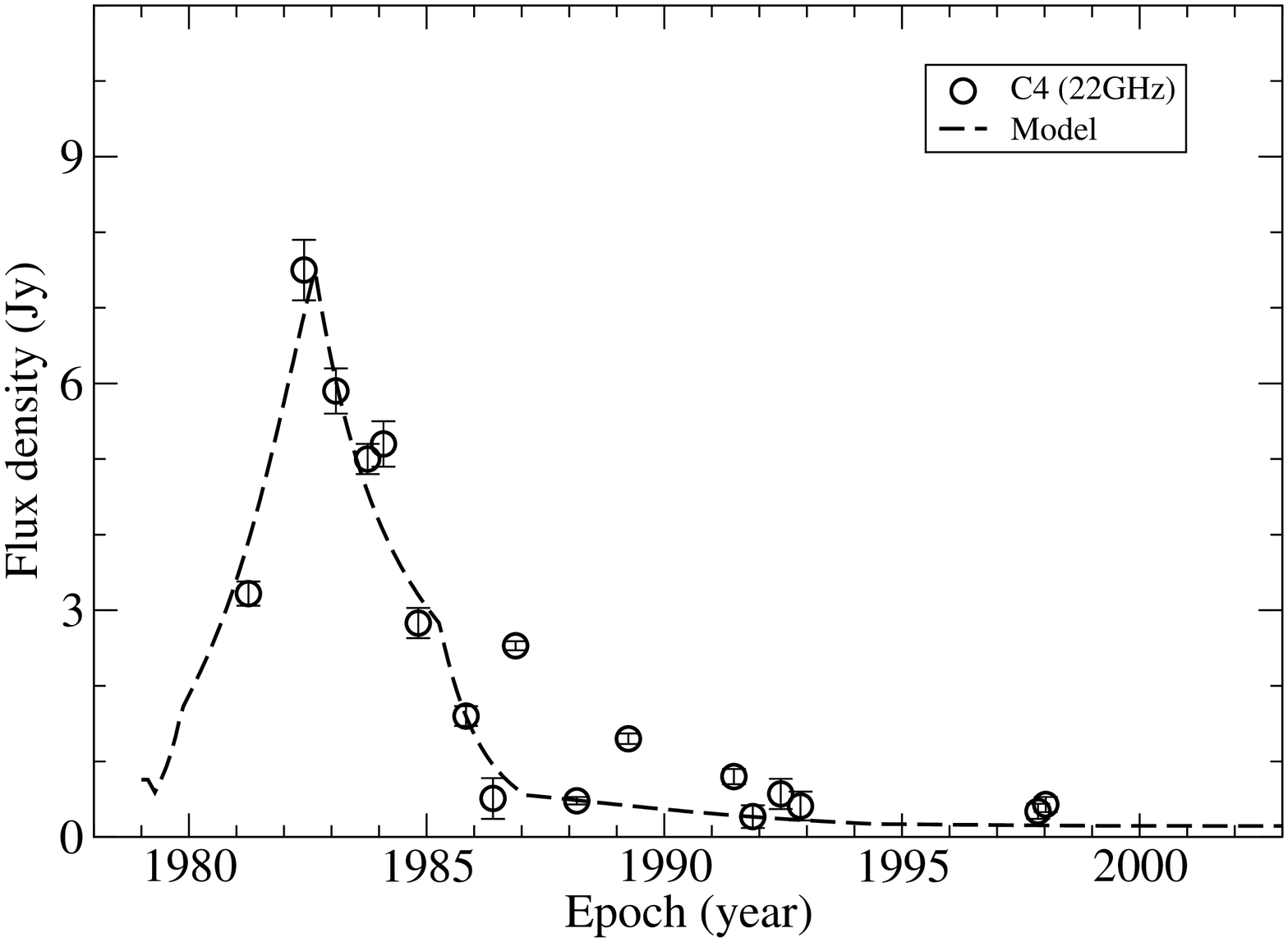}
    \includegraphics[width=5cm,angle=-90]{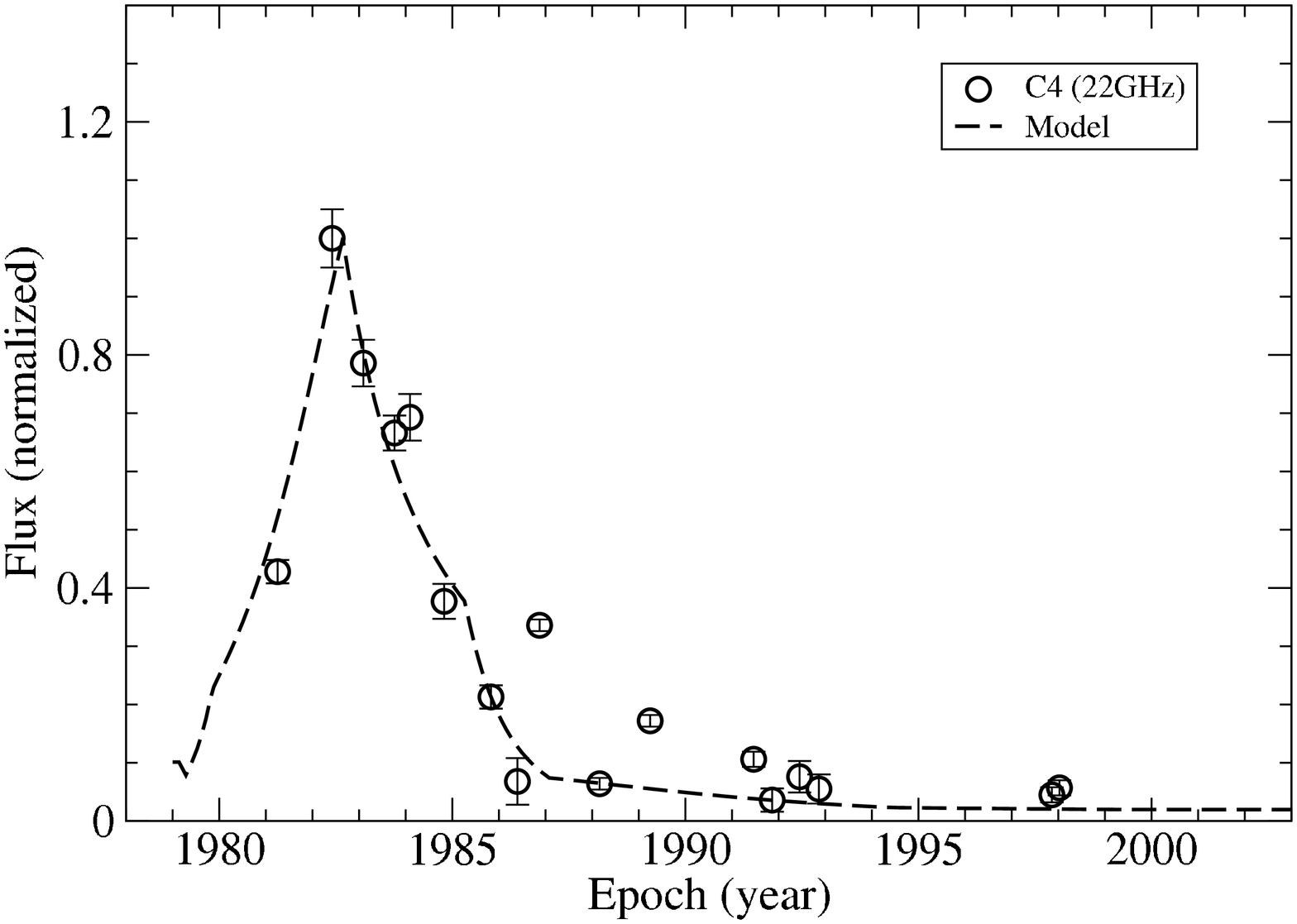}
    \caption{Knot C4: coincidence of the 10\,GHz and 22\,GHz light-curves
    with the Doppler boosting profiles: the observed light-curves (left two 
   panels)  and the normalized light-curves (right two panels).}
    \end{figure*}
     \subsection{Model simulation of kinematics for knot C4}
     In the present model the inner trajectory of knot C4 within 
    $X_n{\leq}$1.14\,mas ($r_n{\leq}$1.15\,mas, traveled distance 
    $Z_c{\leq}$40.0\,mas$\sim$266\,pc)
     could be well fitted by the precessing common trajectory pattern 
    (Fig.5). Its precession phase was modeled  as $\phi_0$=4.28\,rad 
    and ejection epoch $t_0$=1979.0.\\ 
     In Figure 4 are presented the traveled distance Z(t) along the Z-axis,
     and the curves of model parameters $\epsilon(Z)$ and $\psi(Z)$ 
    (right panel), which indicate that within
    Z$\leq$40.0\,mas (t$\leq$1987.08) $\epsilon$=$2.0^{\circ}$ and 
    $\psi$=$7.16^{\circ}$, and knot C4 moved along the precessing common 
    trajectory, while beyond Z=40.0\,mas $\epsilon$ started to increase and 
    $\psi$ stated to decrease, and knot C4 started to move along its own 
    individual trajectory (in its outer trajectory-section). The model-fit
    to its inner trajectory section is shown in Figure 5: the curves in black,
    magenta and blue represents the precessing common trajectories for 
    precession phases $\phi_0$=4.28\,rad and $\phi_0{\pm}$0.31\,rad, 
    respectively. The curve in violet represents the model-fit to the whole
    trajectory.\\
    Correspondingly, its bulk Lorentz factor was adjusted as follows in order
    to take its deceleration after 1982.67 into consideration:
     for Z$\leq$0.5\,mas 
    $\Gamma$=6.0; for Z=0.5-2.0\,mas $\Gamma$ increased from 6 to 8; for
    Z=2-18\,mas $\Gamma$ increased from 8 to 13 ; for Z=18-35\,mas 
    $\Gamma$ decreased from 13 to 10; for Z=35-40\,mas $\Gamma$ decreased 
    from 10 to 5.5; for Z$>$40\,mas $\Gamma$=5.5. $\Gamma_{max}$=13.0 at 
    1982.67, and the corresponding maximum Doppler factor 
    $\delta_{max}$=23.5.\\
     It can be seen from Figure 6 that the entire kinematics of knot C4
     could be well explained in terms of 
    our precessing nozzle scenario, including the whole trajectory $Z_n(X_n)$,
    core separation $r_n(t)$ and coordinate $X_n(t)$ (upper three panels),
     and coordinate $Z_n(t)$, the derived apparent velocity 
    $\beta_{app}(t)$/viewing angle $\theta(t)$ and the derived bulk Lorentz 
    factor $\Gamma(t)$/Doppler factor $\delta(t)$ (bottom three panels).\\
     The apparent speed derived showed two bumps (Figure 6, bottom/middle
     panel): one occurred before 1987.08
    in its inner-trajectory section where knot C4 followed the precessing
     common trajectory with its maximum speed of $\sim$8c at $\sim$1982.7, 
    while the other occurred during the period of 1995--2005 in its outer
    trajectory section where knot C4 moved along its own individual 
    trajectory. The derived viewing angle showed variations along a concave
    curve in the inner trajectory section, while it showed a bump structure
    in its outer trajectory section. This is the first time  to obtain a
    continuous curve for the apparent-velocity/viewing-angle of knot C4 during
    a time-interval of $\sim$20 years.\\
    Both the derived bulk Lorentz factor and Doppler factor showed only
    one-bump structure occurred in its inner trajectory section before 1987.08,
    i.e., it occurred when its motion followed the precessing common 
    trajectory (Figure 6, bottom/right panel), showing its intrinsic 
    acceleration/deceleration along the precessing common helical trajectory.
      \subsection{Doppler boosting effect and flux evolution of knot C4}
     The derived Doppler-factor curve shown in Figure 6 (bottom/right panel)
    reveals a distinct bump structure with its peak at $\sim$1982.6. 
   This is a very good case for yielding the Doppler boosting profile
     (normalized at its peak, $[{\delta(t)}/{\delta_{max}}]^{3+\alpha}$) and 
    studying the relation between its
    flux-density evolution and the Doppler boosting effect.\\
     The observed 10\,GHz and 22\,GHz light-curves of knot C4 are shown 
    in Figure 7. 
    Both the light-curves showed a peak flux density at $\sim$1983, closely
    corresponding to the peak in the Doppler factor curve 
    ($\delta_{max}$=23.5 at $\sim$1982.64). A spectral index
    $\alpha$(22-10GHz)=0.18 was assumed. The intrinsic flux densities were 
    assumed as: 3.54$\times{10^{-4}}$Jy (10\,GHz) and 3.28$\times{10^{-4}}$Jy
    (22\,GHz). \\
     The model-fits to the 10\,GHz and 22\,GHz light-curves in terms of
    Doppler-boosting effect are shown in Figure 8, respectively. It can be
    seen that the 10GHz and 22GHz light-curves were extremely well coincident
    with the Doppler boosting profiles (for both the observed flux
    light-curves and the normalized flux light-curves). 
    This implies that the flux-density variations  observed at 10\,GHz and
    22\,GHz were fully induced by its Doppler boosting effect, and the 
     variations in its intrinsic flux density were not detectable, i.e., the
     emission of the relativistic shock producing the superluminal component 
     C4 was very stable.\\
      The domination of Doppler boosting effect in the emission of knot C4
     strongly justifies the traditional point-view which is supported by
     most blazar physicists and VLBI-observers: superluminal components
     participate relativistic motion toward us along helical trajectories with
     acceleration/deceleration. This result strongly proves that our 
    precessing nozzle  scenario can interpret the entire VLBI-kinematics of
    knot C4 and  correctly derive its Doppler factor as a continuous 
    function of time  which can  be used to  interpret its flux
     density evolution. The explanation of the entire kinematics and emission
   properties for knot C4 in terms of our precessing nozzle scenario is very
   encouraging and instructive.\\
    \begin{figure*}
    \centering
    \includegraphics[width=5cm,angle=-90]{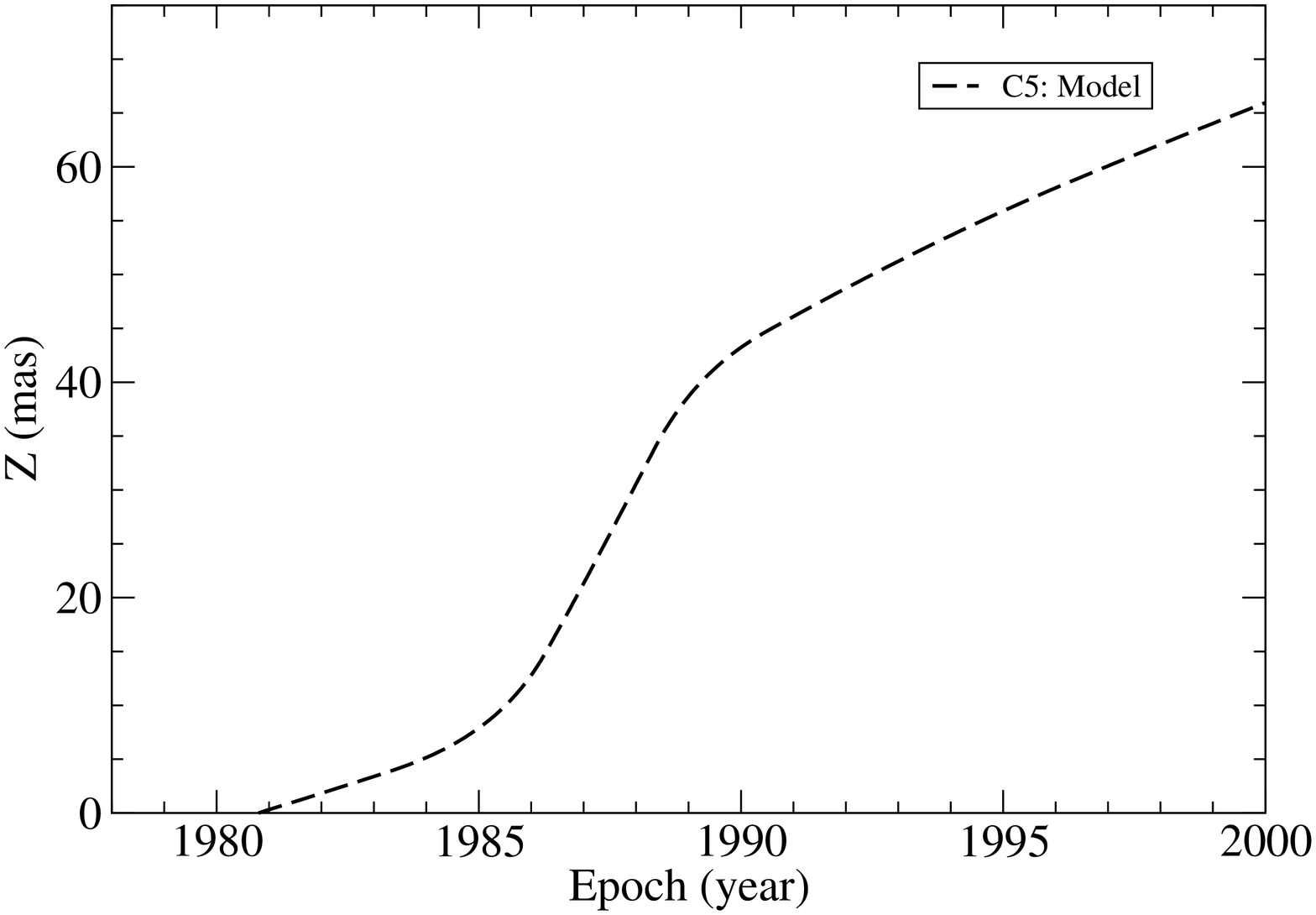}
    \includegraphics[width=5cm,angle=-90]{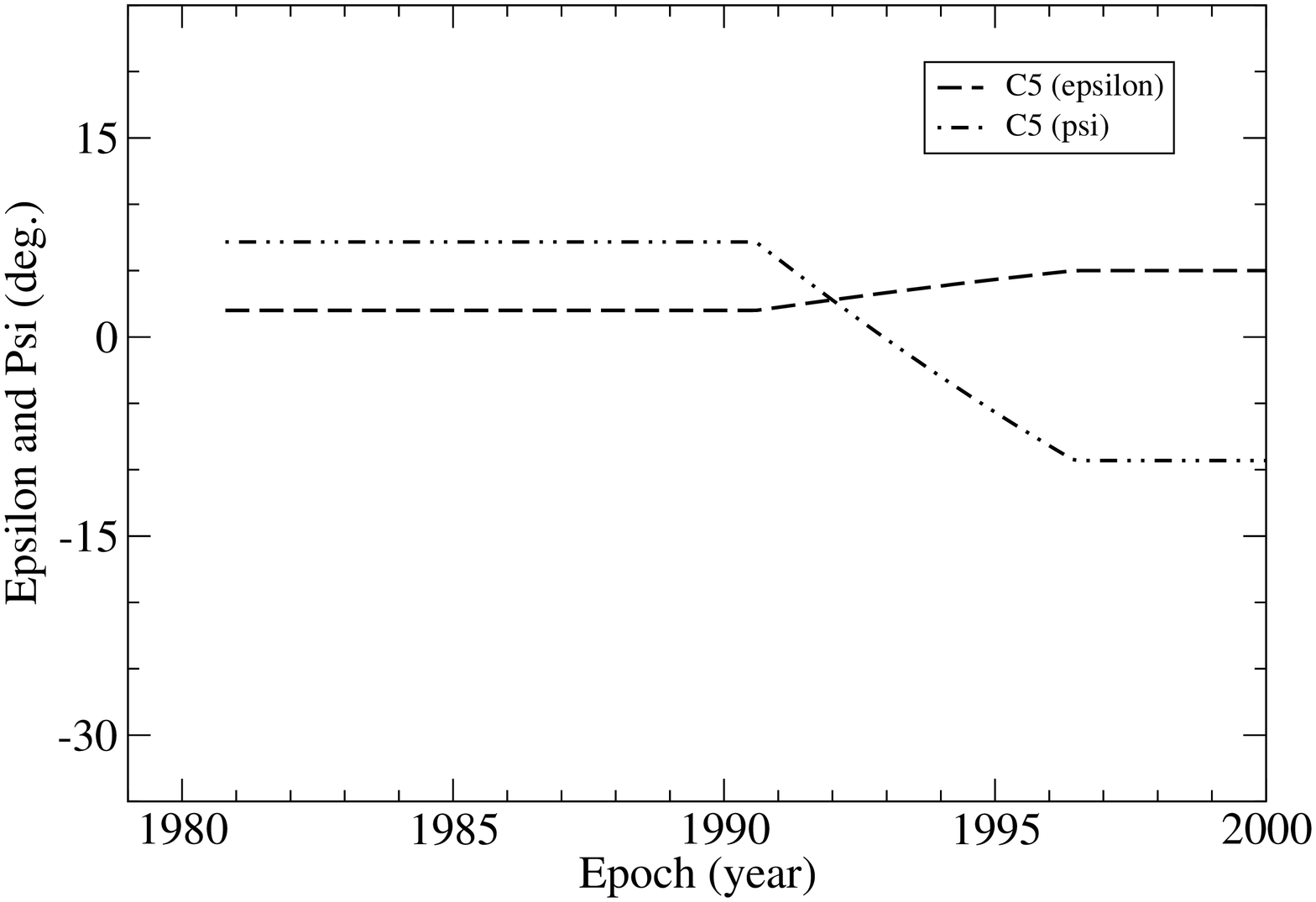}
    \caption{Knot C5. The traveled distance Z(t) along the Z-axis (left panel),
    and the curves of ${\epsilon}$(t) and ${\psi}$(t) (right panel).
    Within Z=44.8\,mas (t=1990.4) $\epsilon$=$2^{\circ}$ and 
    $\psi$=$7.16^{\circ}$, and knot C5 moved along the precessing common
    trajectory. Beyond Z=44.8\,mas it moved along its own individual 
    trajectory.}
    \end{figure*}
    \begin{figure*}
   \centering
   \includegraphics[width=6cm,angle=-90]{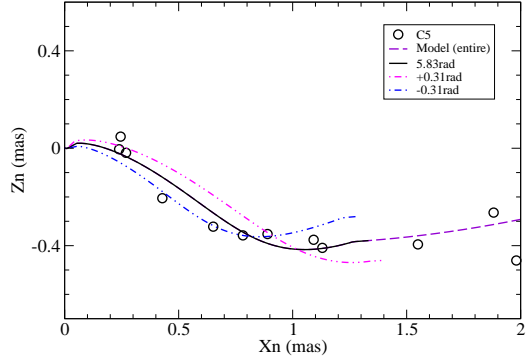}
    \caption{Knot C5: Model-fit to its inner trajectory section (within
    $r_n$=1.25\,mas (or $X_n$=1.21\,mas, t=1990.4). The curve in black
    represents the precessing common trajectory for precession phase
     $\phi_0$=5.83\,rad  (corresponding ejection time $t_0$=1980.80).
    The curves in magenta and blue represent the precessing common trajectories
     for precession phases $\phi_0$+0.31\,rad and $\phi_0$-0.31\,rad, 
    respectively.}
    \end{figure*}
   \begin{figure*}
   \centering
   \includegraphics[width=4.5cm,angle=-90]{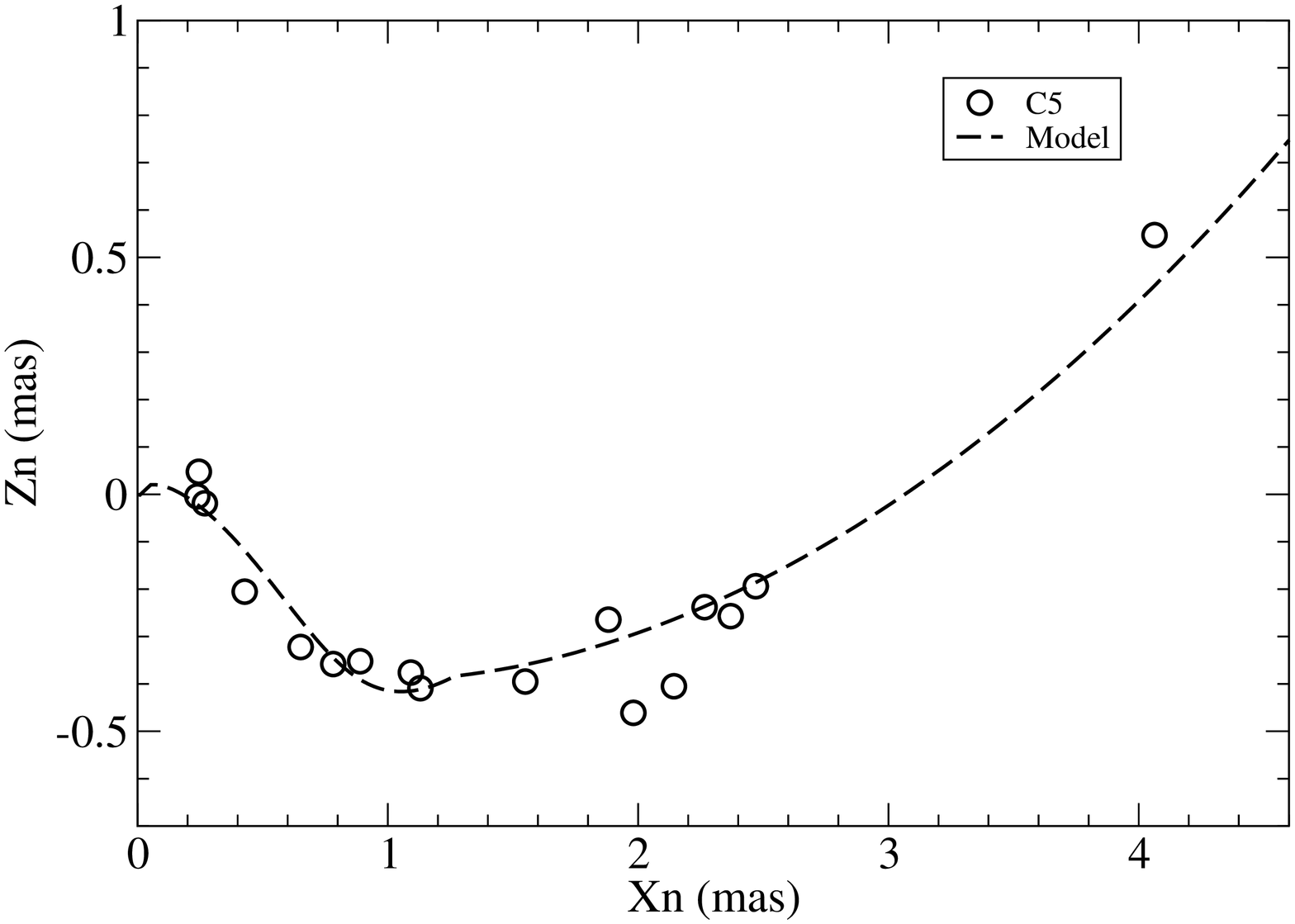}
   \includegraphics[width=4.5cm,angle=-90]{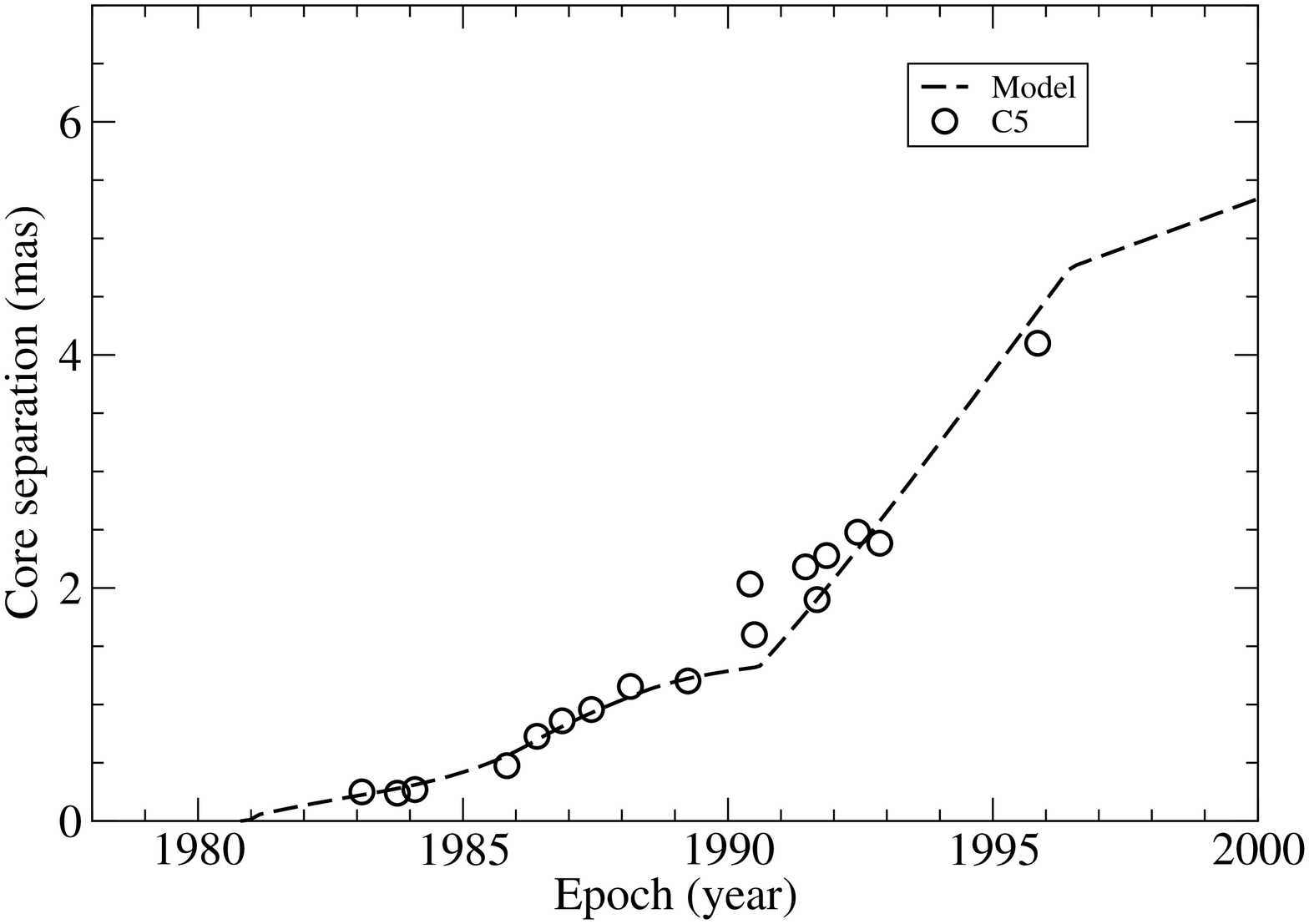}
   \includegraphics[width=4.5cm,angle=-90]{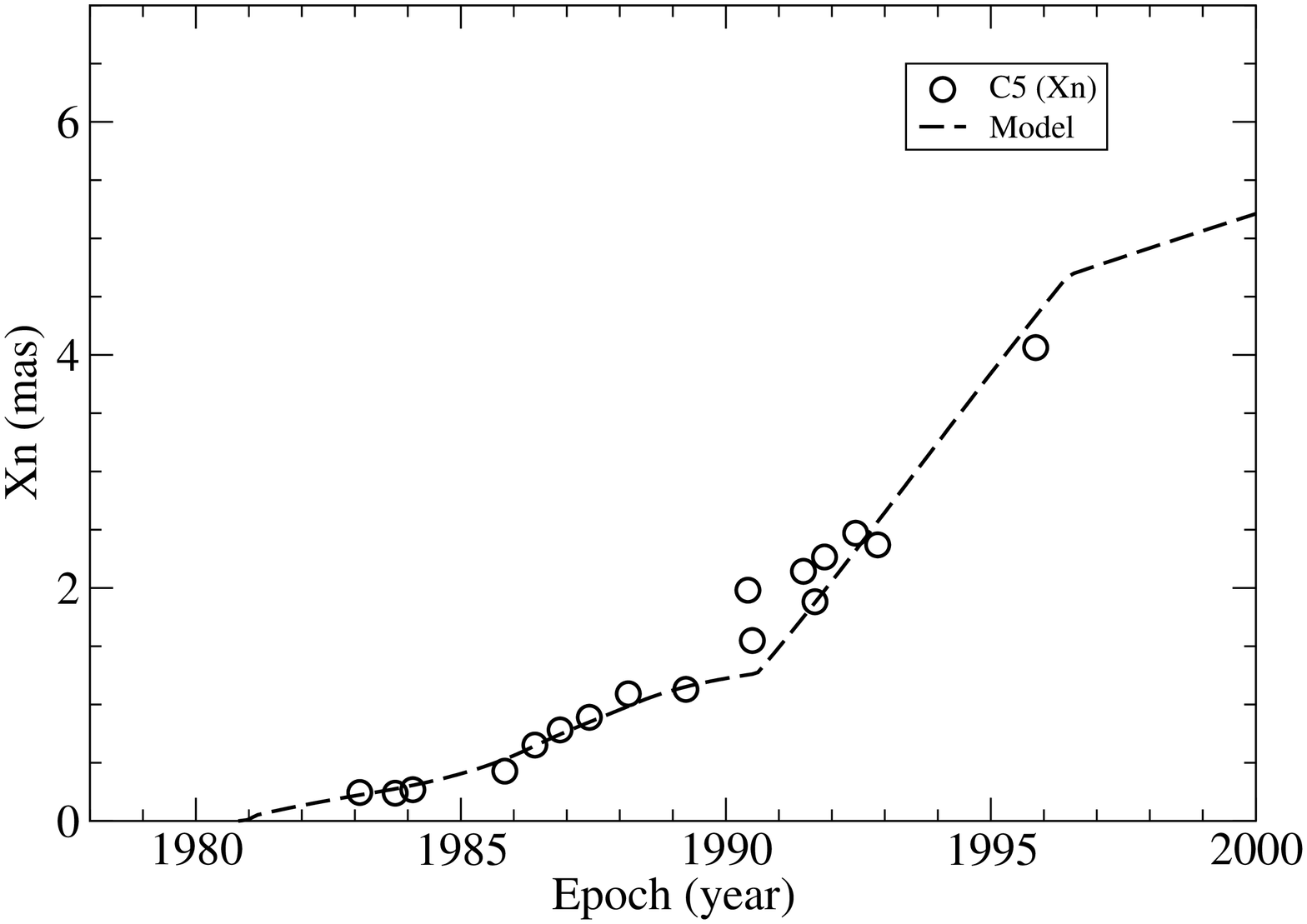}
   \includegraphics[width=4.5cm,angle=-90]{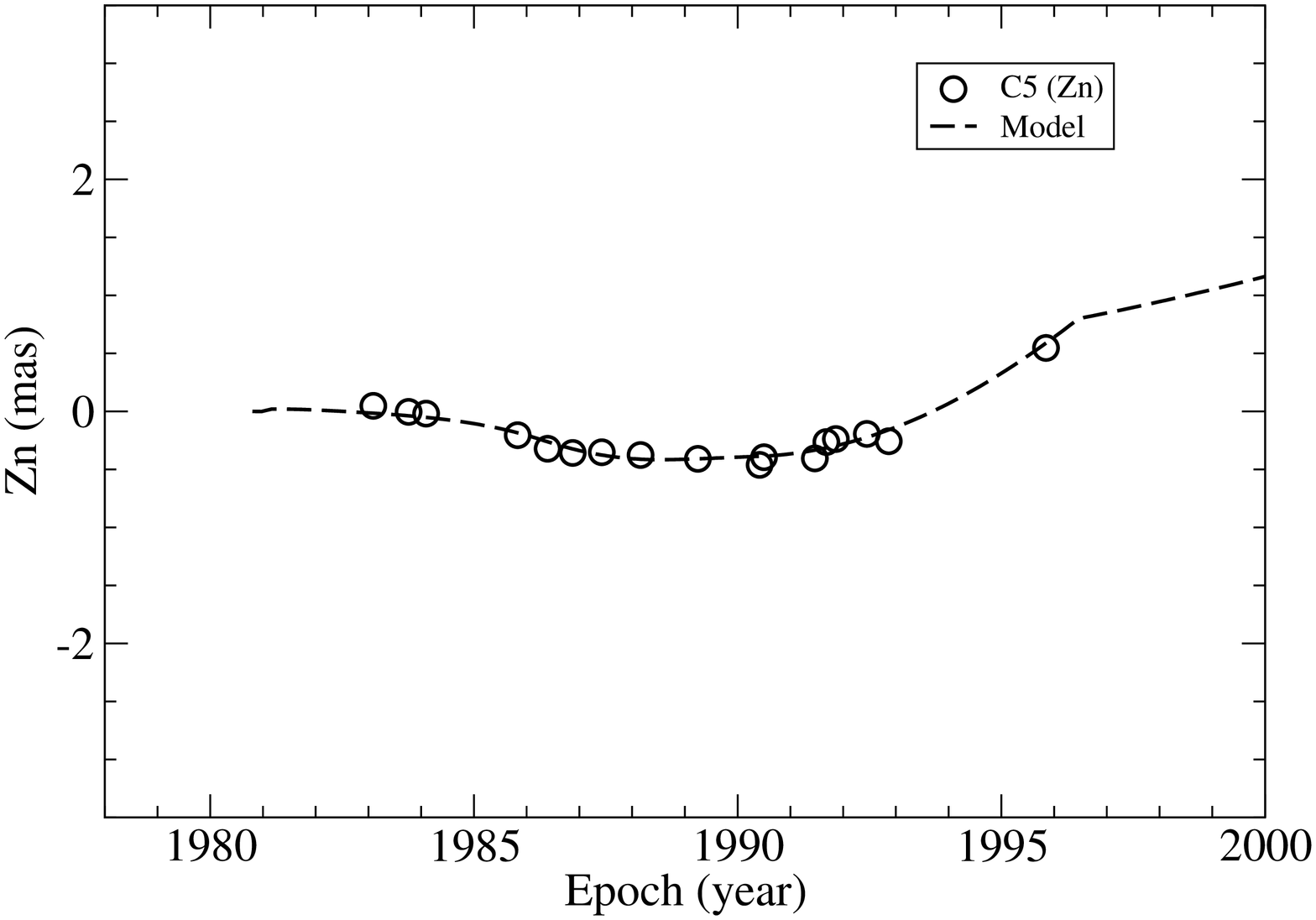}
   \includegraphics[width=4.5cm,angle=-90]{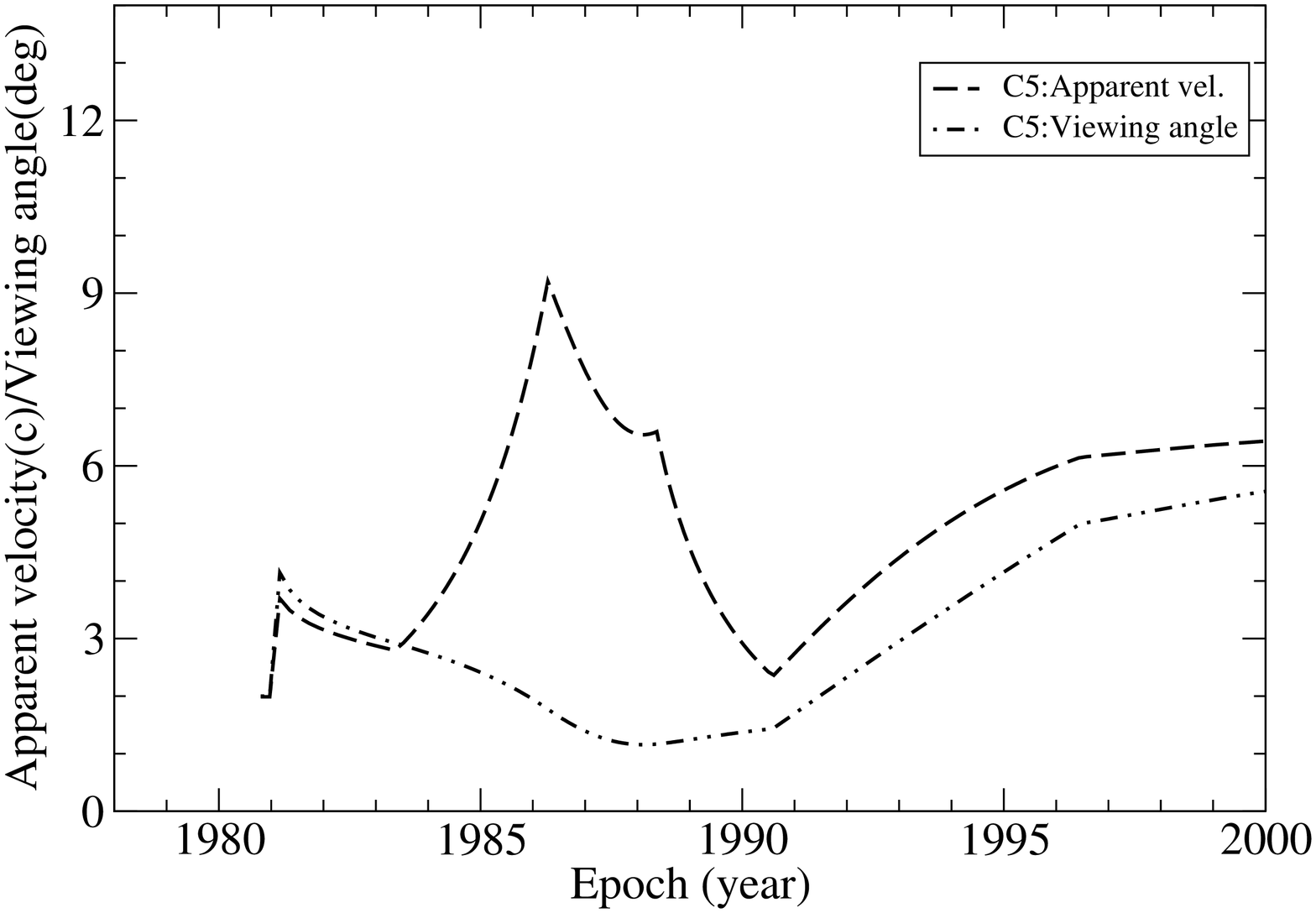}
   \includegraphics[width=4.5cm,angle=-90]{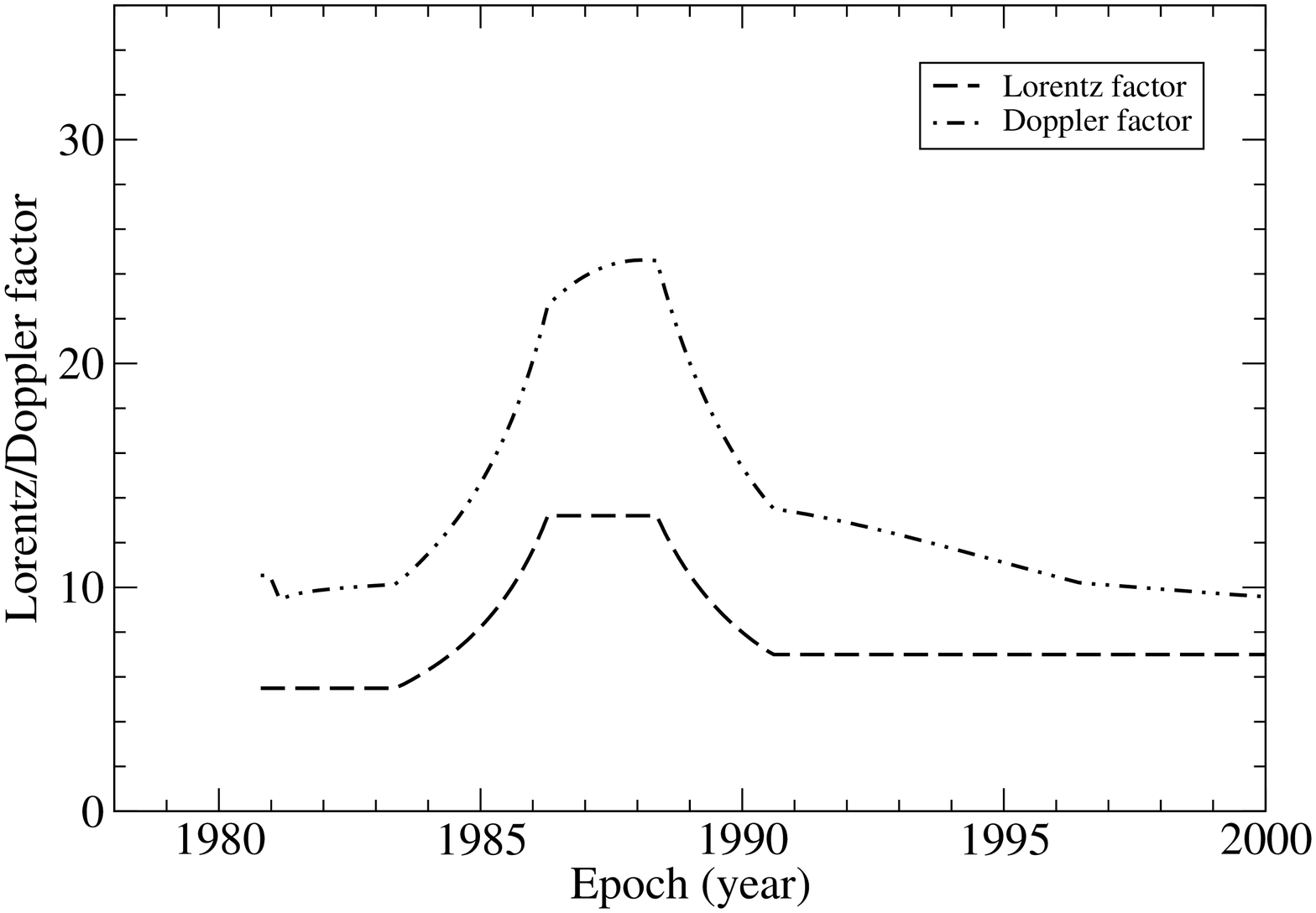}
   \caption{Knot C5. Model-fitting results: trajectory $Z_n$($X_n$), core 
    separation $r_n$ and coordinate $X_n$(t) (upper three panels), and 
    coordinate $Z_n$(t), the modeled apparent velocity ${\beta}_a$(t) and 
   viewing angle $\theta$(t), bulk Lorentz factor $\Gamma$(t) and Doppler
    factor $\delta$(t) (bottom three panels).
    Precession phase $\phi_0$=5.83\,rad and ejection time $t_0$=1980.80.}
  \end{figure*}
   \section{Interpretation of kinematics and flux evolution for knot C5}
   \subsection{Model simulation of kinematics for knot C5}
   For the interpretation of the kinematics of superluminal components in jet-A
   of 3C345 in terms of the precessing nozzle scenario the kinematic behavior
   of knot C5 is quite significant, because
   its  ejection epoch  $t_0$=1980.80 and  corresponding
   precession phase $\phi_0$=5.83\,rad were different from those of knot C9
   ($t_0$=1995.06 and $\phi_0$=5.54+4$\pi$) by about two precession periods
   (Comparing Figure 10 with Figure 15 in the next section for knot C9).
   Moreover, its apparent trajectory was very similar to that of knot C9, 
   demonstrating the recurrence of the precessing common trajectory and 
   confirming the basic assumption in our scenario: the superluminal components
   move along the  common trajectory which precesses to give rise to the
   individual trajectories of the superluminal knots at corresponding
    precession phases.\\
   In Figure 9 are shown the traveled distance of knot C5 along the Z-axis
   (left panel) and the curves of parameters $\epsilon$ and $\psi$, which
   indicate: during the period 1980.80-1990.42 (Z$\leq$44.8\,mas, 
   $X_n{\leq}$1.21\,mas, $r_n{\leq}$1.25\,mas) knot C5 moved along 
    the precessing common
   trajectory. After 1990.42 $\epsilon$ started to increase and $\psi$ started 
   to decrease, and knot C5 started to move along its own individual trajectory
   (in its outer trajectory section).\\ 
   The model-fit to its inner trajectory section is shown in Figure 10: the 
   curves in black, magenta and blue represent the precessing common 
   trajectories for precession phases $\phi_0$=5.83\,rad and 
   5.83$\pm$0.31\,rad ($\pm$5\% of the precession period), respectively.
   The curve in violet represents the fit to the whole trajectory.\\
     The model fits to its whole kinematic behavior are shown in Figure 11.
   It can be seen that its entire 
   trajectory, core separation, two coordinates are well model-simulated. The
   derived apparent speed showed a distinct bump during the period 
   $\sim$1983--1990, while the viewing angle varied along a depressed  curve.
   The derived curves of bulk Lorentz factor and Doppler factor had distinct
   bumps during 1986.2--1988.3. Thus the intrinsic 
   acceleration/deceleration and Doppler boosting effect occurred in 
   its inner trajectory section where knot C5 moved along the precessing 
   common trajectory.\\
    \begin{figure*}
    \centering
    \includegraphics[width=6cm,angle=-90]{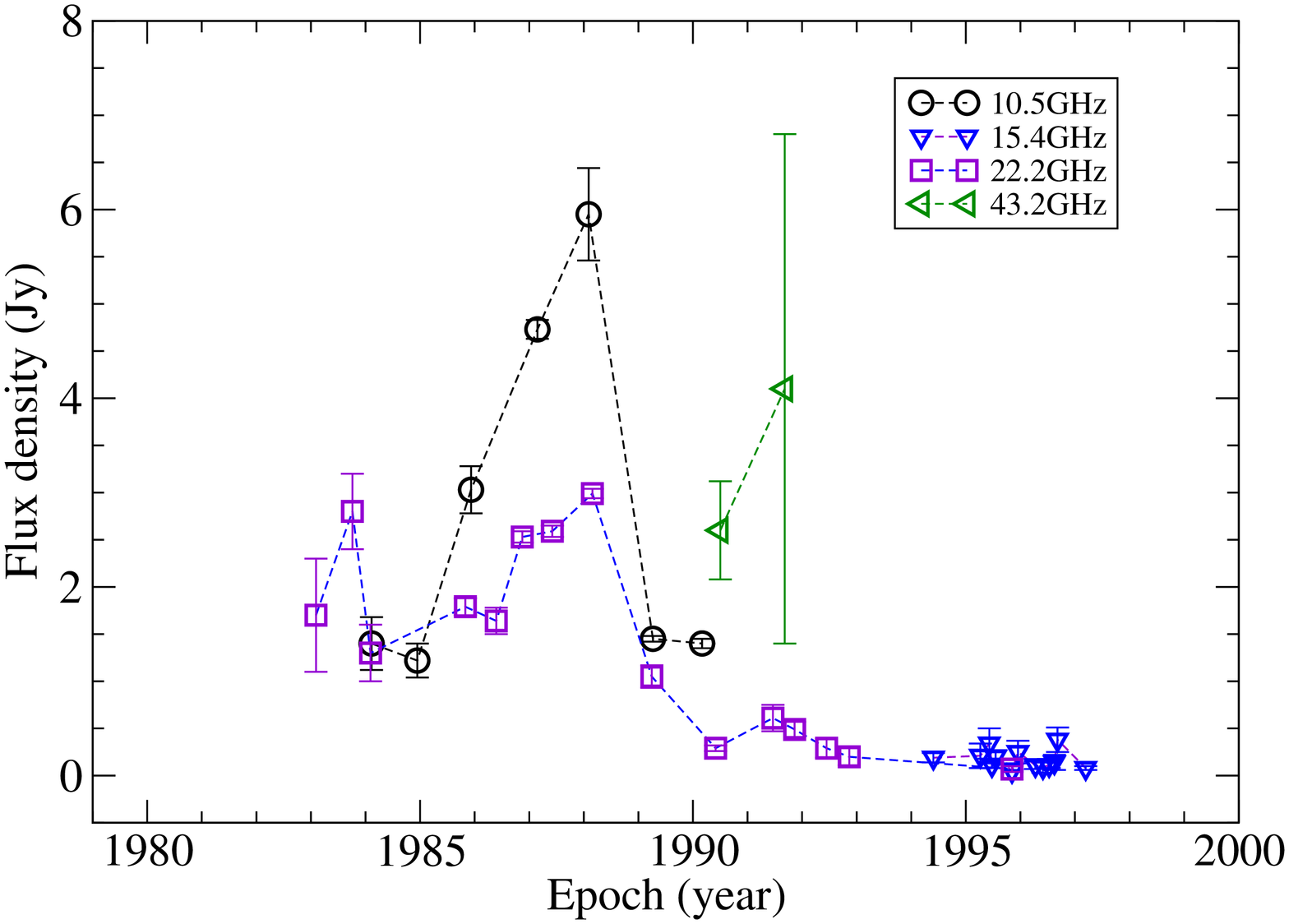}
    \caption{Knot C5: the light-curves observed at 10\,GHz and 22\,GHz with 
    some data-points observed at 15\,GHz and 43\,GHz.}
    \end{figure*}
    \begin{figure*}
    \centering
    \includegraphics[width=5cm,angle=-90]{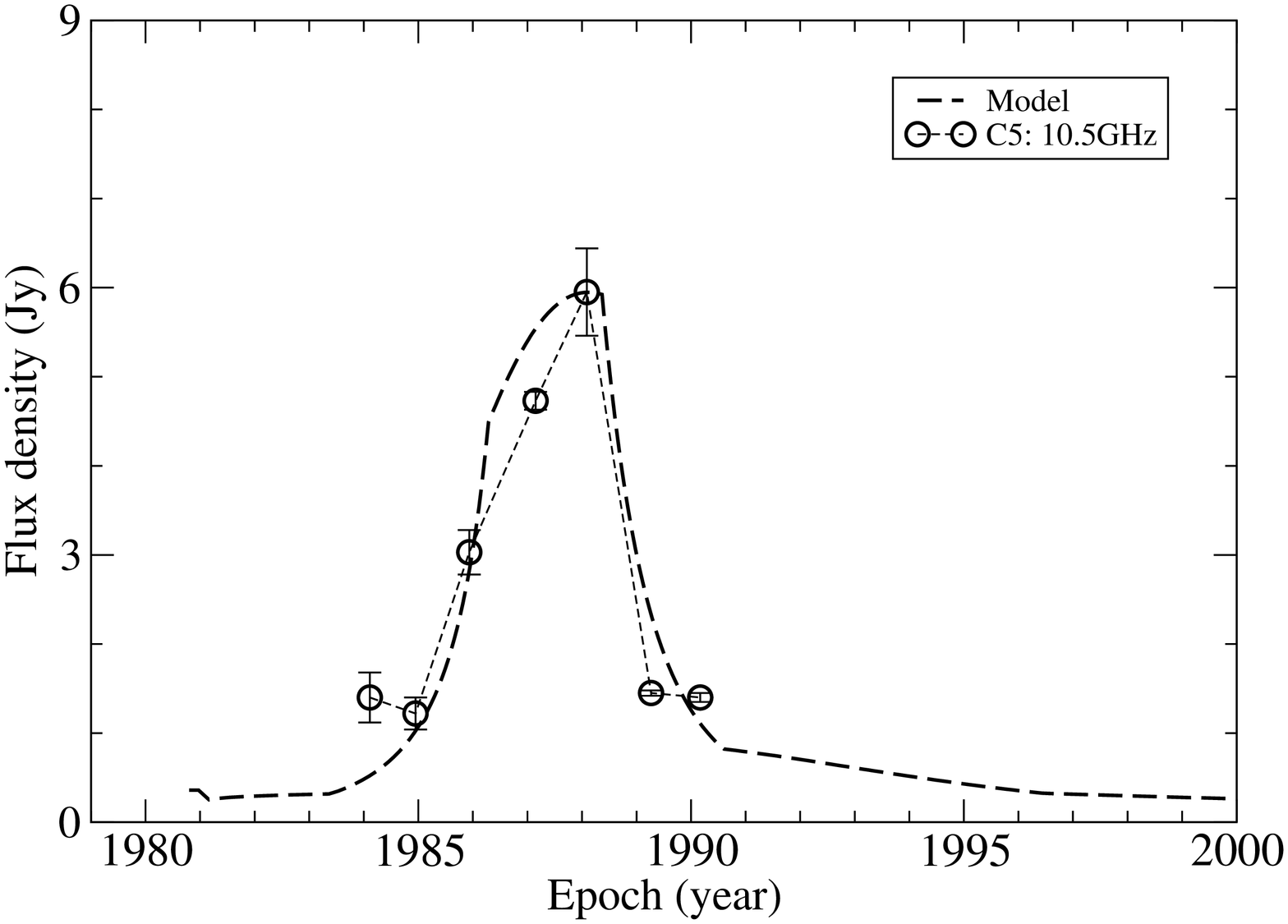}
    \includegraphics[width=5cm,angle=-90]{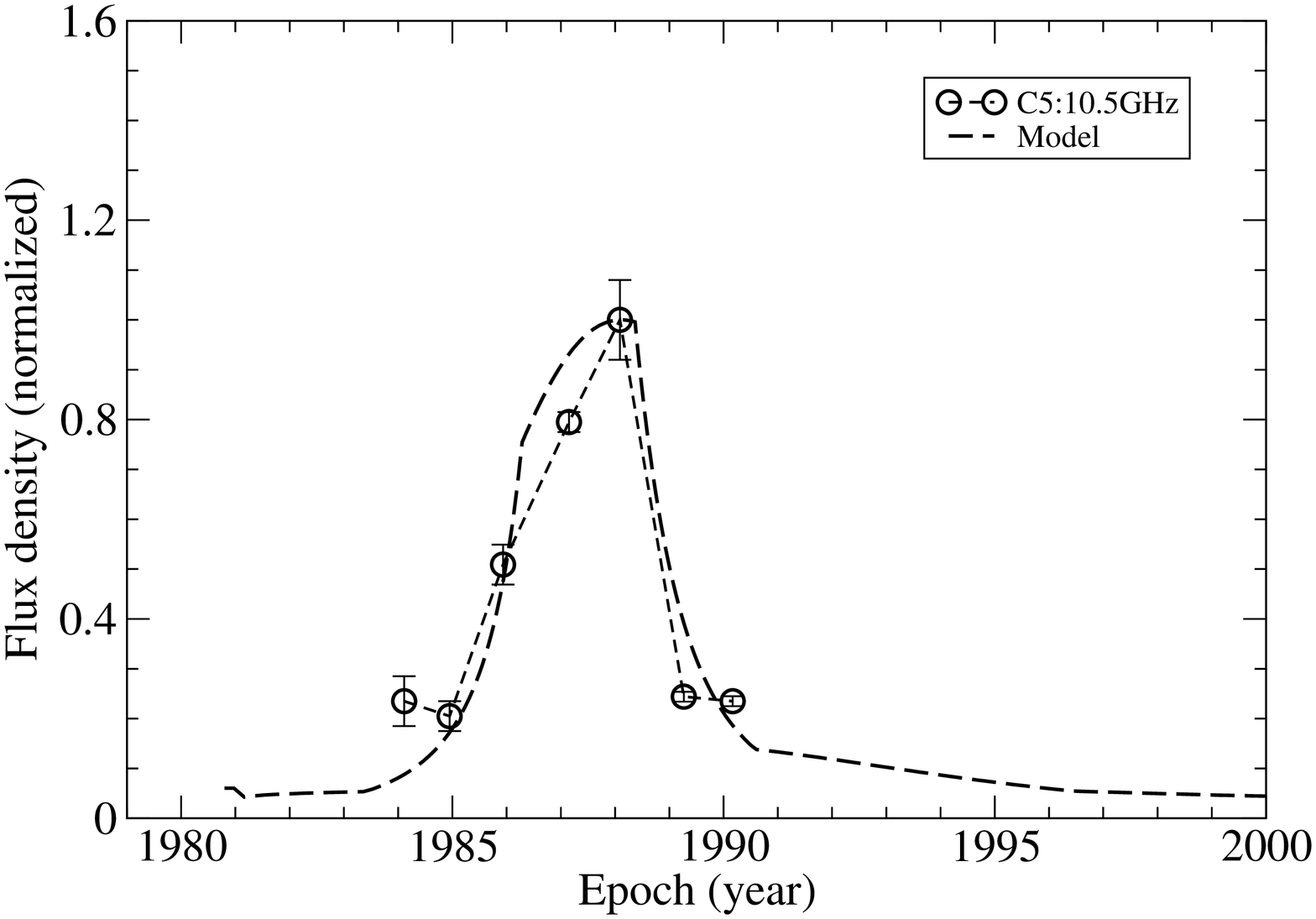}
    \includegraphics[width=5cm,angle=-90]{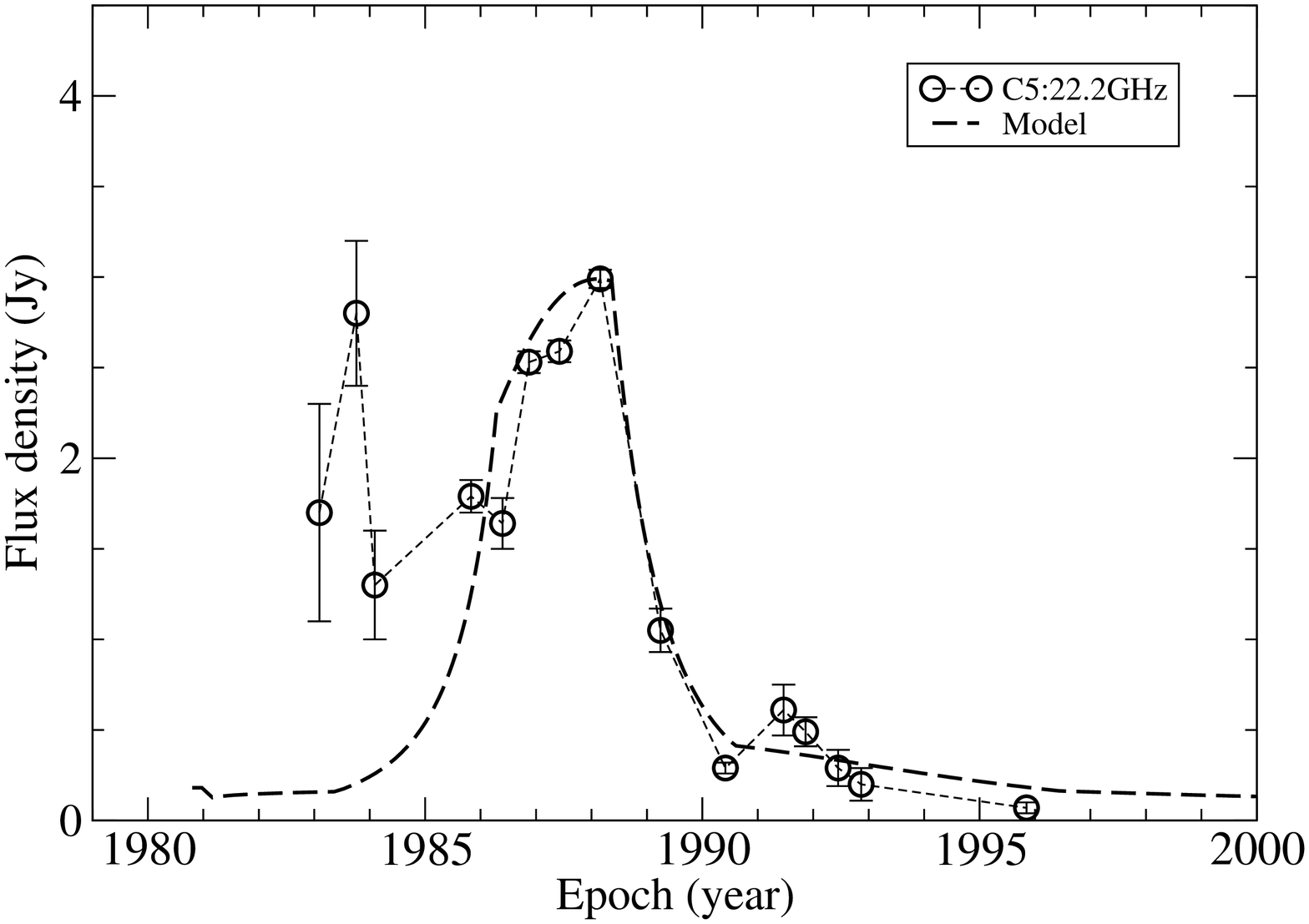}
    \includegraphics[width=5cm,angle=-90]{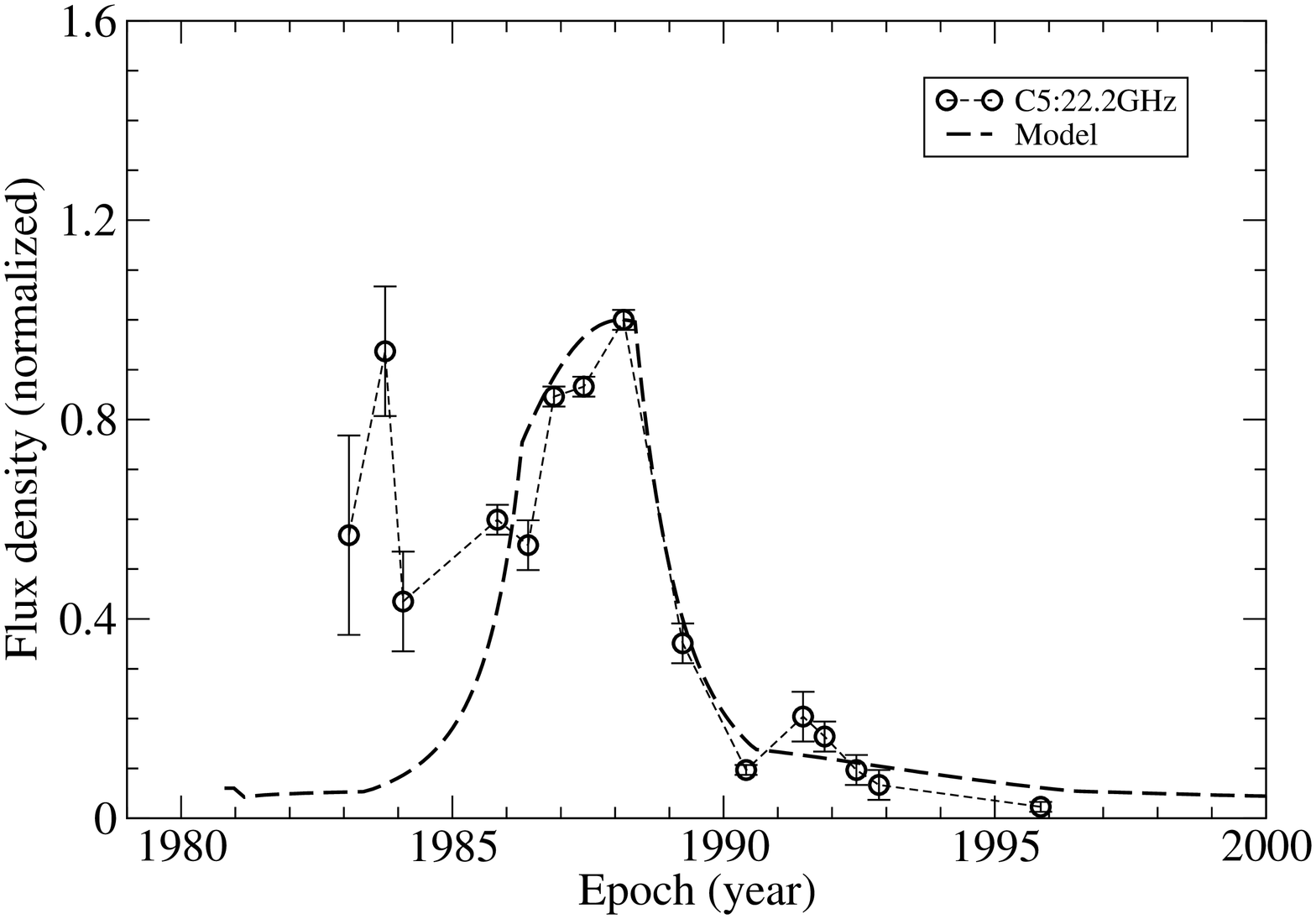}
    \caption{Knot C5. The 10\,GHz  and 22\,GHz light-curves 
    were well coincident
    with the Doppler boosting profiles for both the observed flux light-curve
    (left two panels) and the  normalized flux light-curve (right two panels).}
    \end{figure*}
     Bulk acceleration and deceleration were required and the Lorentz factor
    was modeled as:
    for Z$\leq$4.0\,mas $\Gamma$=5.5; for Z=4.0-15\,mas 
    $\Gamma$ increased from 5.5 to 13.2 ($\Gamma_{max}$=13.2 at 1986.21);
    for Z=15--34\,mas,  $\Gamma$=const.=13.2; 
    for Z=34-45\,mas $\Gamma$ decreased from 13.2 to 7.0 (1990.53); 
    for Z$>$45\,mas $\Gamma$=7.0.\\
     It can be seen  that the entire kinematic behavior within 
    core separation ${r_n}\stackrel{<}{_\sim}$1.25\,mas can be  well 
    fitted, implying that its observed
     precessing common trajectory could extend to a spatial distance of 
   $Z_c$=44.8\,mas (or $\sim$298\,pc) from the core.\\
    \subsection{Doppler boosting effect and flux evolution of knot C5}
    The bump structure derived for the Doppler factor of knot C5 provided
    an very good explanation of the flux evolution of knot C5 during its motion
    along the precessing common trajectory. The observed 10\,GHz and 22\,GHz
    light-curves  are shown in Figure 12, clearly revealing a flare event during
    $\sim$1985--1991 with a peaking flux at $\sim$1988.0, which was closely
     coincident with  the Doppler-boosting profile. 
     We assumed the spectral index
    $\alpha$(10--22GHz)=0.30, the intrinsic flux densities were assumed to be
    1.5$\times{10^{-4}}$Jy (10\,GHz) and 7.7$\times{10^{-5}}$Jy (22\,GHz).
     The model-derived maximum Doppler factor $\delta_{max}$=24.6 
    at $\sim$1987.4.\\
    The model-fits to the observed 10\,GHz and 22\,GHz light-curves are 
    shown in Figure 13 (left two panels), while their normalized light-curves
     were fitted by the normalized Doppler-boosting profiles 
    ($[\delta(t)/\delta_{max}]^{3+\alpha}$) in the right two panels. It can be
     seen that the  15\,GHz and 22\,GHz light-curves were very well coincident
    with the Doppler boosting  profiles which were calculated by using the
    model-predicted Lorentz  factor curve $\Gamma(t)$  and the model-predicted
   viewing angle curve $\theta(t)$ (Fig.11, bottom/right and bottom/middle
    panels, respectively).\\
    The shorter time-scale variations at 22\,GHz during $\sim$1983--1984 could
    be due to the intrinsic flux variations of knot C5 (see the discussion  on
    the flux variations of knots C9 and C10 below).\\
    \begin{figure*}
    \centering
    \includegraphics[width=5cm,angle=-90]{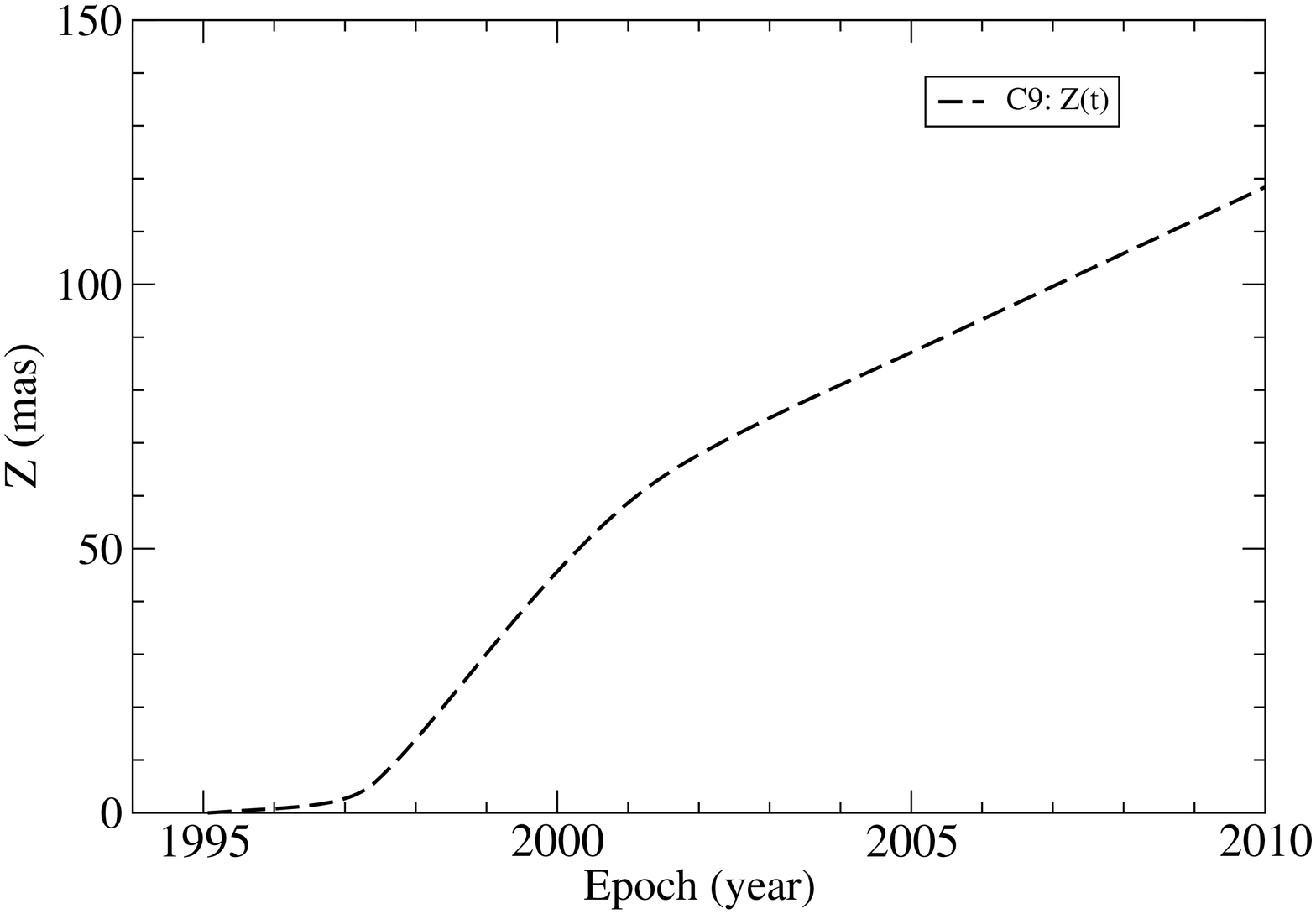}
    \includegraphics[width=5cm,angle=-90]{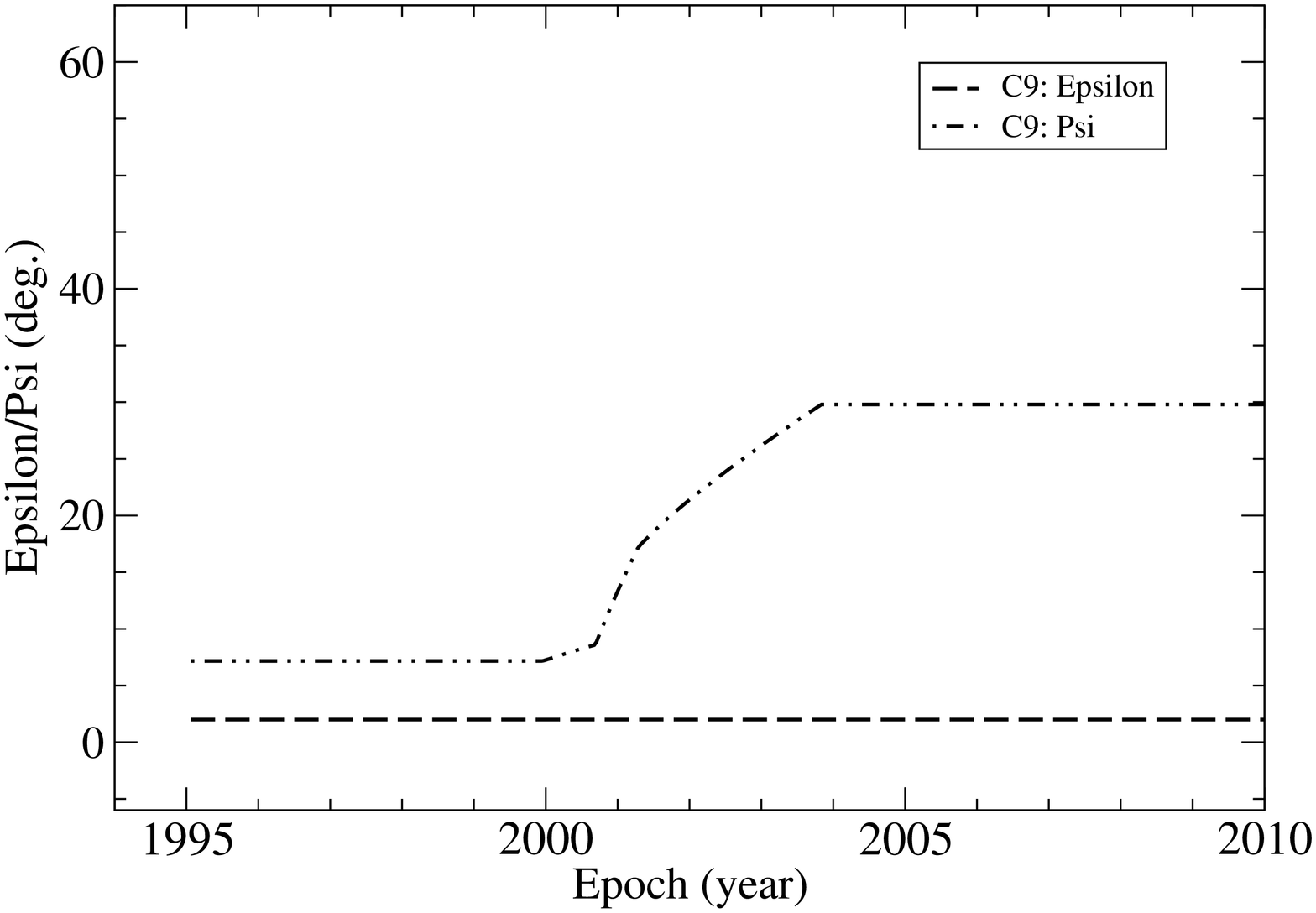}
    \caption{Knot C9. Left panel: the traveled distance Z(t) along the Z-axis.
    The right panel shows $\epsilon$=const.=$2^{\circ}$ and 
    $\psi$=$7.16^{\circ}$ before 1999.94 ($r_n{\leq}$1.25\,mas or
    $X_n{\leq}$1.22\,mas), when it moved along the precessing common trajectory
     (in its inner-trajectory section). After 1999.94
     $\psi$ increased and knot C9 started to move along its own individual 
    trajectory (in its outer trajectory section). }
    \end{figure*}
   \begin{figure*}
   \centering
   \includegraphics[width=6cm,angle=-90]{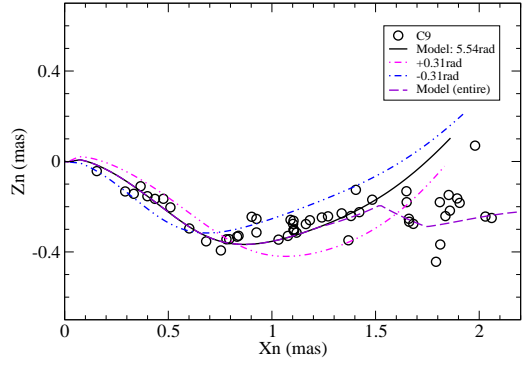}
   \caption{Knot C9. Model fit to the inner trajectory section ($Z_c{\leq}$
   44.8\,mas, $X_n{\leq}$1.21\,mas, t$\leq$1999.94) where it moved along the
     precessing common 
    trajectory. The black curve represents the precessing common trajectory for 
    precession phase $\phi_0$=5.54+4$\pi$ (ejection time $t_0$=1995.06),
     while curves in magenta and blue represent those for precession phases 
    $\phi_0{\pm}$0.31\,rad ($\pm$5\% of the precession period).
    The curve in violet represents the model-fit to its whole trajectory.}
     \end{figure*}
   \begin{figure*}
   \centering
   \includegraphics[width=4.5cm,angle=-90]{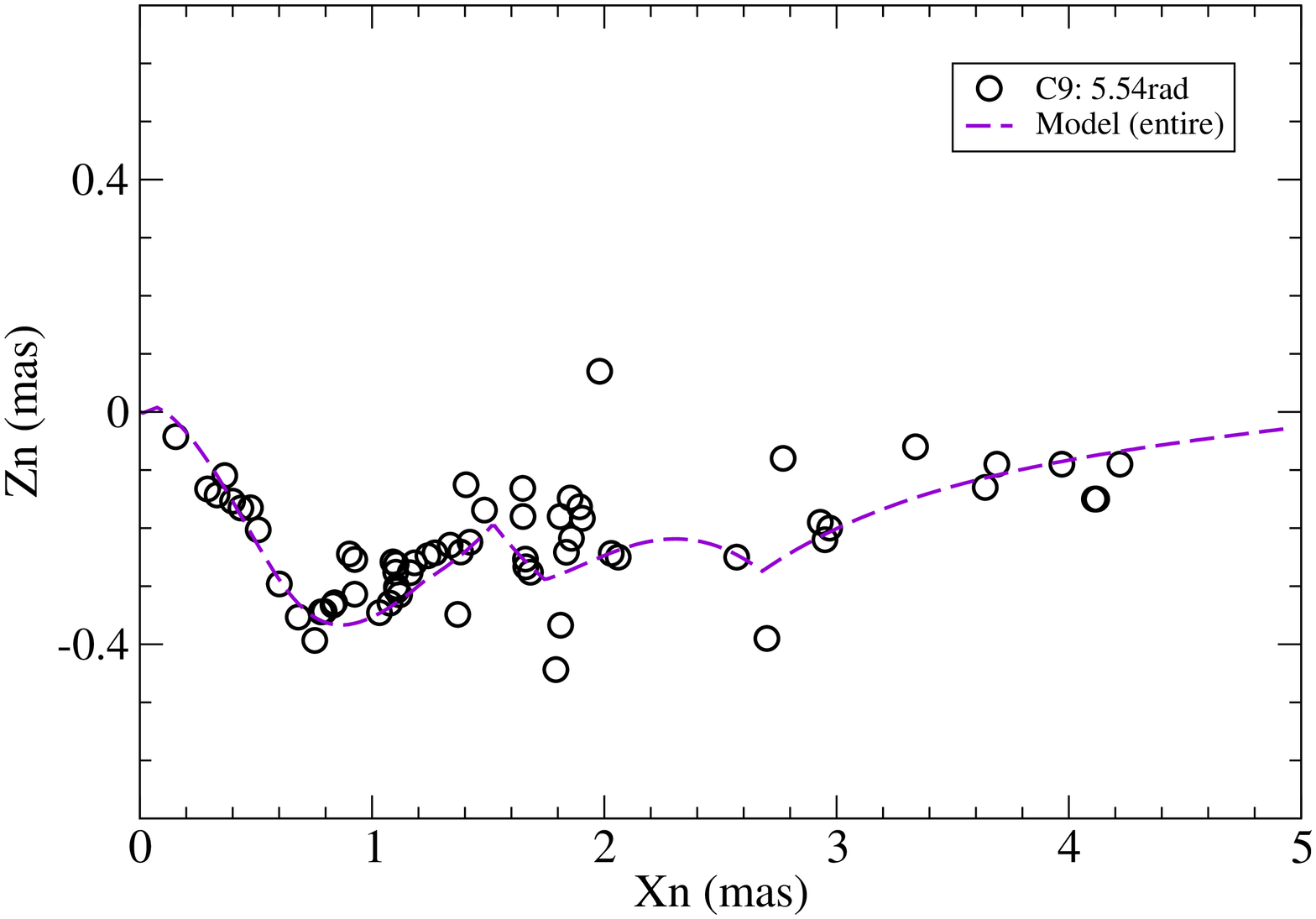}
   \includegraphics[width=4.5cm,angle=-90]{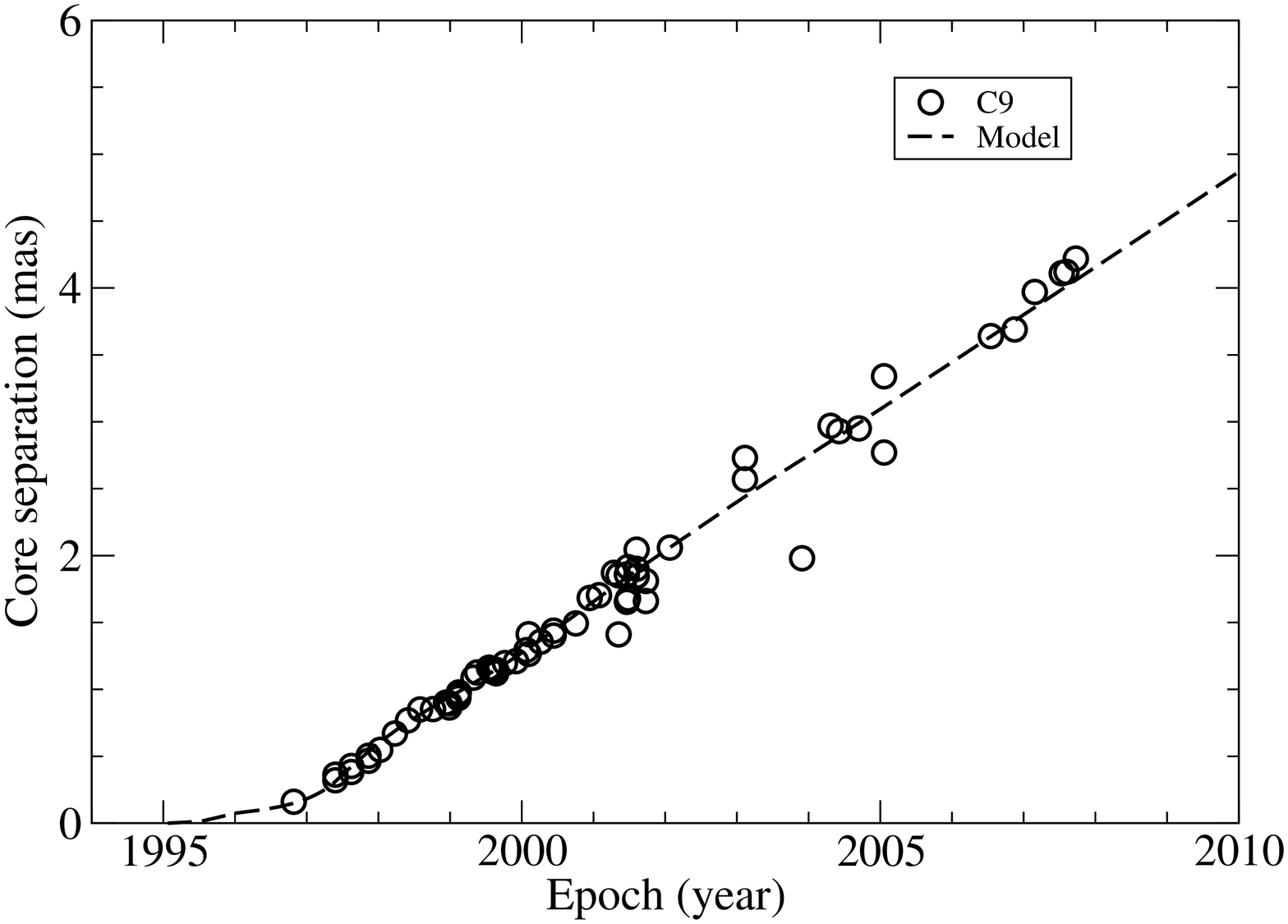}
   \includegraphics[width=4.45cm,angle=-90]{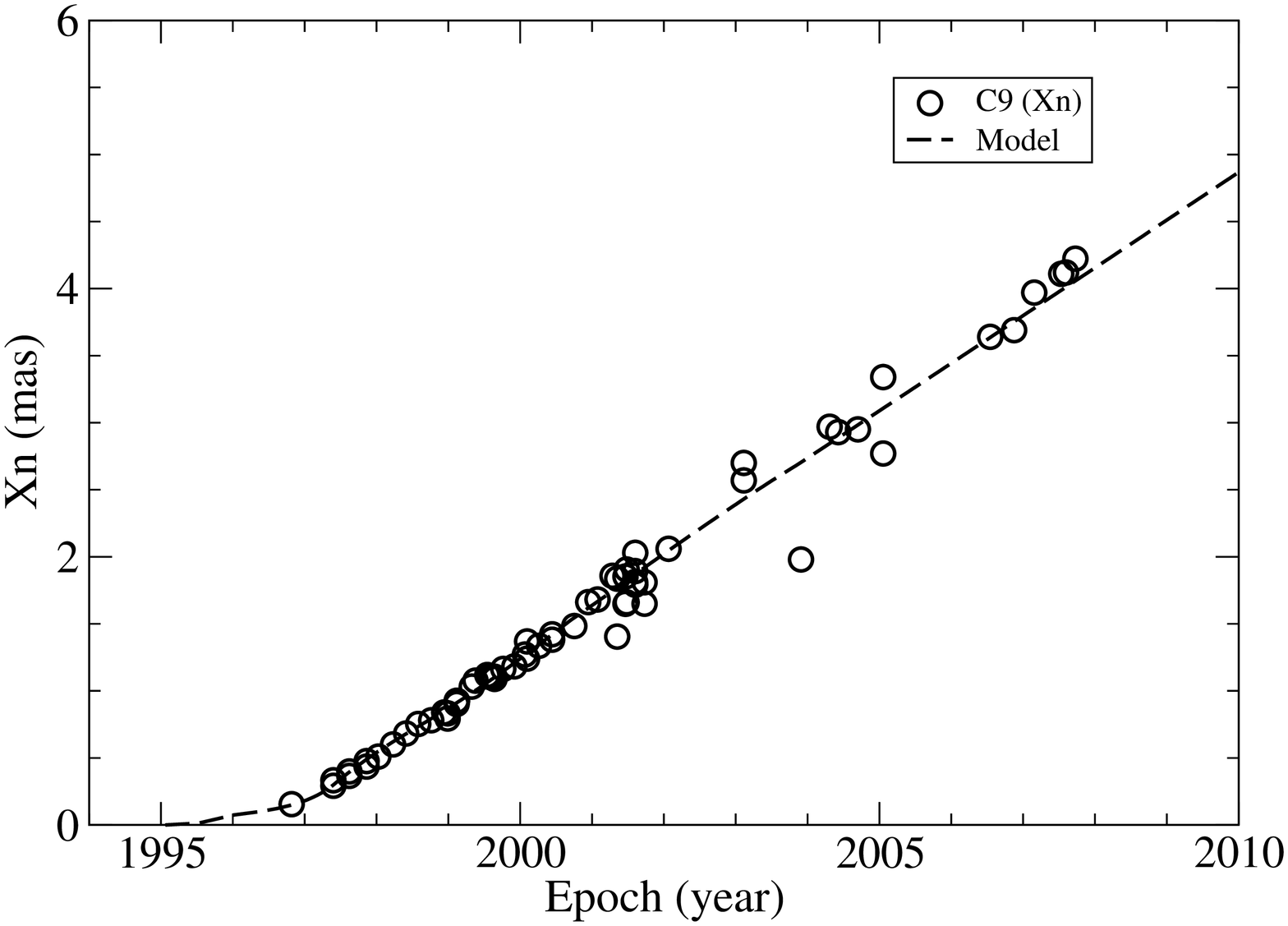}
   \includegraphics[width=4.5cm,angle=-90]{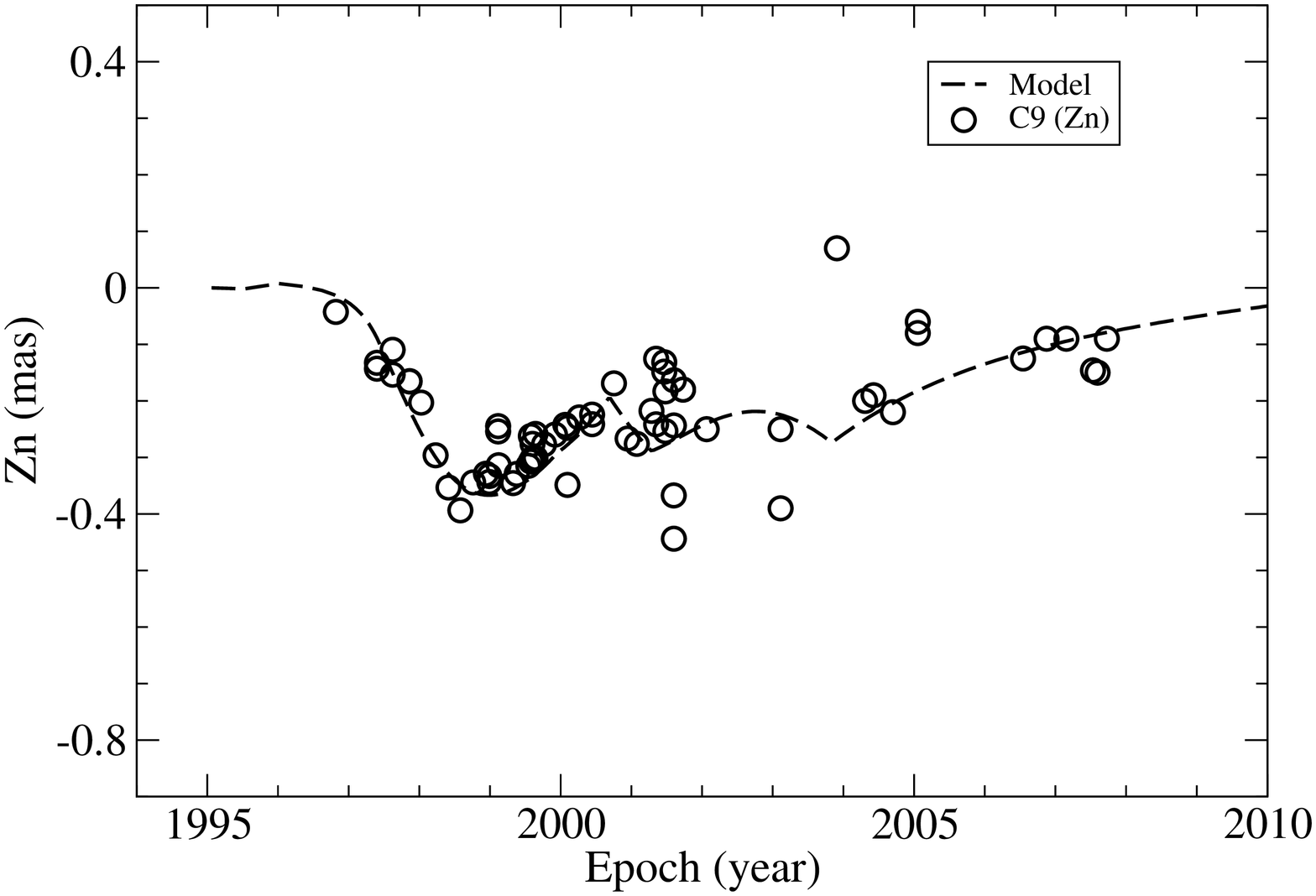}
   \includegraphics[width=4.5cm,angle=-90]{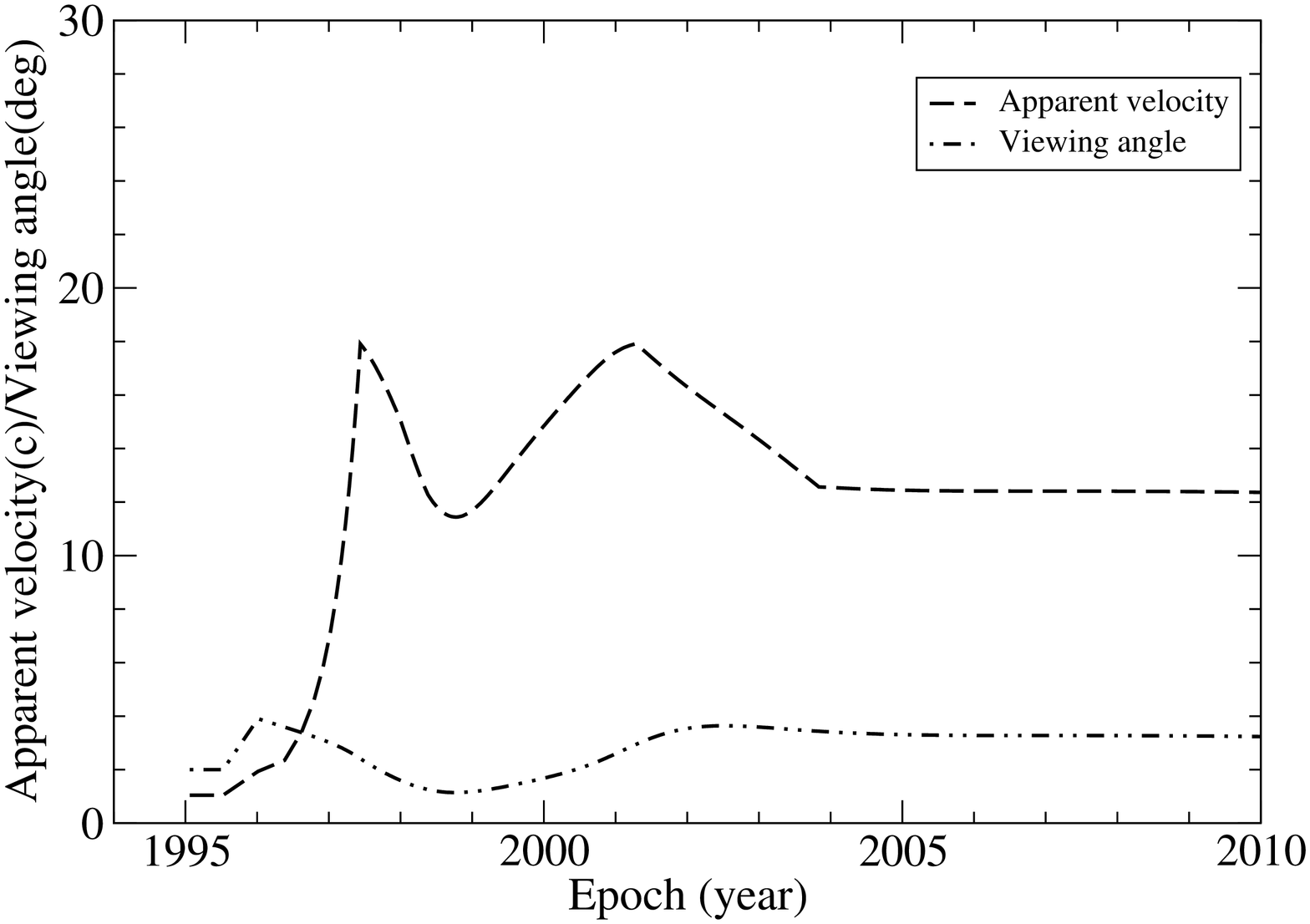}
   \includegraphics[width=4.5cm,angle=-90]{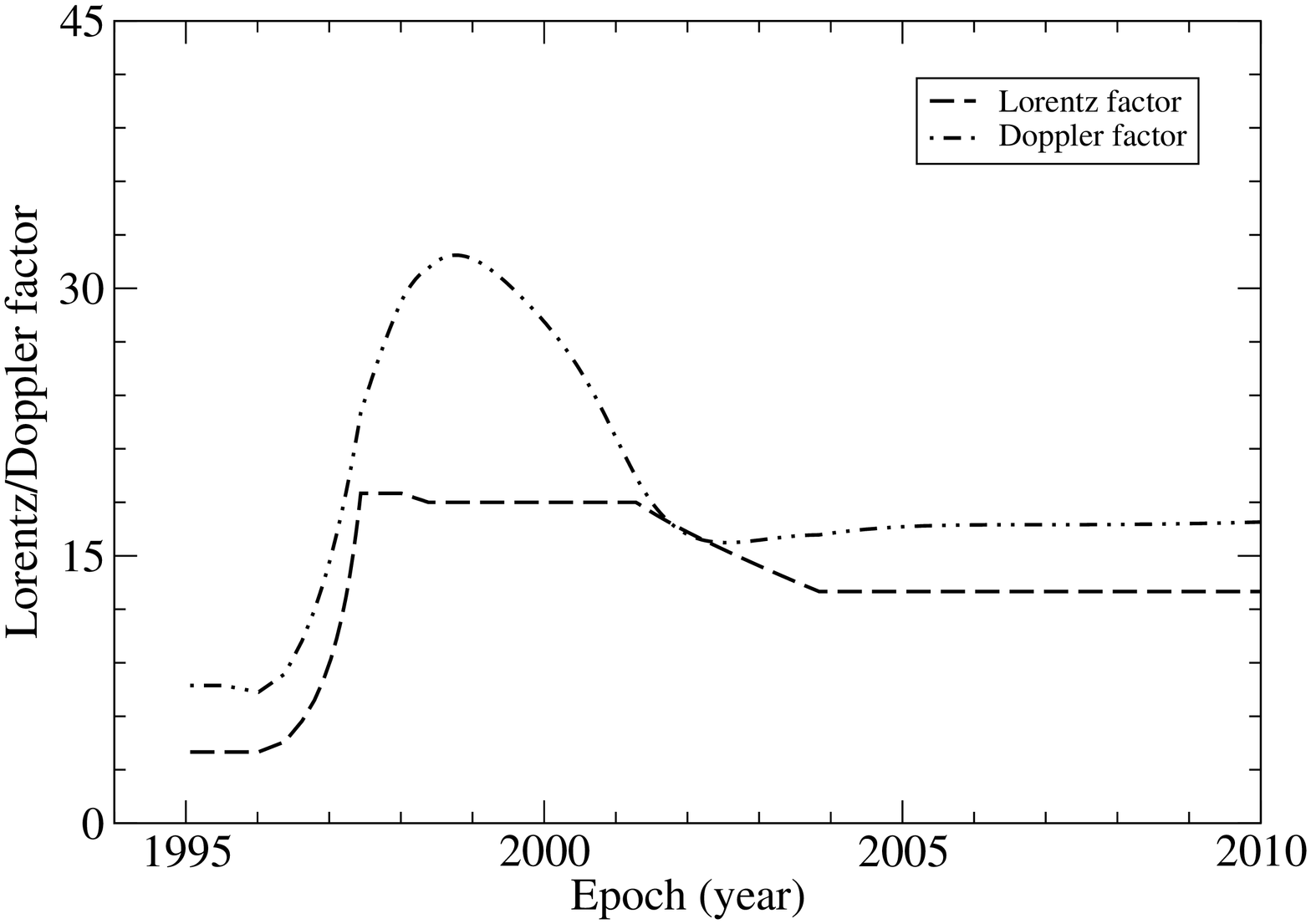}
   \caption{Knot C9. Model-fitting results for the whole trajectory 
    $Z_n$($X_n$), core separation $r_n(t)$ and coordinate $X_n$(t) (upper
     three panels); model-fitting results for 
    coordinate $Z_n$(t), model-derived apparent-velocity $\beta_{app}(t)$ and
    viewing-angle $\theta(t)$, and  model-derived bulk Lorentz-factor
     $\Gamma(t)$ and Doppler-factor $\delta(t)$ (bottom three panels).
     $\Gamma(t)$ has a platform of maximum ($\simeq$18.5--18.0) during
      1997.4-2001.3.}
       \end{figure*}
   \section{Interpretation of kinematics and flux evolution for knot C9}
    As shown in the previous paper (Qian \cite{Qi22}),
    The results of model-fitting of its kinematics and the explanation of 
    its flux  evolution in terms of Doppler boosting effect are very important
    and encouraging for the application of our precessing nozzle scenario to
    study the phenomena in  3C345 and other blazars. Knot C9 was a typical
    and exceptionally instructive example of applying
    the scenario for a satisfying interpretation of its VLBI-kinematics
     and flux evolution. Here  we recapitulate the main  results 
   obtained in Qian 
    (\cite{Qi22}) and supplement some new results on its flux evolution
    and spectral features.
     \subsection{Model simulation of kinematics for knot C9}
      In the left panel of Figure 14 the traveled distance Z(t)  
     along the Z-axis from the core of knot C9 is shown. In the right panel 
     is shown the parameters $\epsilon$
     and $\psi$ as functions of time. It can be seen that $\epsilon(t)$ is a
    constant (0.0349\,rad=$2^{\circ}$), but $\psi(t)$ varied with time:
    before 1999.94 $\psi$=0.125\,rad=$7.16^{\circ}$ and its trajectory 
    followed the precessing common trajectory for the precession phase 
    $\phi_0$=5.54+4$\pi$ (corresponding to
    its ejection time $t_0$=1995.06). After 1999.94 $\psi$ increased and its
    trajectory deviated from the precessing common trajectory and knot C9 moved
    along its own individual trajectory (in its outer trajectory section).\\
     The model-fit to its inner trajectory section (or precessing common 
    trajectory) is shown in Figure 15, where the curve in black represents the
    precessing common trajectory section for the precession phase $\phi_0$=
    5.54+4$\pi$.  The curves in magenta  and blue represent
    the model-trajectories for precession phases $\phi_0$$\pm$0.31rad, showing
     its ejection epoch being fitted with an accuracy of $\pm$0.36\,yr (5\% of 
    the precession period). The curve in violet represents the model-fit to
    its entire trajectory. Its precessing common trajectory section
     extended to $r_n{\sim}$1.25\,mas or the traveled distance 
    $Z_c{\sim}$44.8\,mas=298\,pc. \\
     In Figure 16 the model-fitting results of its entire kinematics are 
    presented: the entire trajectory $Z_n(X_n)$, core separation $r_n(t)$,
     coordinate $X_n(t)$ (upper three panels), and coordinate $Z_n(t)$,
    the modeled  apparent velocity $\beta_{app}(t)$,
    viewing angle $\theta(t)$, bulk Lorentz factor $\Gamma(t)$ and Doppler
    factor $\delta(t)$ (bottom three panels).\\
     It can be seen that  the kinematic properties  of knot C9 during the 
    entire observing period $\sim$1997--2008 (within core separation
     $r_n{\sim}$4.1\,mas) were very well model-fitted. \\
    Its apparent speed, viewing angle, bulk Lorentz factor and Doppler factor 
    as functions of time were derived through the model-fitting of its 
    entire kinematics. The accelerated  motion observed in knot C9 during
    1986--1988 could be well explained by the increase 
    in its Lorentz factor and its motion along the helical trajectory. \\
    In order to explain its apparent motion and flux evolution 
     its bulk acceleration and deceleration were required and its 
    bulk Lorentz factor was modeled as (bottom/right panel in Figure 16):
    For Z$\leq$1.0\,mas $\Gamma$=4; for Z=1--6\,mas
    $\Gamma$ increased from 4 to 18.5 ($\Gamma_{max}$=18.5 at 1997.44); 
     for Z=6--14\,mas $\Gamma$=18.5; For Z=14--20\,mas
     $\Gamma$ decreased from 18.5 to 18.0; For
     Z=20-61.7\,mas (t=2001.28) $\Gamma$=18.0; For Z=61.7--80\,mas (t=2003.84)
     $\Gamma$ decreased from 18.0 to 13.0; for Z$>$80\,mas $\Gamma$=13.0.\\
    The modeled Doppler-factor curve showed a smooth bump structure during
    1996--2002, providing a very determinative Doppler-boosting effect to 
    explain its flux variations observed at 43\,GHz, 22\,GHz and 15\,GHz.\\
    The derived  apparent velocity $\beta_a(t)$ showed two peaks while the
     viewing angle $\theta(t)$ varied along a depressed curve (bottom/middle
    panel in Figure 16).\\
    \begin{figure*}
   \centering
   \includegraphics[width=6cm,angle=-90]{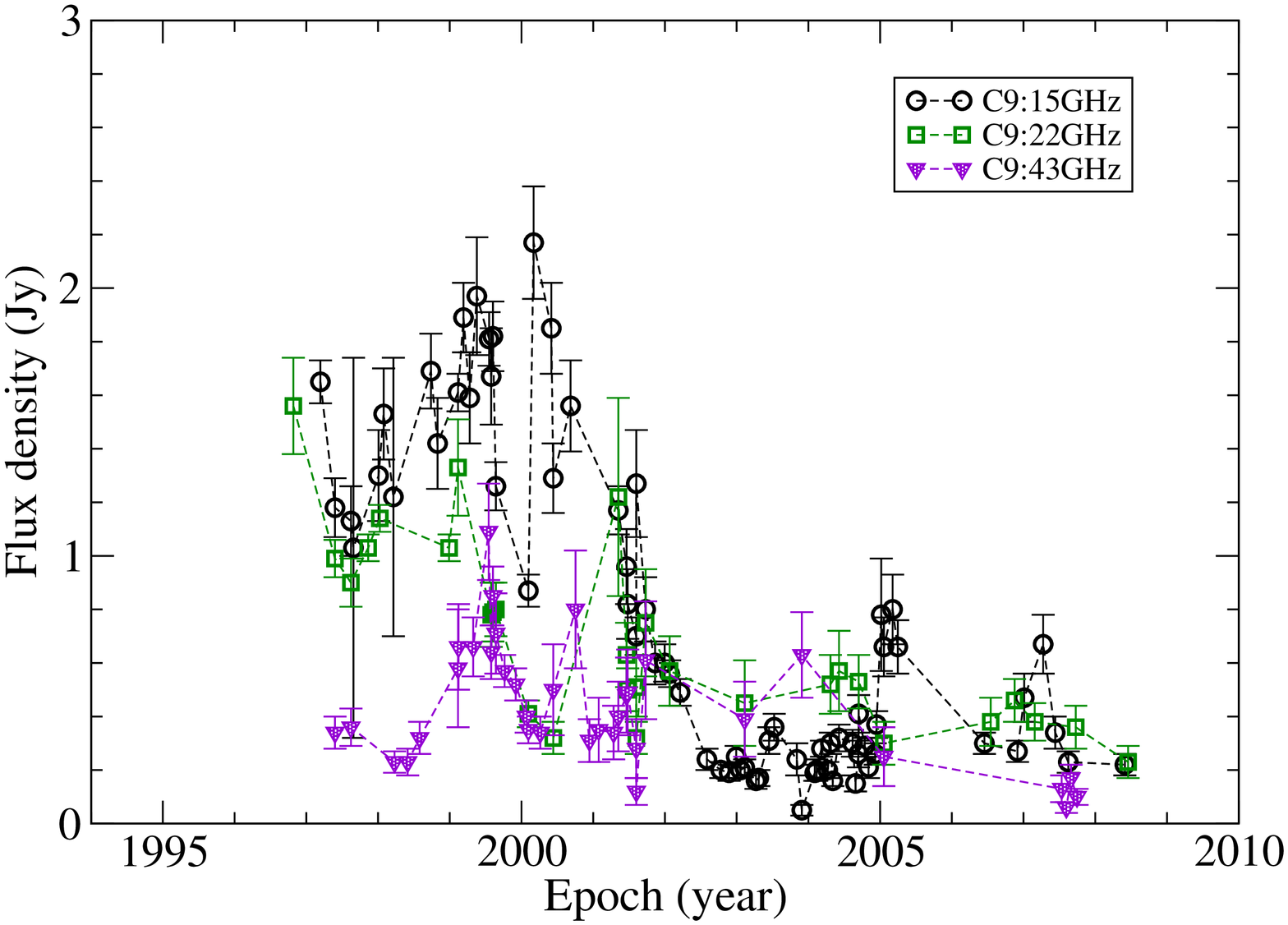}
    \caption{Knot C9. Lght-curves observed at 15\,GHz, 22\,GHz and 43GHz.}
   \end{figure*}
    \begin{figure*}
    \centering
    \includegraphics[width=5cm,angle=-90]{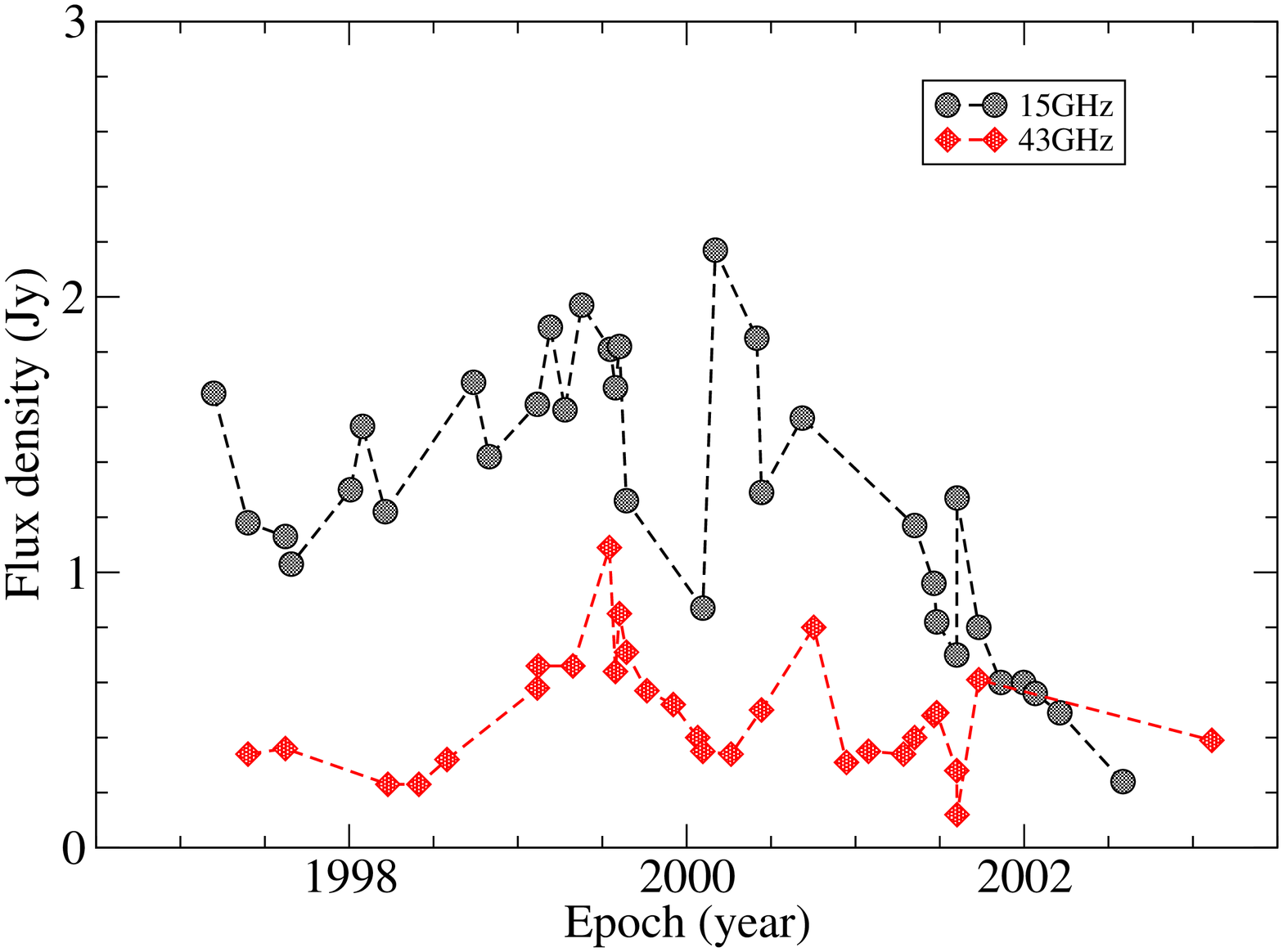}
    \includegraphics[width=5cm,angle=-90]{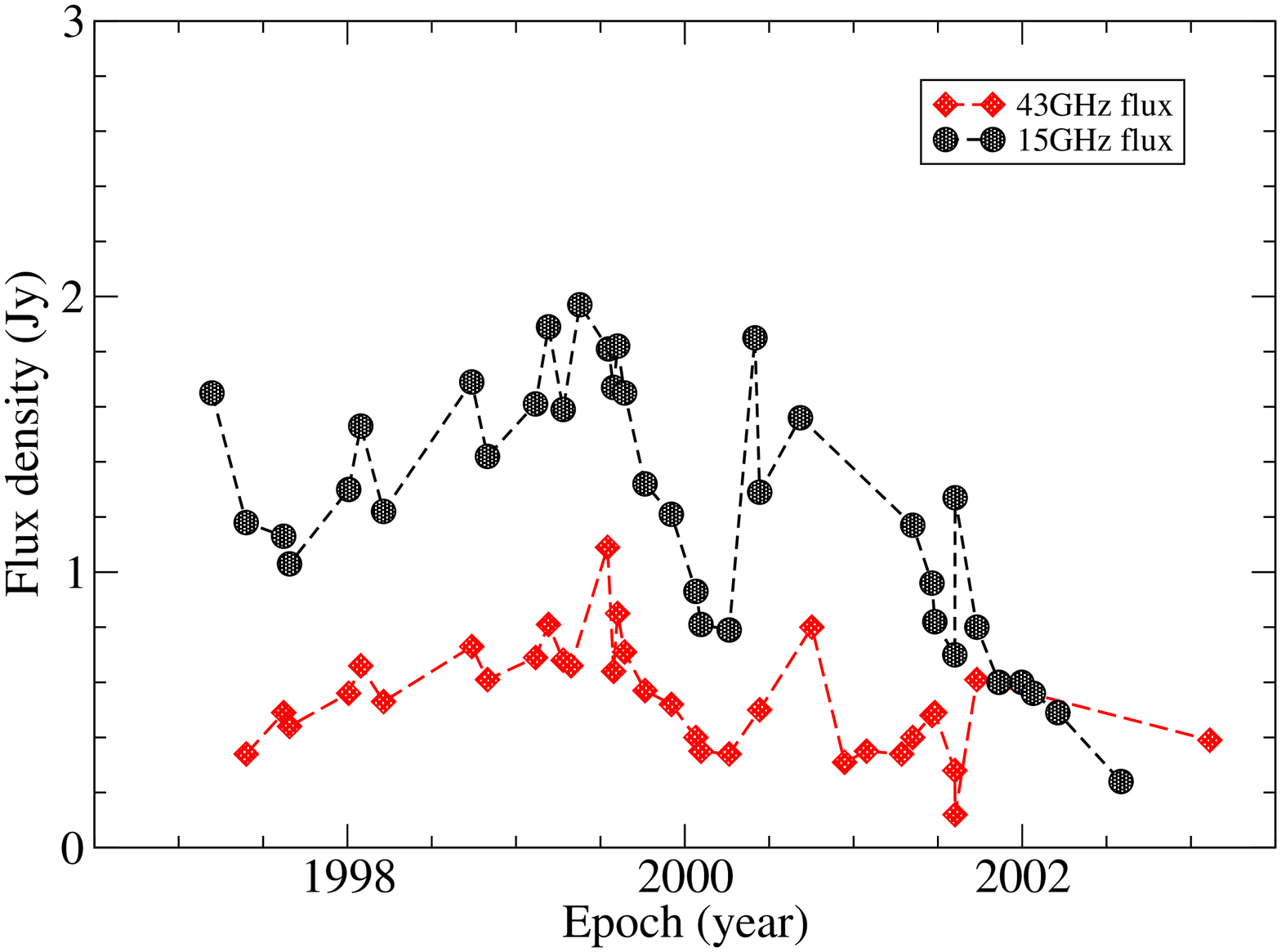}
    \caption{Knot C9. Light-curves observed at 15\,GHz and 43\,GHz (left panel)
     and the reformed light curves with an assumed spectral index 
     $\alpha$(15-22-43GHz)=0.80.}
    \end{figure*}
    \begin{figure*}
    \centering
    \includegraphics[width=5cm,angle=-90]{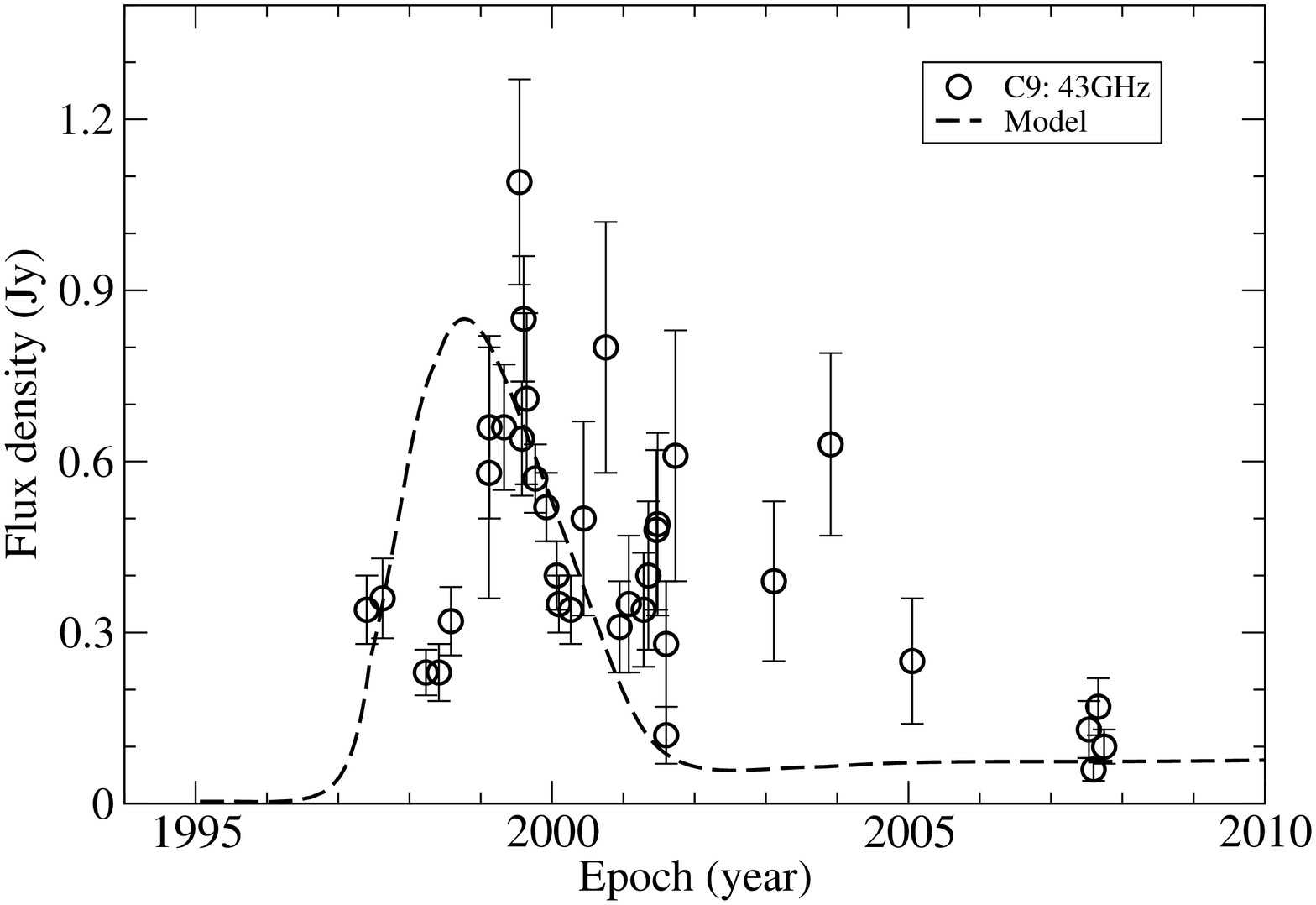}
    \includegraphics[width=5cm,angle=-90]{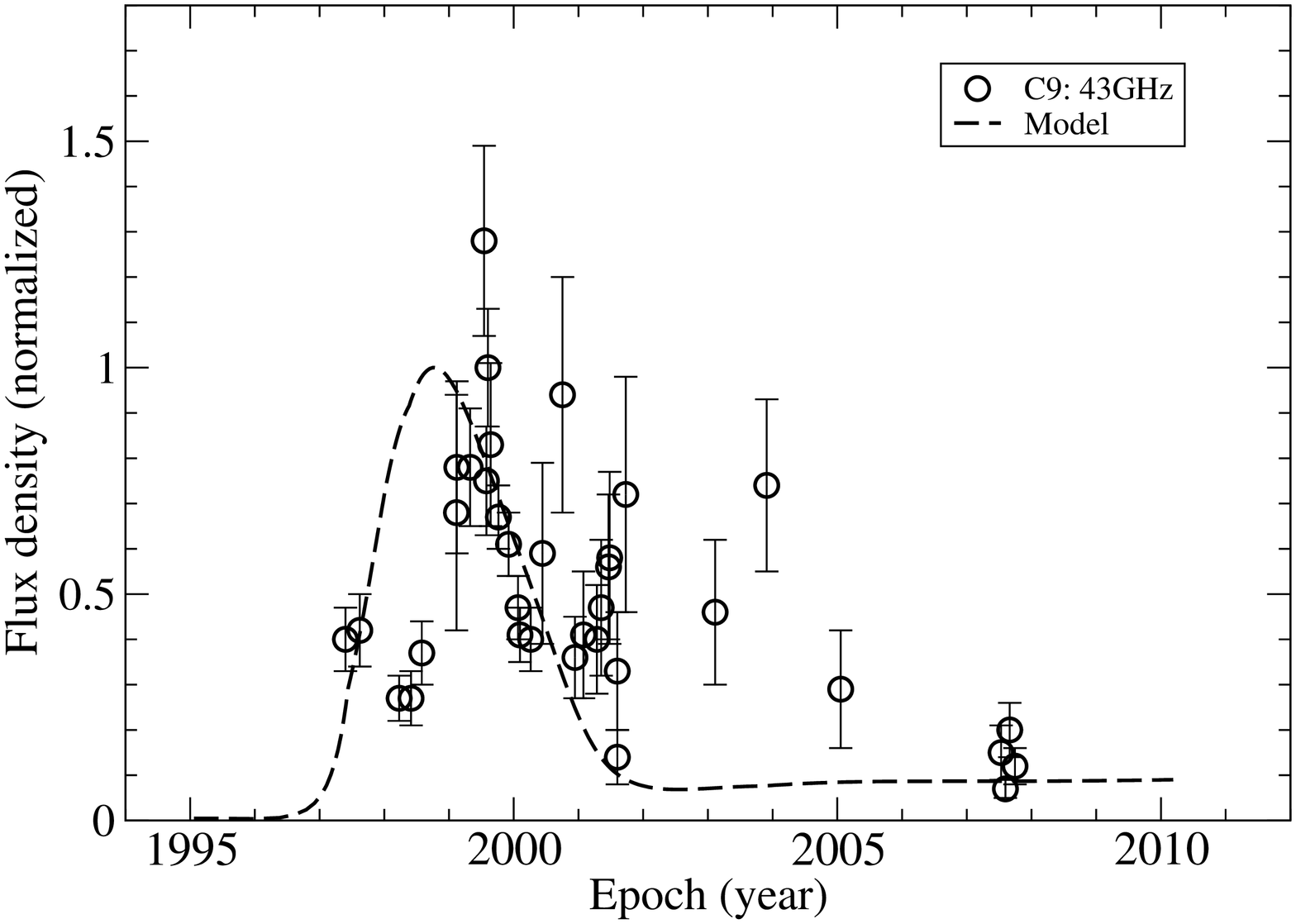}
    \includegraphics[width=5cm,angle=-90]{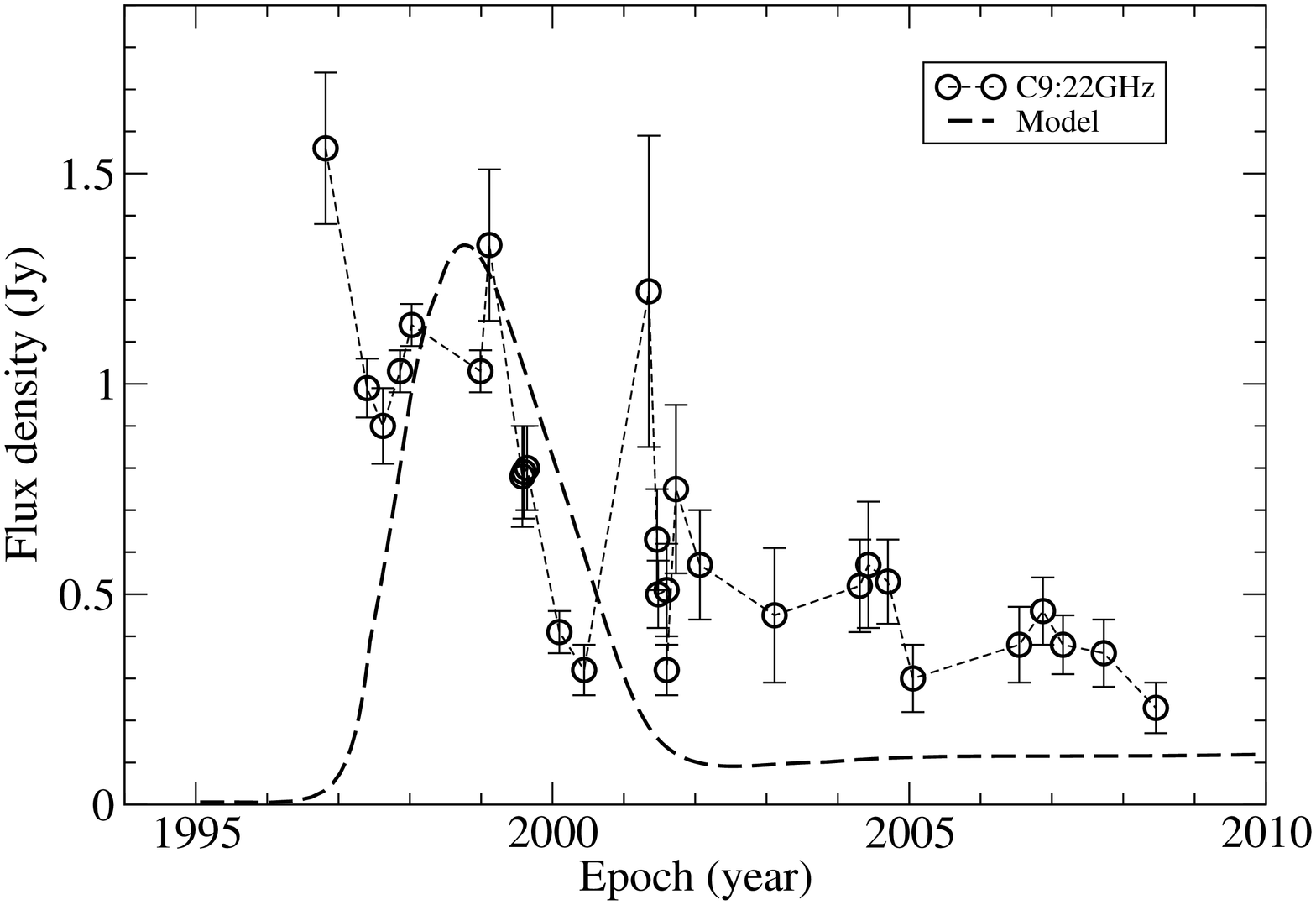}
    \includegraphics[width=5cm,angle=-90]{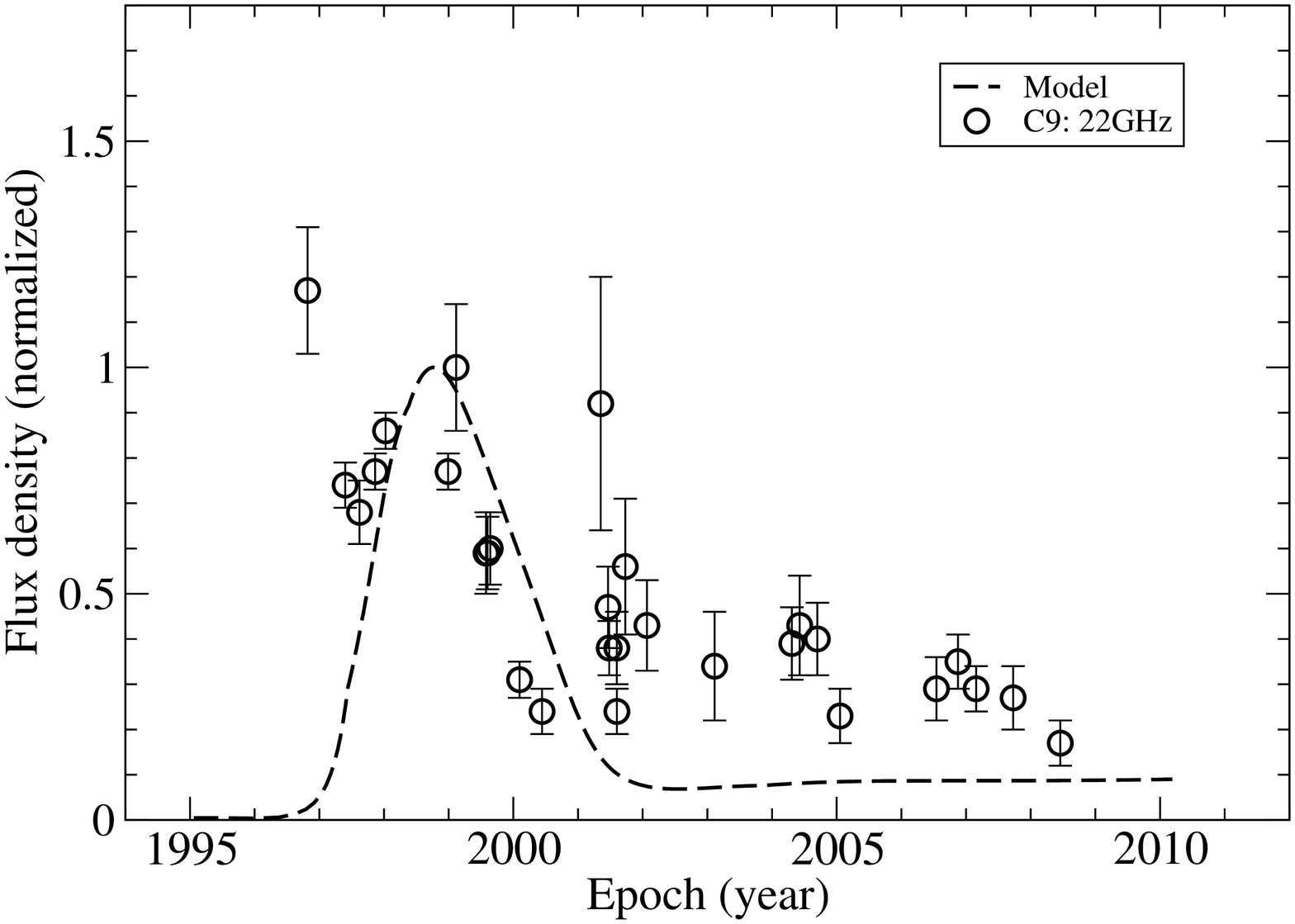}
    \includegraphics[width=5cm,angle=-90]{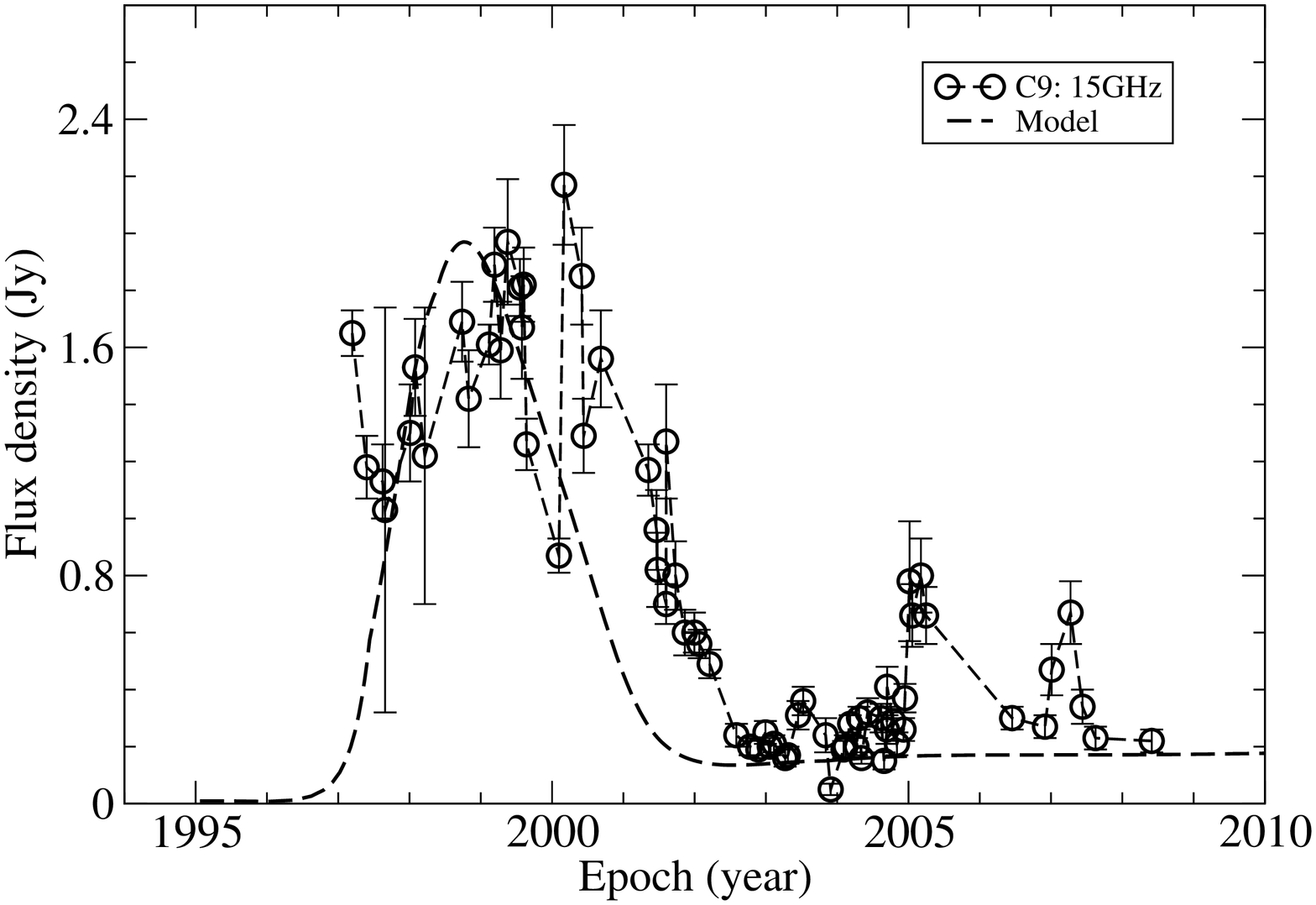}
    \includegraphics[width=5cm,angle=-90]{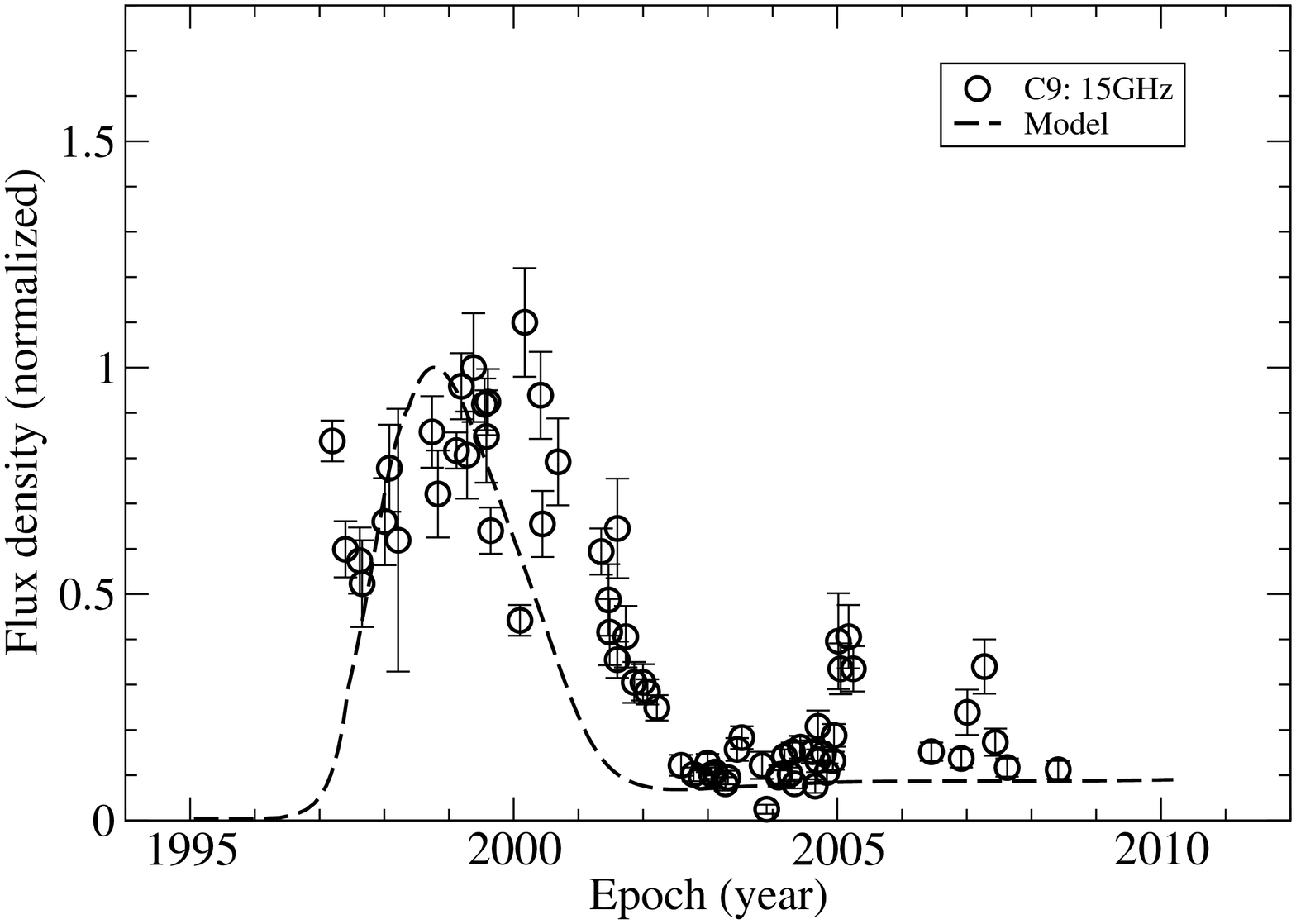}
    \caption{Knot C9. Top two panels: the observed 43GHz light-curve and
    its normalized light-curve fitted by the Doppler-boosting profiles.
    Middle two panels: the observed 22\,GHz light-curve and its normalized 
    light-curve fitted by the Doppler-boosting profiles.
    Bottom two panels: the observed 15\,GHz light-curve and its normalized 
    light-curve matched by the Doppler-boosting profiles.}
    \end{figure*}
   \begin{figure*}
    \centering
    \includegraphics[width=5cm,angle=-90]{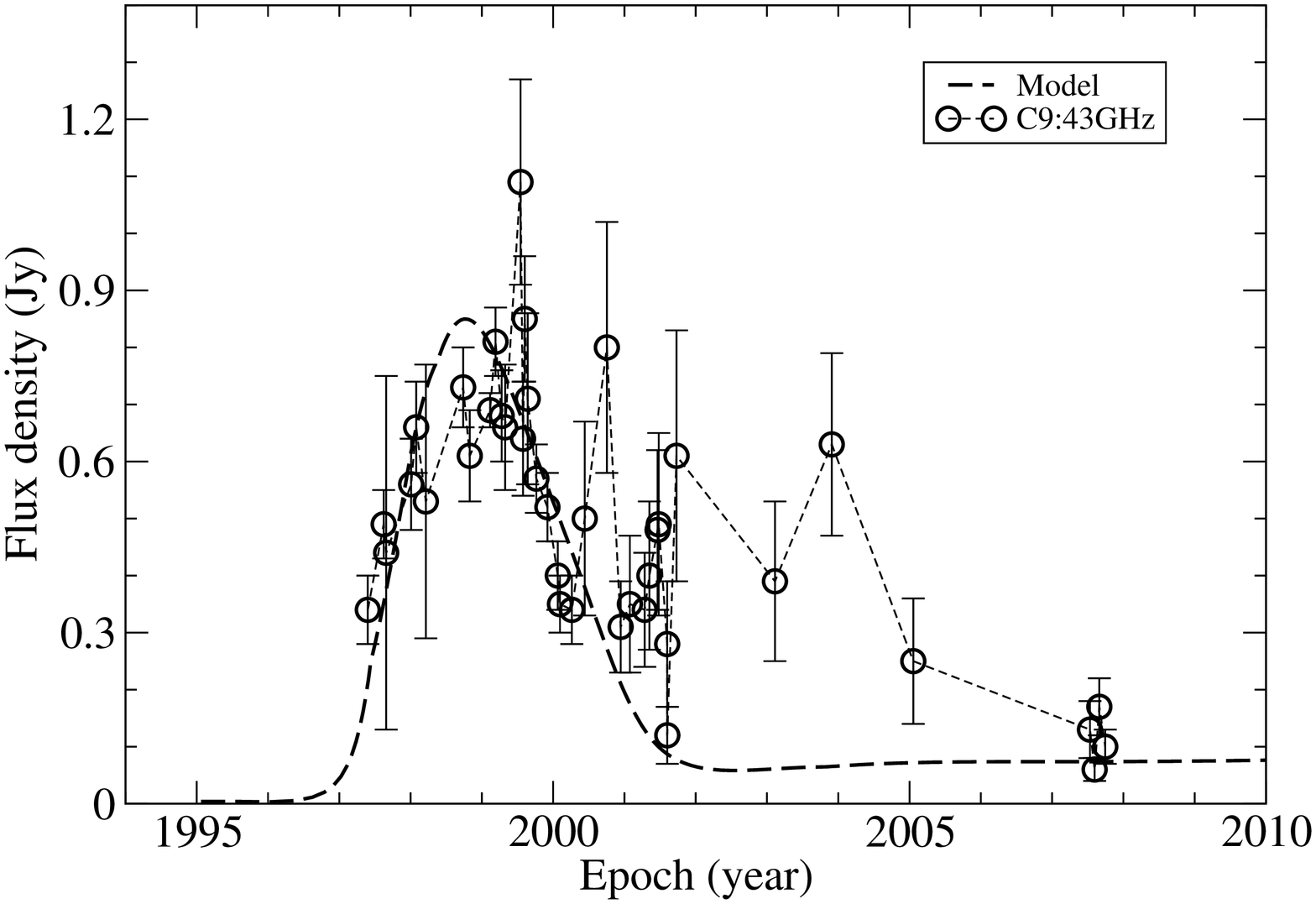}
    \includegraphics[width=5cm,angle=-90]{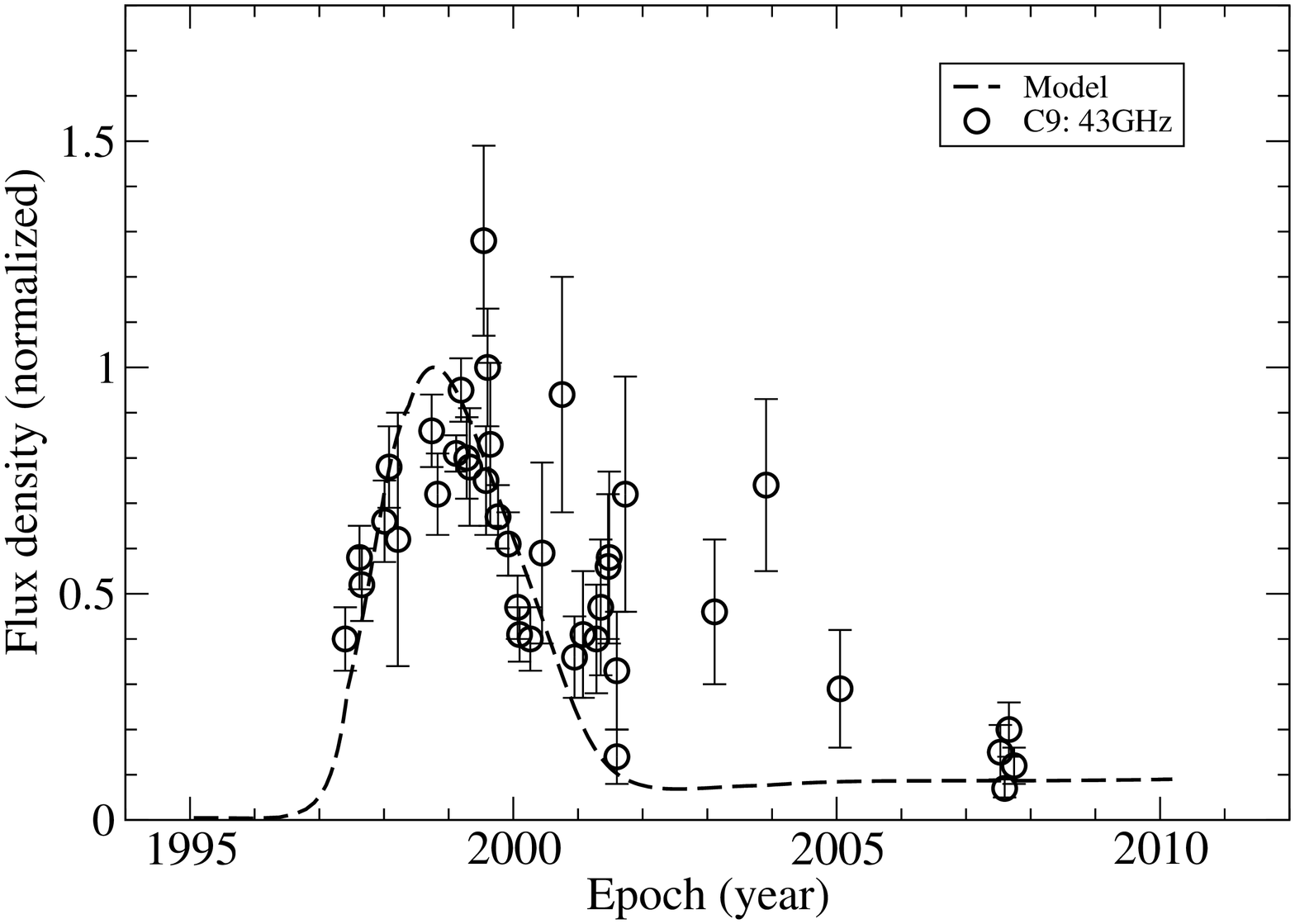}
    \includegraphics[width=5cm,angle=-90]{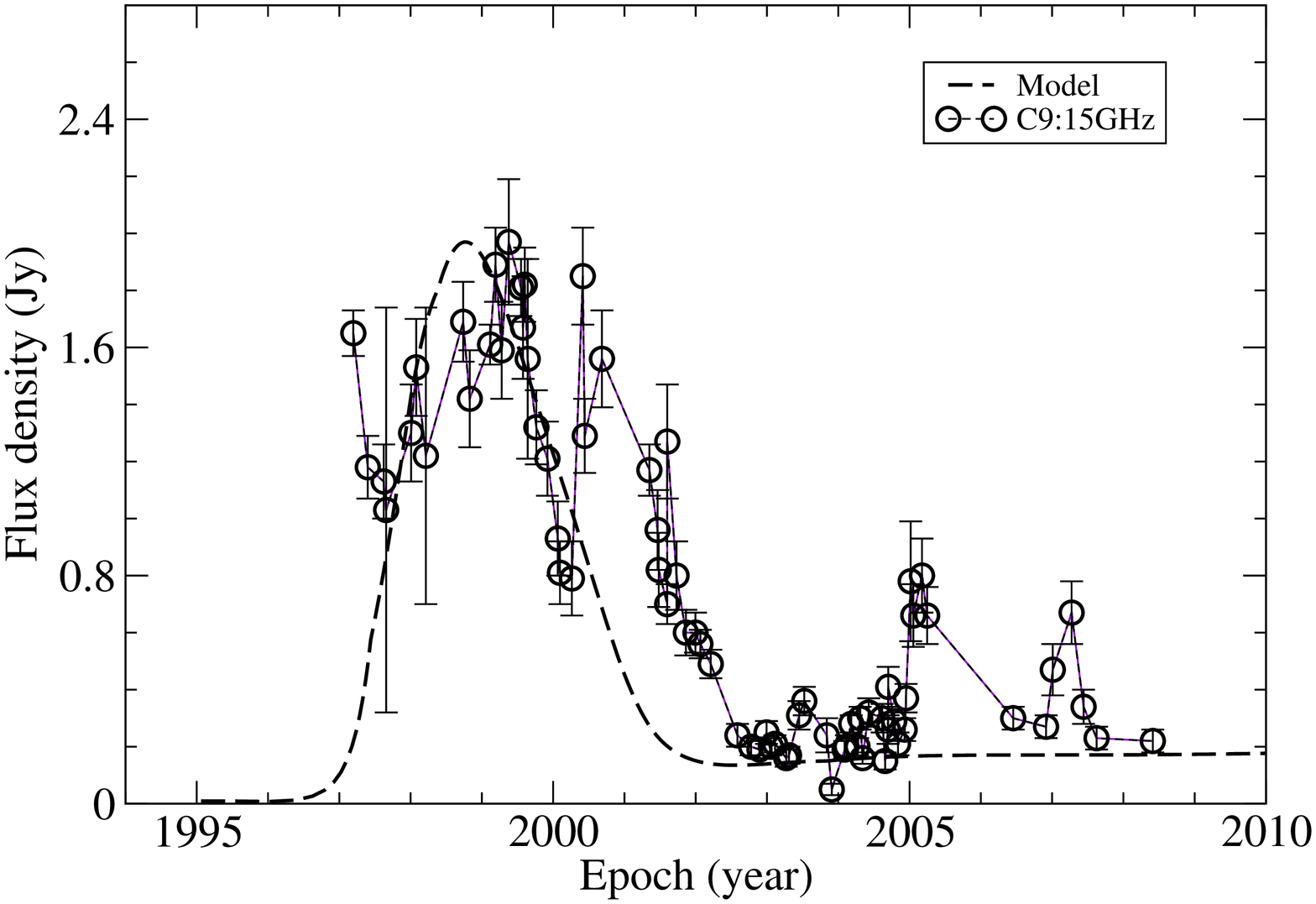}
    \includegraphics[width=5cm,angle=-90]{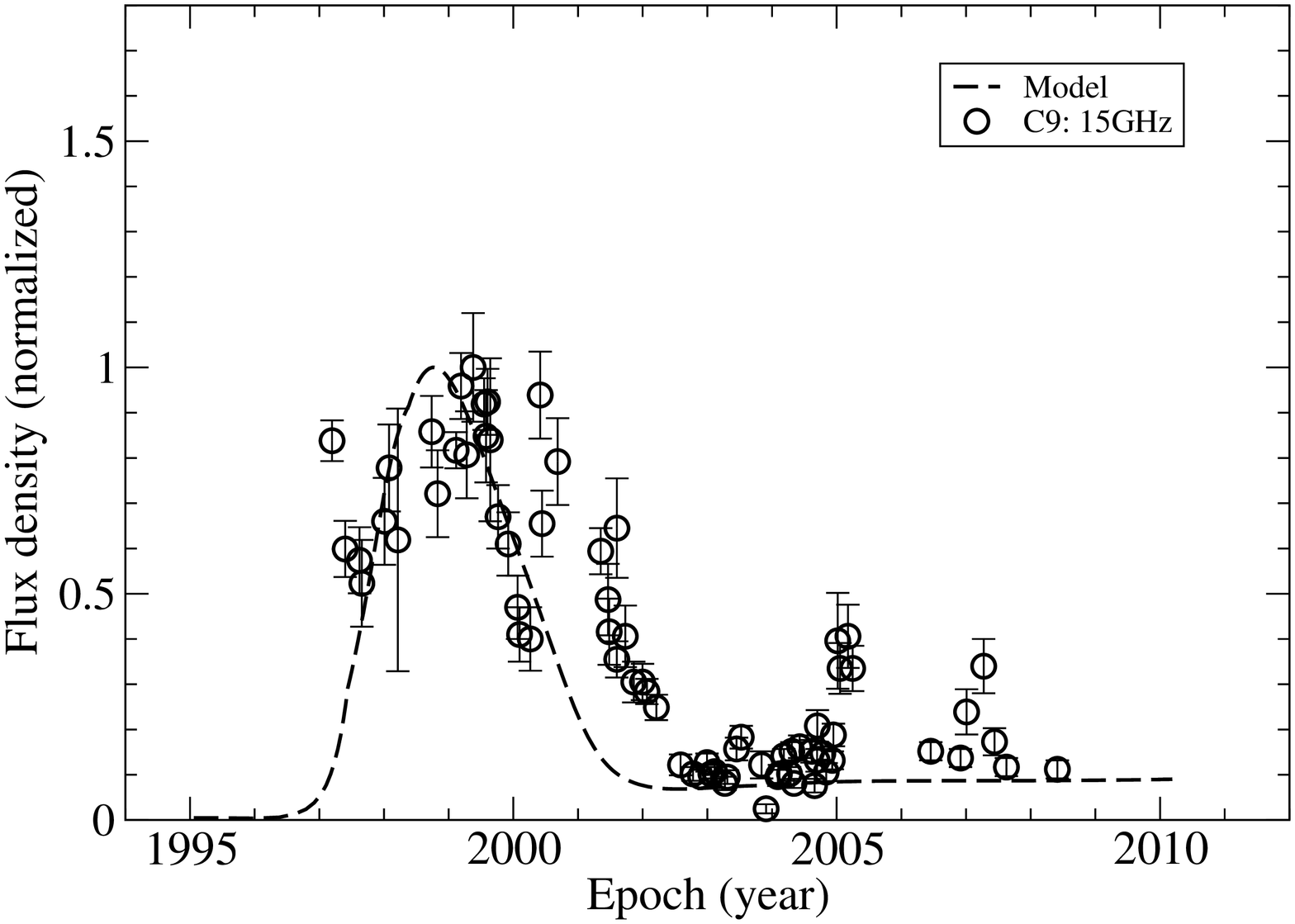}
    \caption{Knot C9. Upper panels: the reformed 43GHz light-curve and its
    normalized light-curve well matched with the Doppler boosting profiles.
    Bottom panels: the reformed 15\,GHz light-curve and its normalized 
    version well coincident with its Doppler boosting profiles. These reformed
    curves more appropriately matched the Doppler-boosting profiles.}
    \end{figure*}
    \subsection{Doppler-boosting effect and flux evolution of knot C9}
     The modeled Doppler factor $\delta$(t) as a continuous function 
     of time shown in Figure  16 (bottom/right panel) had a smooth bump
     structure during 1997--2002, thus providing a distinctly smooth
     Doppler-boosting profile to study the Doppler-boosting effect in the
     flux variations of knot C9. This was a rare and extremely valuable 
     opportunity to test our precessing nozzle scenario and investigate the
     relation between its flux evolution and Doppler-boosting effect.\\
      The light-curves observed at 43\,GHz, 22\,GHz and 15\,GHz  of knot C9 
     are shown in Figure 17.
     In Figure 18 are presented the reformed 15GHz and 43GHz light-curves:
     the rising phase of the observed 43GHz light-curve was reformed by adding 
     the 43GHz flux calculated from the 15\,GHz rising-phase data-points with a
     spectral index $\alpha$(15-43GHz)=0.80. Similarly, the decaying phase
      of the observed 15GHz light-curve was reformed by adding the 15GHz flux 
      calculated from the 43\,GHz decaying-phase data-points.\\
      It can be seen that  the observed and reformed light-curves reveal
     a flaring behavior during  $\sim$1997.50-2000.25, having a close 
     correlation by visual inspection.\\
     Based on the datasets observed at 43\,GHz, 22\,GHz and 15\,GHz we obtained
     $\alpha$(22-43\,GHz)$\simeq$0.70$\pm$0.59 and 
     $\alpha$(15-43\,GHz)$\simeq$0.81$\pm$0.25. 
     In the model-fits to the flux evolution the spectral index was  assumed 
     $\alpha$(15-22-43\,GHz)=0.80.\\
     The modeled maximum Doppler factor $\delta_{max}$=31.86 occurred 
     at $t_{max}$=1998.78. The intrinsic flux densities were assumed to be:
     3.82$\times{10^{-6}}$Jy
     (15\,GHz), 2.58$\times{10^{-6}}$Jy (22\,GHz) and 1.65$\times{10^{-6}}$Jy
     (43\,GHz), corresponding the observed maximum flux densities: 1.97\,Jy 
     (1999.38), 1.33\,Jy (1999.12) and 0.85\,Jy (1999.60), respectively.\\
     In Figure 19 are presented the model-fits to the  light-curves observed 
     at 43\,GHz, 22\,GHz and 15\,GHz (left three panels) and their normalized
     light-curves (right three panels).\\
     In Figure 20 are presented the model-fits to the reformed light-curves
     observed at 43\,GHz and 15\,GHz (left two panels) and their normalized
     light-curves (right two panels). Obviously, the reformed light-curves
     were fitted more appropriately by the Doppler-boosting profiles.\\
   It can be seen from Figures 19 and 20 that both the observed and reformed 
    light-curves  during  1997.5--2000.3 were very well coincident with the 
    Doppler-boosting  profiles
   which was anticipatively-determined through the model-simulation of the
    kinematics of knot C9.\\
     However, there were some shorter time-scale flux
    variations during 2000.5--2008.5 remained to be explained. Since these 
    variations were not related to the Doppler-boosting effect induced by the
    accelerated/decelerated motion of knot C9, they could be due to:
    (i) the variations in intrinsic flux density $S_{int}$ of knot C9 
    and (ii) the variations in its spectral index $\alpha$ (see equation (22)
     in section 3). For example, assuming that the  shorter time-scale 
    variations at 15\,GHz were due to the variations in its intrinsic flux
    density with the spectral index unchanged ($\alpha$(15\,GHz)=0.80), then 
   the variations at 22\,GHz and 43\,GHz could be due to the variations in 
    spectral index ($\alpha$(22\,GHz) and $\alpha$(43\,GHz)) at the two 
    frequencies which should be  different from that at 15\,GHz.\\ 
    Obviously, the interaction of the traveling relativistic shocks 
    (superluminal components) with the surrounding evironments could produce
    these intrinsic variations.\\   
   \section{Interpretation of kinematics and flux evolution for knot C10}
    The flux evolution of knot C10 is a very instructive one, providing another
    valuable opportunity to test our precessing nozzle scenario.
    Its flux variations consisted  of two main flaring events caused by the
     Doppler boosting effect, involving its complex behavior of bulk 
    acceleration/deceleration and change in its trajectory pattern:
    one occurred in the inner precessing common trajectory section, while 
     the  other in the outer individual trajectory section.\\
    We first discuss the model-simulation of its kinematics and then its 
   Doppler-boosting effect and flux evolution.\\ 
    Its precession phase  was modeled as $\phi_0$(rad)=6.14+4$\pi$,
    corresponding to 
    the ejection time $t_0$=1995.76. Its precessing common trajectory 
   apparently extended to $X_n{\sim}$0.35\,mas (or the traveled distance
     ${Z_c}\sim$6.13\,mas along the Z-axis, or $\sim$40.8\,pc from the core).\\
    \subsection{Model simulation of kinematics for knot C10}
    In Figure 21 are presented the curves for the following functions:
    (1) Z(t) describes its traveled distance along the Z-axis from the core
    as a function of time (left panel); (2) functions $\epsilon(t)$ and
    $\psi(t)$ (right panel) describe the jet-axis direction of inner and 
    outer trajectory sections, showing only
    in the inner trajectory section (during the time-interval of 
    1995.76-1999.14 (corresponding to $X_n{\leq}$0.35\,mas) where
     knot C10 moved along the precessing common trajectory.\\ 
     Figure 22 shows the transition from the precessing common 
    trajectory-section to the outer trajectory-section
    at $X_n\sim$0.35\,mas. The curves in black, magenta and blue describe the
    precessing common trajectories for precession phases 
    $\phi_0$=6.14+4$\pi$ and 
    ${\phi_0}\pm$0.31\,rad, respectively. The curve in violet describes the
     model-fit to the whole trajectory.\\
     The results of model-fitting of the whole kinematics for knot C10 
    are presented in Figure 23, including the model-fits of entire trajectory
    $Z_n(X_n)$, core separation $r_n(t)$, coordinates $X_n(t)$ and $Z_n(t)$,
    and the derived apparent velocity $\beta_{app}(t)$ and viewing angle
    $\theta(t)$, the derived bulk Lorentz factor $\Gamma(t)$ and Doppler factor
    $\delta(t)$. It can be seen by visual inspection that the trajectory, core
     separation, coordinates are well fitted by our precessing nozzle model.\\
    As a distinct feature we noticed that, at $\sim$1999.14 its core 
    separation $r_n$ suddenly increased (Fig.23, top/middle panel), which
    was possibly due to the change in the modeled jet-axis direction (see
    Fig.21, right panel), where its motion along the precessing common 
    trajectory ended and transited to its own individual trajectory (Fig.22).
    \subsection{Doppler-boosting effect and flux evolution of knot C10}
     In order to consider the Doppler-boosting effect in the flux-evolution of
      knot 10 we needed to know its spectral index in radio bands.
      Based on the data observed
     at three frequencies (15, 22 and 43\,GHz) we obtained: spectral index
      $\alpha$(22-43GHz)$\simeq$0.64$\pm0.4$ and 
    $\alpha$(15-43GHz)$\simeq$0.90$\pm$0.27.
   Thus in the following model-fitting of the Doppler-boosting effect we
    adopted a spectral index $\alpha$=0.80 in the whole band of 
    15 to 43\,GHz. The Doppler-boosting profile was then defined by
     $[\delta(t)/{\delta}_{max}]^{3.80}$ (Figs.25 and 26),
     where $\delta_{max}$=14.11 was the maximum Doppler factor of the first
     flare event at 1998.36 in our fitting model.\\
     The radio light-curves observed at 15, 22 and 43\,GHz are shown 
    in Figure 24. The observed light-curves at 15, 22 and 43\,GHz were 
     normalized by using the observed maximum fluxes of the first flare,
     taken as 2.17, 1.60 and 0.94\,Jy, respectively. Thus
     the corresponding intrinsic fluxes at 15, 22 and 43\,GHz were chosen as
     $S_{in}$=9.3$\times{10^{-5}}$Jy, 6.8$\times{10^{-5}}$Jy and 
     4.0$\times{10^{-5}}$Jy, respectively. \\
       As shown in Figure 23 the anticipatively derived bulk Lorentz factor 
     showed two-bumps: one occurred in the inner trajectory-section (or 
     the precessing common trajectory before 1999.14) with its 
     $\Gamma_{max}$=8.5 at epoch 1998.36), while the other occurred
     in the outer trajectory section with its $\Gamma_{max}$=17.0 at epoch 
     2000.92. Specifically, its bulk Lorentz factor was modeled to indicate
     its behavior of acceleration/deceleration to re-acceleration/deceleration
     : for Z$\leq$0.6\,mas (1995.76-1997.00) $\Gamma$=3.0; for Z=0.6-4.2\,mas 
     (1997.00-1998.38) $\Gamma$ increased from 3.0 to 8.5; for Z=4-6.4\,mas
      (1998.38-1999.41) $\Gamma$ decreased from 8.5 to 4.0; for Z=6.4-12.1\,mas
    (1999.41-2000.92) $\Gamma$ increased from 4.0 to 17.0; for Z=12.1-23.2\,mas
    (2000.92-2003.87) $\Gamma$ decreased from 17.0 to 7.0 and then kept to be
    const.=7.0 (Fig.23, bottom/right panel).\\
     Correspondingly, the derived  apparent velocity $\beta_{app}$(t) also
     showed a double-bump structure as shown in Figure 23 (bottom/middle 
     panel).\\
      However, the derived Doppler factor showed three bumps, the first one
      occurred in the precessing common 
     trajectory section ($\delta_{max}$=14.11 at epoch 1998.36), while 
     the other two  occurred in the outer trajectory section with their 
    $\delta_{max}$=11.35 at 1999.98 and $\delta_{max}$=13.89 at 2002.00,
    respectively.\\
      It can be seen from Figures 25 and 26 that the observed flux light-curves 
      were similar to the Doppler-boosting profiles, but had much complex
    structures. Most interestingly, the 22\,GHz light-curves (both observed
    and normalized) were extremely well fitted by the Doppler-boosting profiles
    as shown  in Figure 25.\\
   The 15\,GHz and 43\,GHz light-curves shown in Figure 26 were also fitted 
   quite well by the Doppler-boosting profiles. However, at the two frequencies
   a number of data-points largely deviated  from the modeled Doppler-boosting
    profiles. This phenomenon is quite similar to that observed in knot C9 
   (Figs.19 and 20).\\
   As an example, during $\sim$2000.0-2001.5 the data-points observed at 15
    and 43\,GHz distributed much higher than  the modeled Doppler-boosting
    profiles (Fig. 26). This could be due to the following effects: 
    (1) at 15\,GHz knot C10 had its spectral index to be larger than 0.80 
    or its intrinsic flux density being larger
     than the adopted value  9.3${\times}10^{-5}$Jy; (2) at 43\,GHz knot
      C10  had its spectral index to be smaller than 0.80 or its intrinsic
      flux being larger than the adopted value 4.0${\times}10^{-5}$Jy. 
     Additionally, both the spectral indexes and intrinsic fluxes at 15 and 
     43\,GHz were variable with time.\footnote{No such behavior was clearly
     observed at 22\,GHz: knot C10 has a spectral index 0.80 and a steady 
     intrinsic flux 6.8${\times}10^{-5}$Jy at this frequency.}\\
      It worths noting that for knot C10 the selection of the model-parameters
    was quite difficult, because the model-fits involved multiple functions 
    simultaneously (requiring model-fits to its trajectory, 
    coordinates, core separation, apparent velocity and flux evolution 
    (complex light-curves) at multi-frequencies. All these observed properties 
    were functions of time. Thus the successful interpretation of its 
    kinematics and flux evolution seems very encouraging for validating our
    precessing nozzle scenario.   
    \begin{figure*}
    \centering
    \includegraphics[width=5cm,angle=-90]{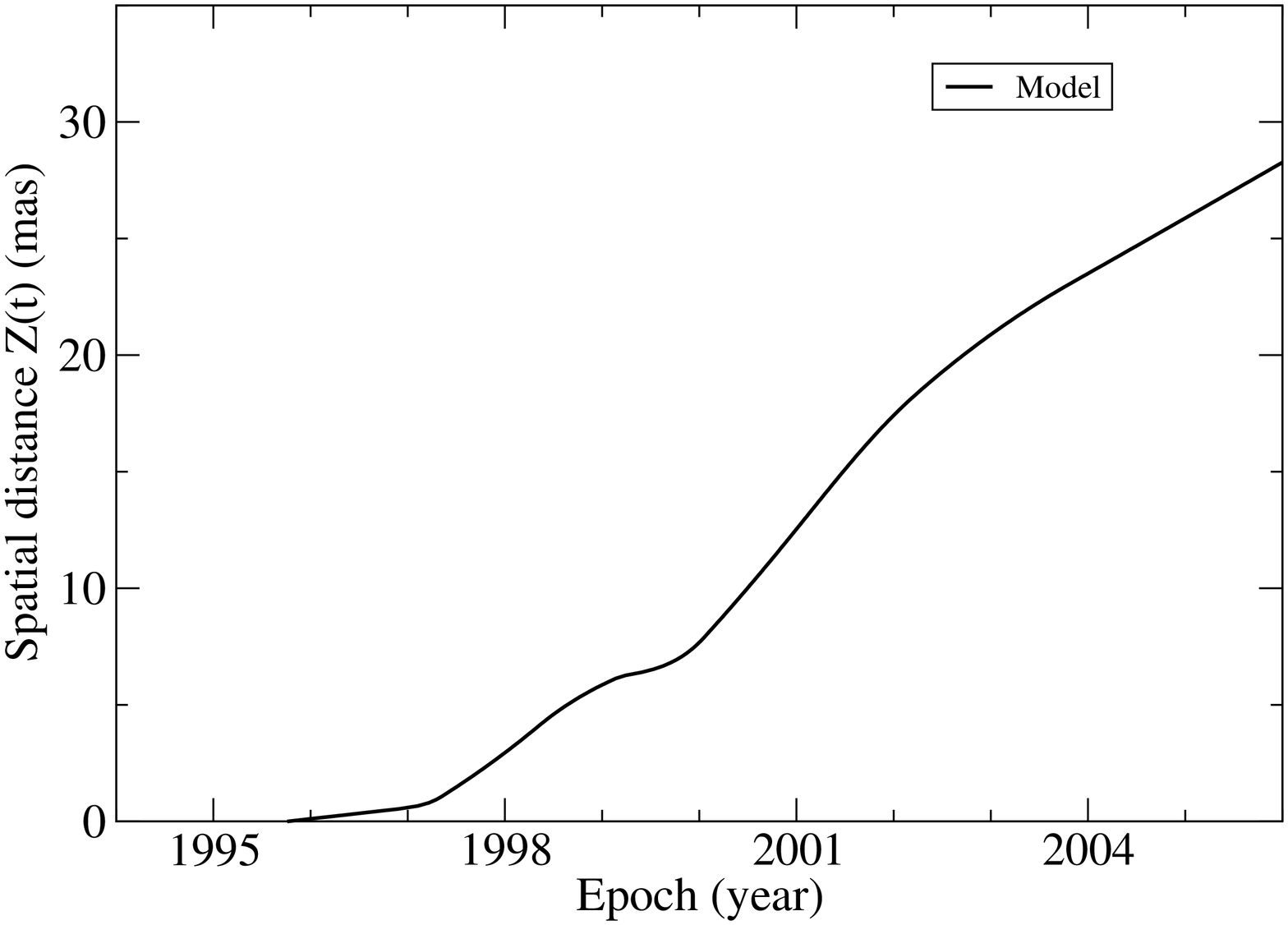}
    \includegraphics[width=5cm,angle=-90]{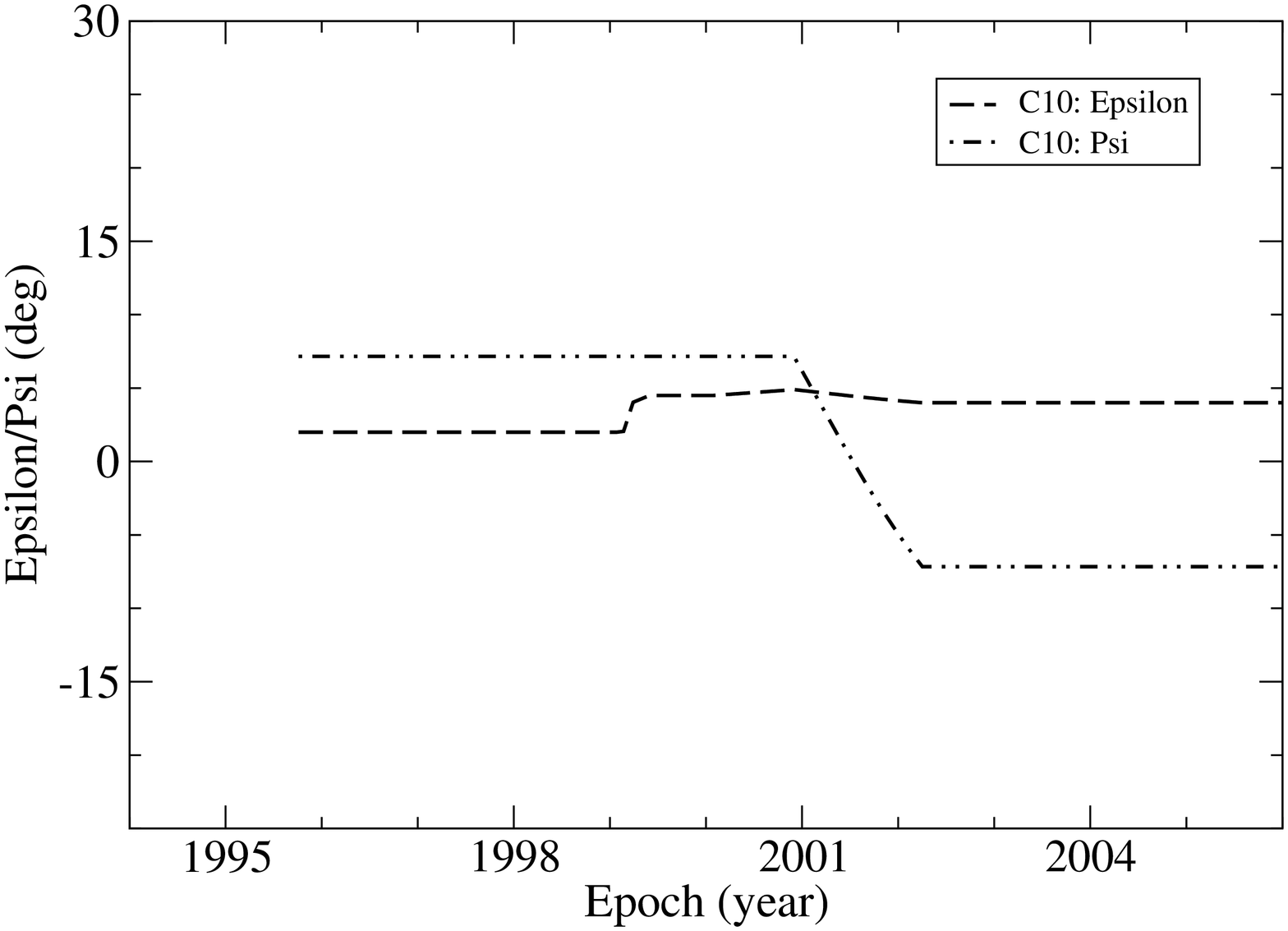}
    \caption{Knot C10. Left panel: the traveled distance Z(t) from the core
    (or spatial distance) along the Z-axis. Right panel: parameters
     $\epsilon(t)$ and $\psi(t)$ which define the modeled direction
    of jet-axis. Before 1999.14 ($X_n{<}$0.35\,mas) $\epsilon$=$2.0^{\circ}$
     and $\psi$=$7.16^{\circ}$, and knot C10 moved along
    the precessing common trajectory. After 1999.14 ($X_n{>}$0.35\,mas) 
    $\epsilon$ started to increase and knot C10 started to move
     along its own individual trajectory in its outer trajectory-section.
    Parameter $\psi$ started to decrease from 2000.92 ($X_n{>}$1.19\,mas)
    and its outer trajectory pattern changed once again. }
    \end{figure*}
     \begin{figure*}
    \centering 
    \includegraphics[width=6cm,angle=-90]{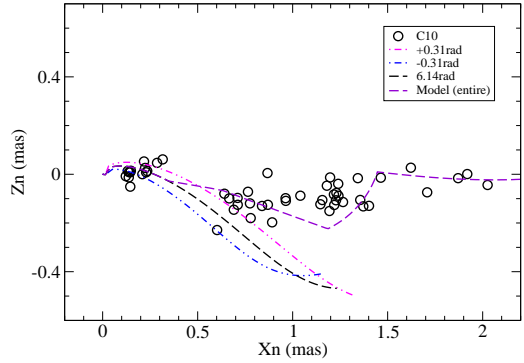}
    \caption{Knot C10. Model-fit to the inner trajectory section in terms of
    the  precessing  common trajectory which apparently extended to
     $X_n{\sim}$0.35\,mas (corresponding spatial distance ${Z_c}\sim$6.13\,mas
    or 40.8\,pc). The curve 
     in violet represents the  model-fit to its whole trajectory, while
    the curves in black, magenta and blue represent its precessing common 
    trajectories for precession phases $\phi_0$(rad)=6.14+4$\pi$ and
     6.14$\pm$0.31+4$\pi$, respectively.}
    \end{figure*}
    \begin{figure*}
     \centering
     \includegraphics[width=4.5cm,angle=-90]{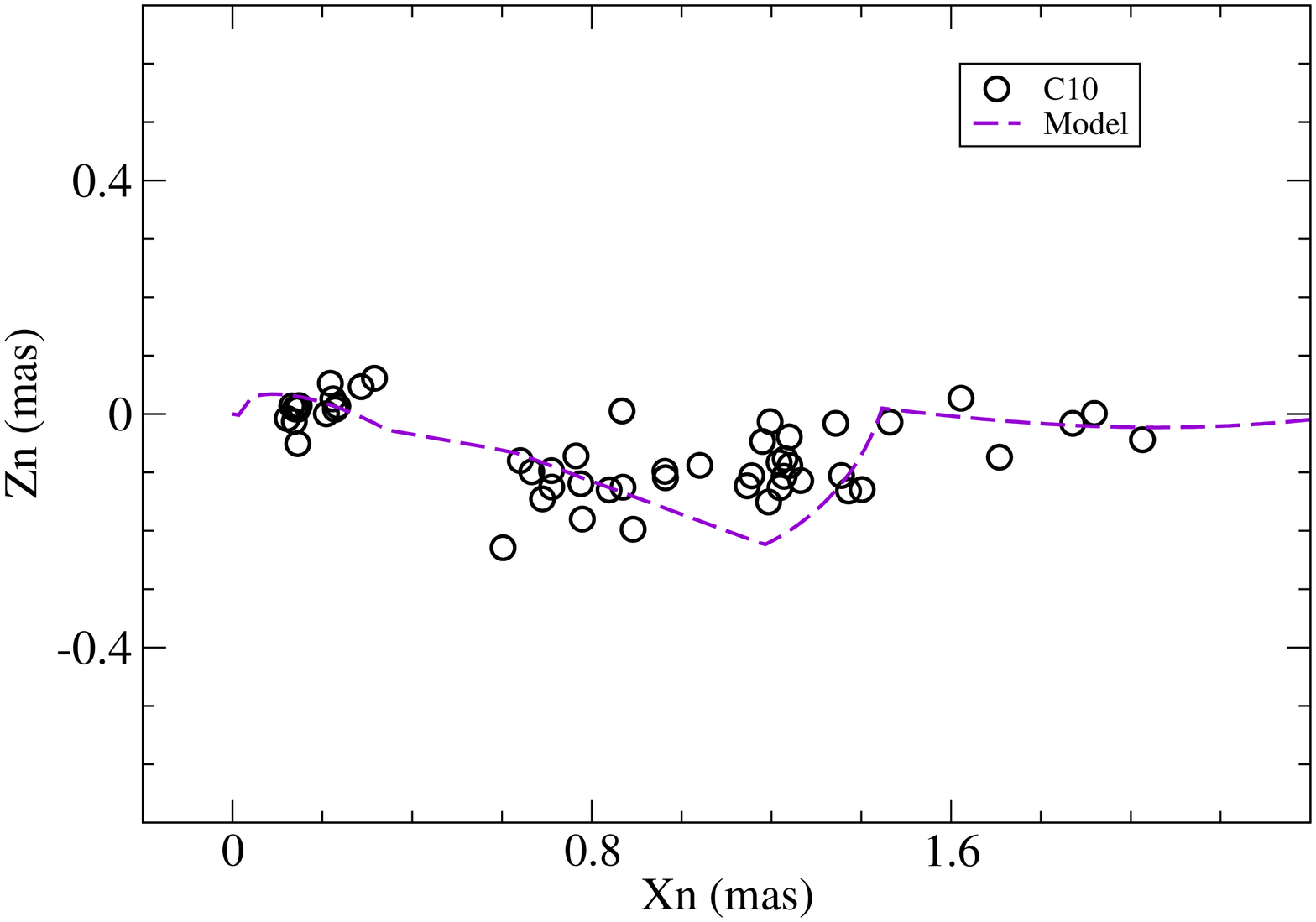}
     \includegraphics[width=4.5cm,angle=-90]{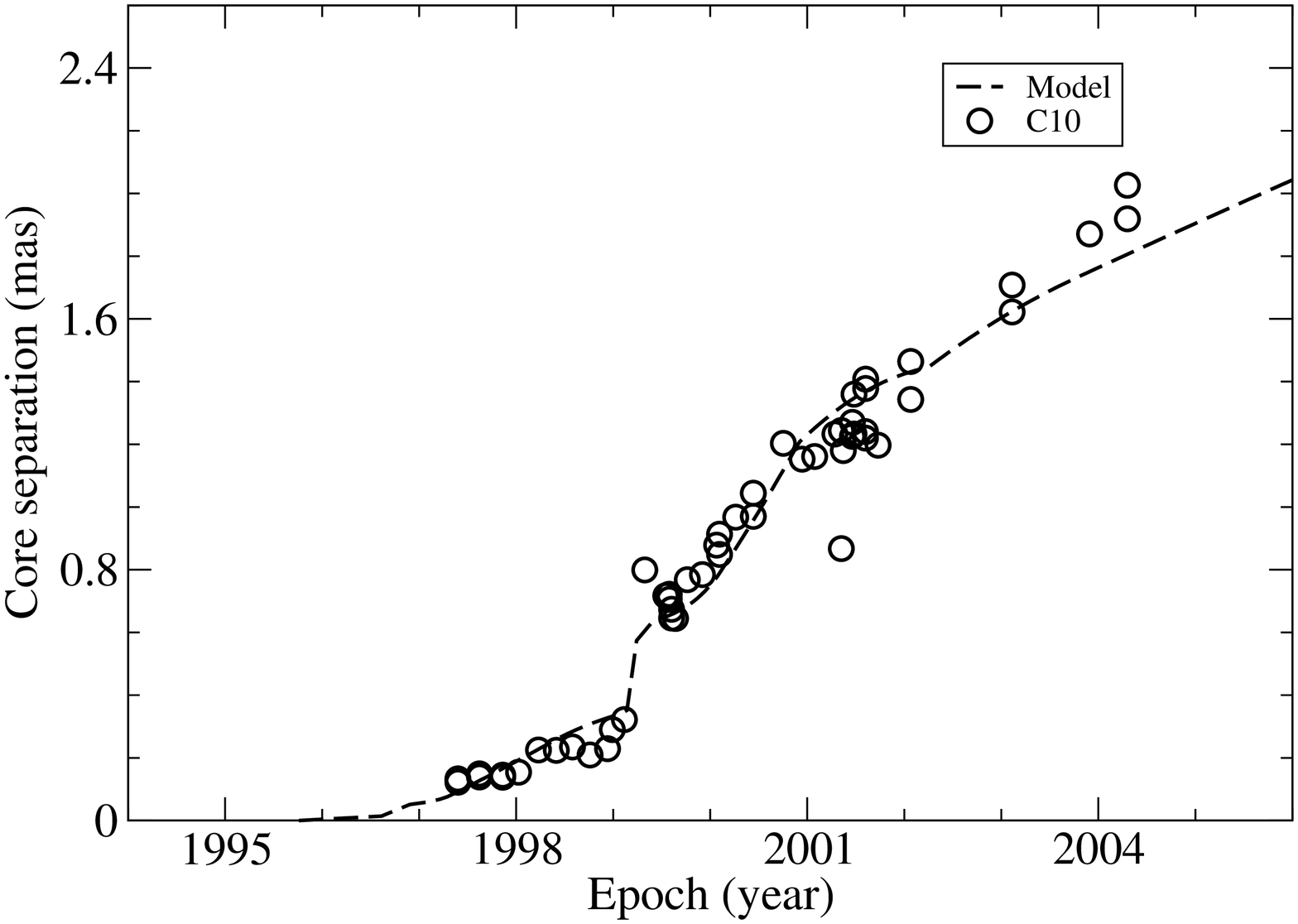}
     \includegraphics[width=4.5cm,angle=-90]{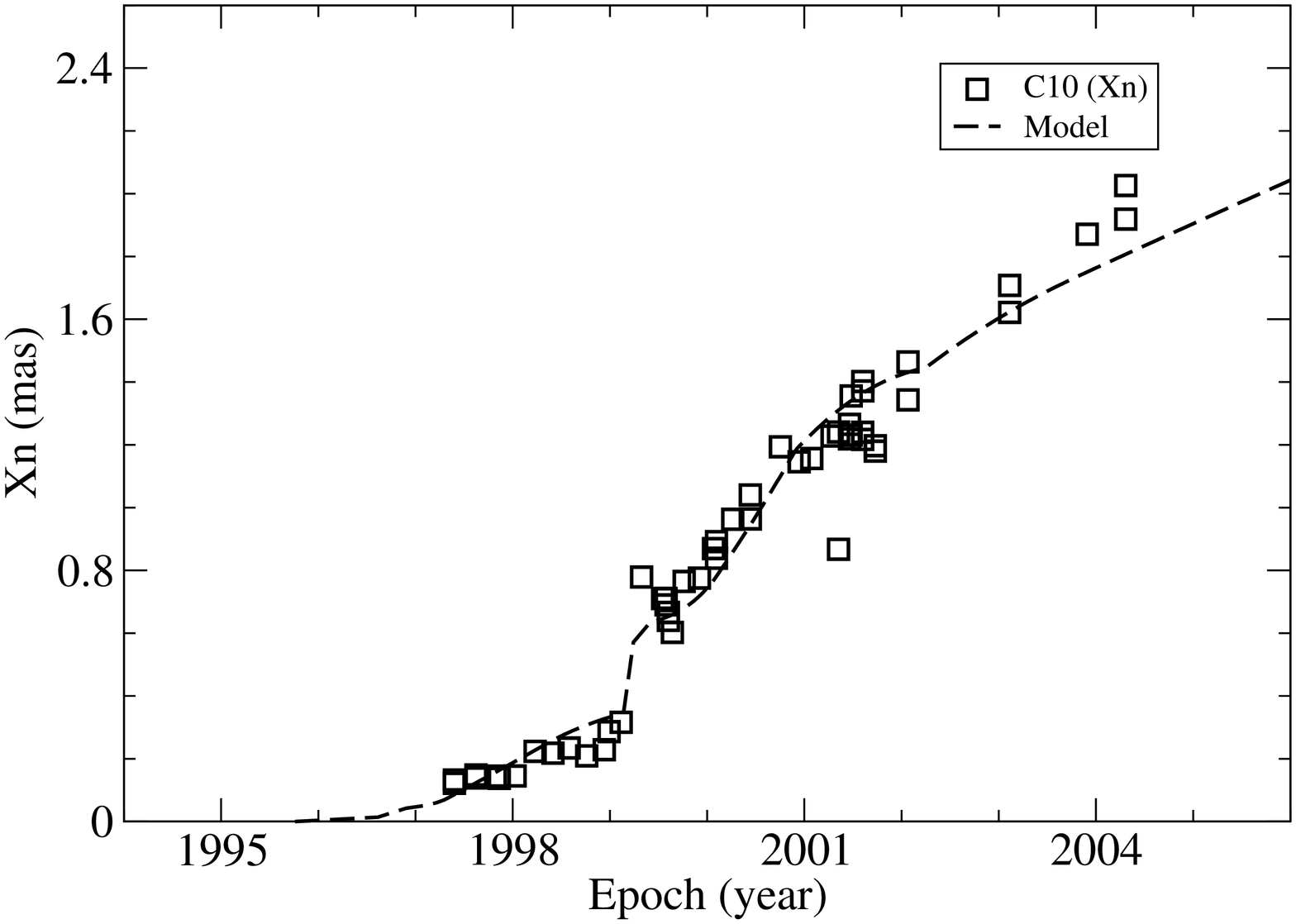}
     \includegraphics[width=4.5cm,angle=-90]{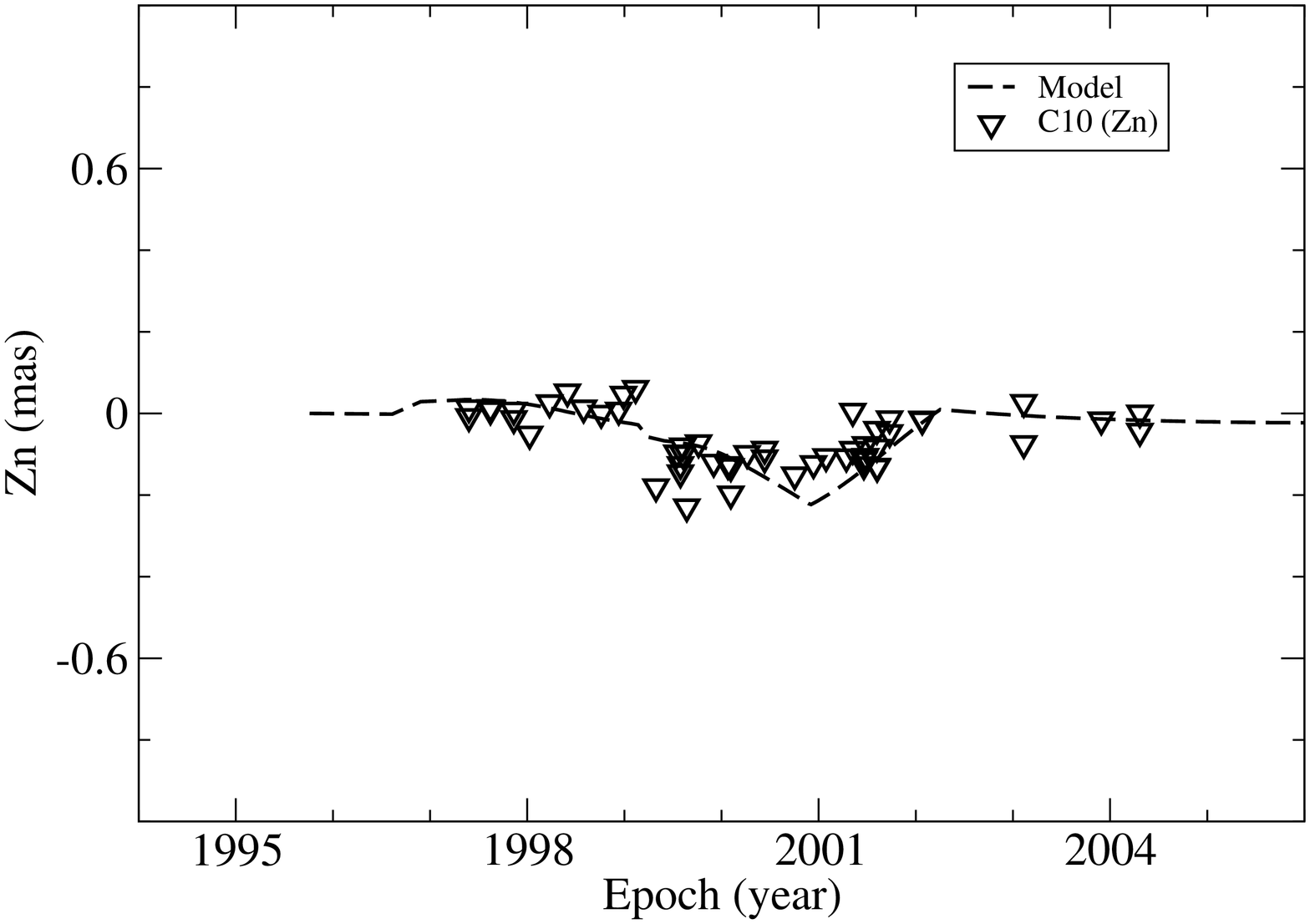}
     \includegraphics[width=4.5cm,angle=-90]{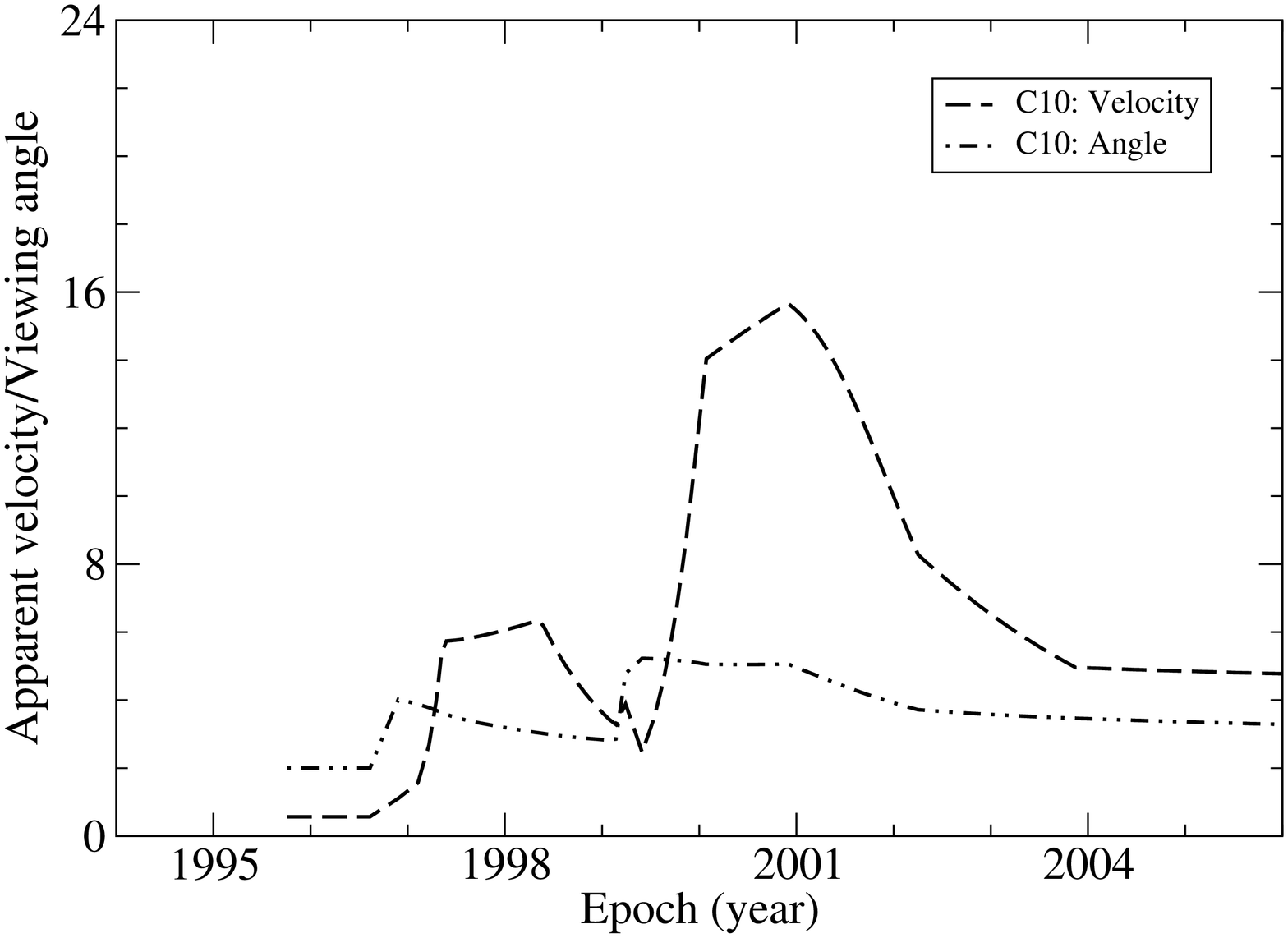}
     \includegraphics[width=4.5cm,angle=-90]{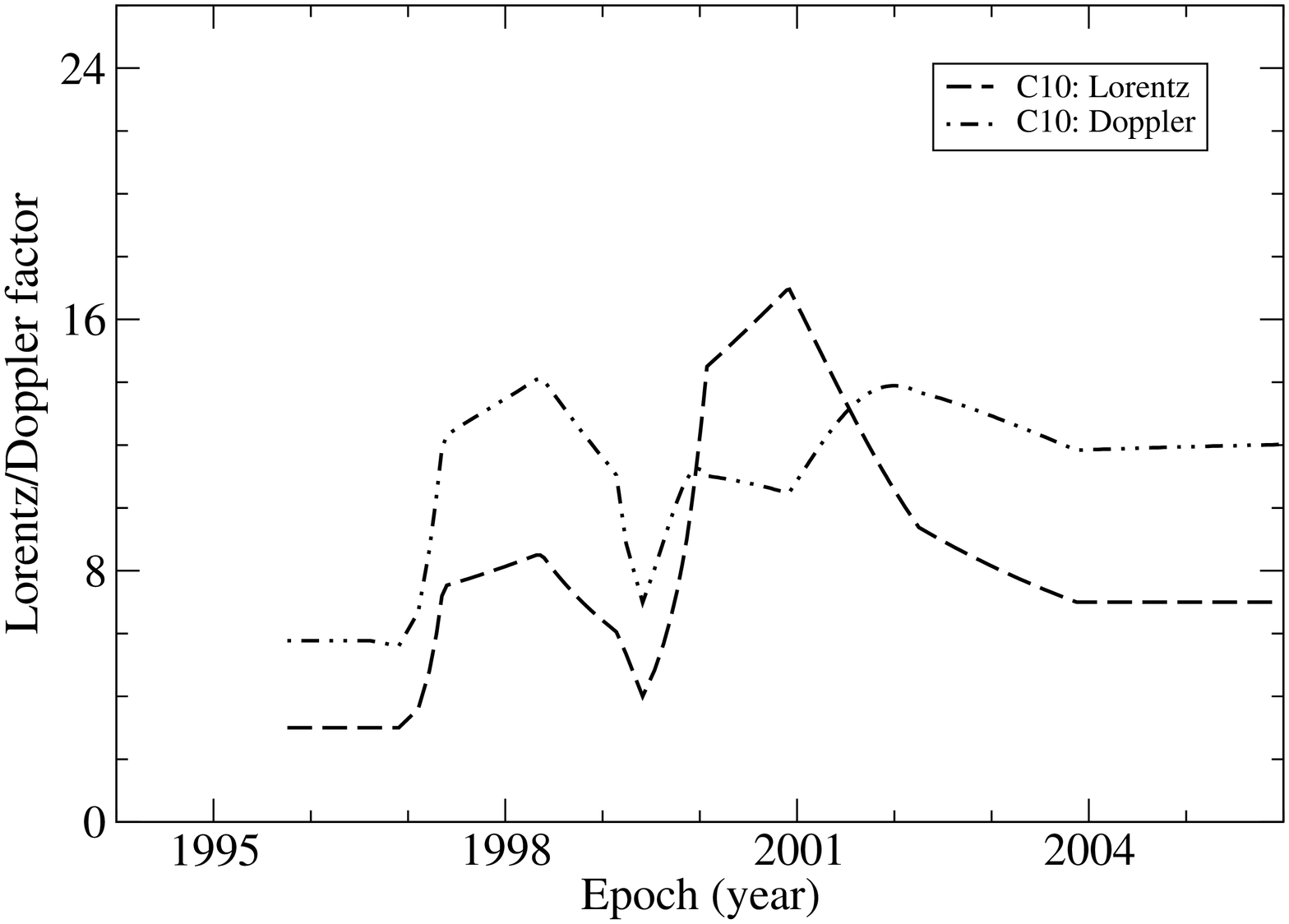}
     \caption{Knot C10: Model simulation  of the VLBI-kinematics.
     Upper panels: apparent trajectory $Z_n(X_n)$ , core separation $r_n(t)$
     and coordinate $X_n(t)$. Bottom panels: coordinate $Z_n(t)$, 
    apparent speed $\beta_{app}(t)$/viewing angle $\theta(t)$  and bulk
     Lorentz factor $\Gamma(t)$/Doppler factor $\delta(t)$. It is noted that
     the sudden increase in core separation $r_n$ at 1999.14 was due to a 
     sudden change in the modeled jet-axis direction and its trajectory 
     started to deviate from the
     precessing common trajectory. Both apparent velocity and bulk Lorentz
     factor show a two-bump structure, while the modeled Doppler factor shows
     a three-bump structure with peaks at 1998.36, 1999.98 and 2002.00,
      respectively.}
     \end{figure*}
    \begin{figure*}
    \centering
    \includegraphics[width=7cm,angle=-90]{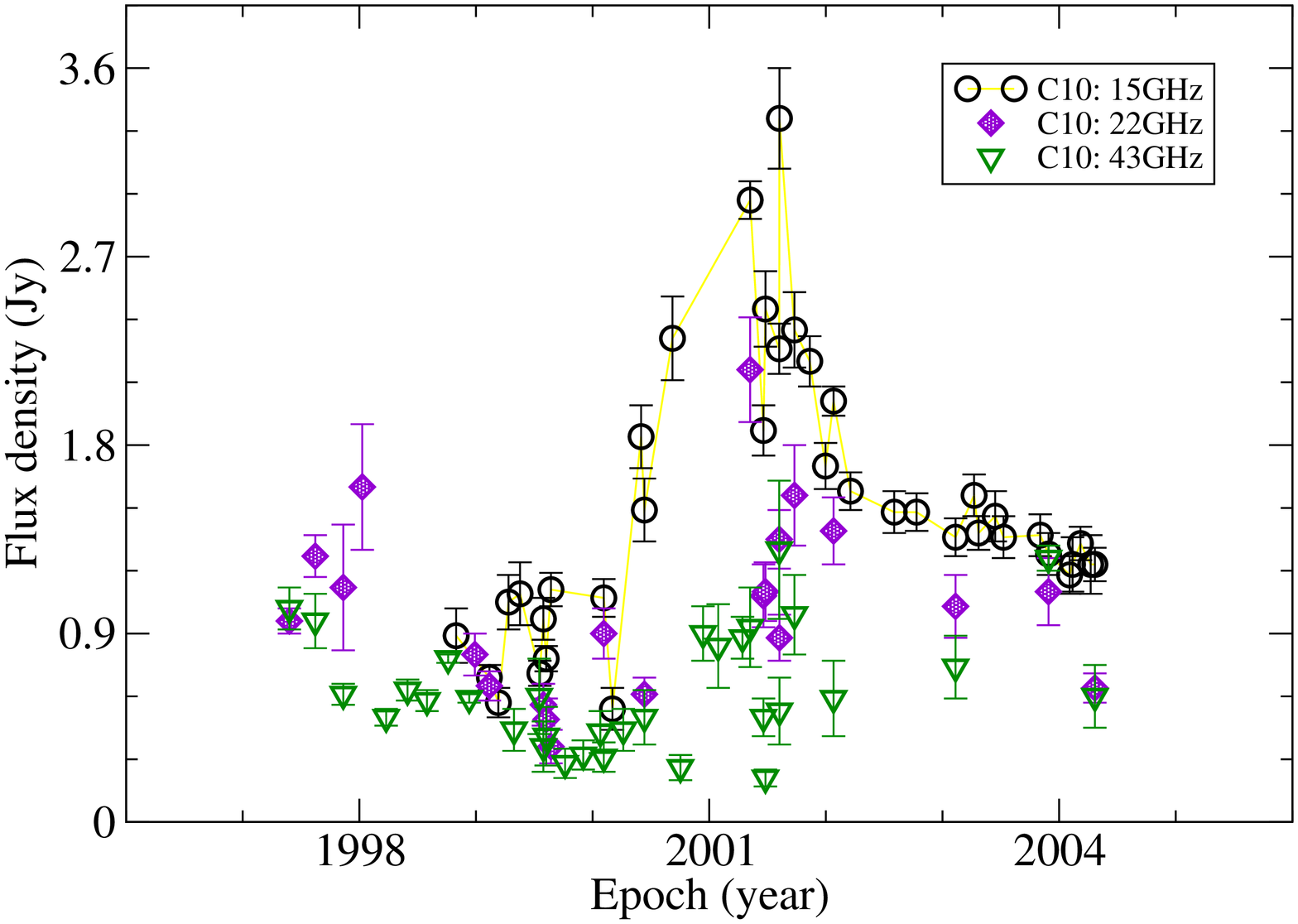}
    \caption{Knot C10: light-curves observed at 15\,GHz, 22\,GHz and 43\,GHz.}
    \end{figure*}
   \begin{figure*}
   \centering
   \includegraphics[width=5cm,angle=-90]{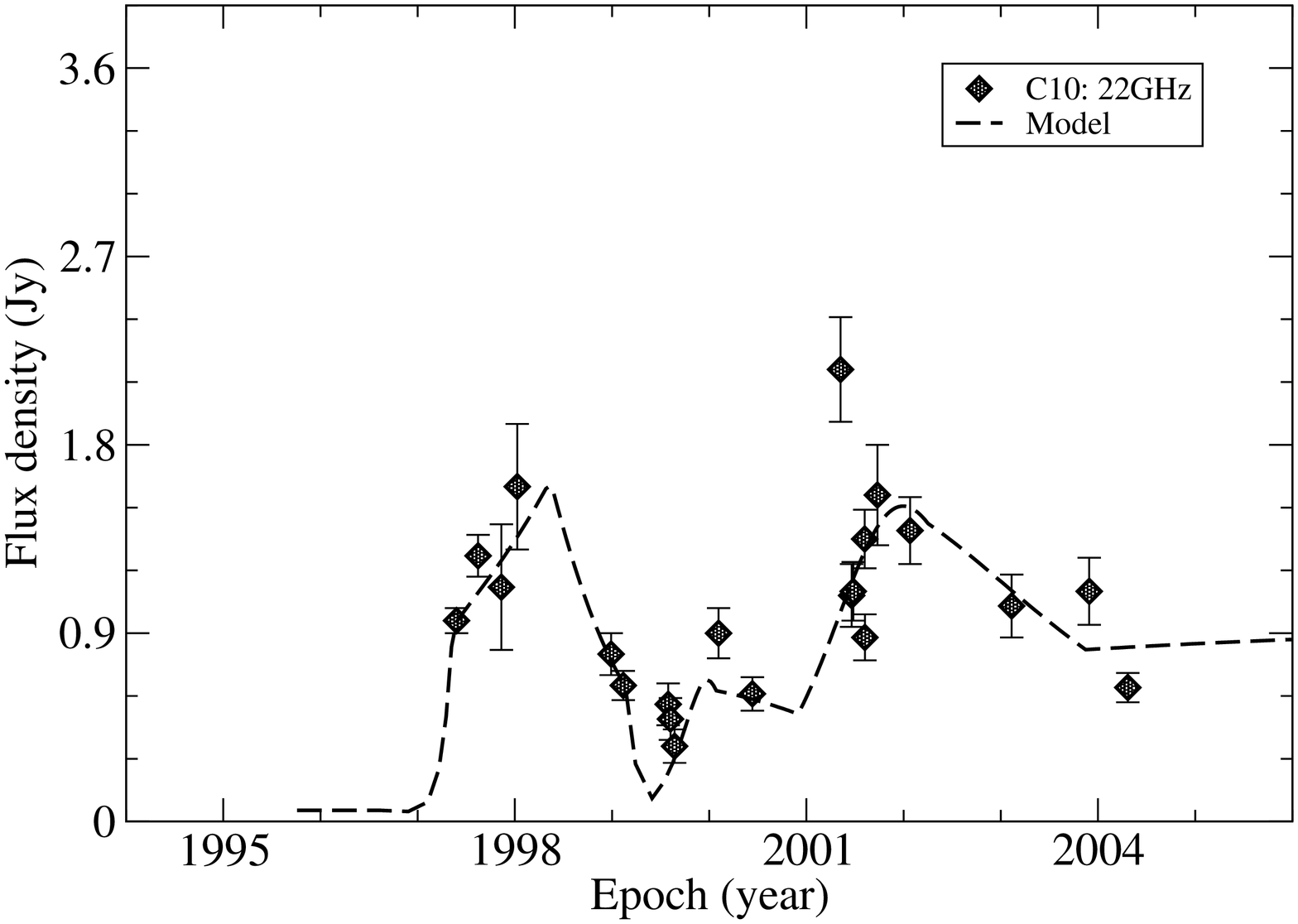}
   \includegraphics[width=5cm,angle=-90]{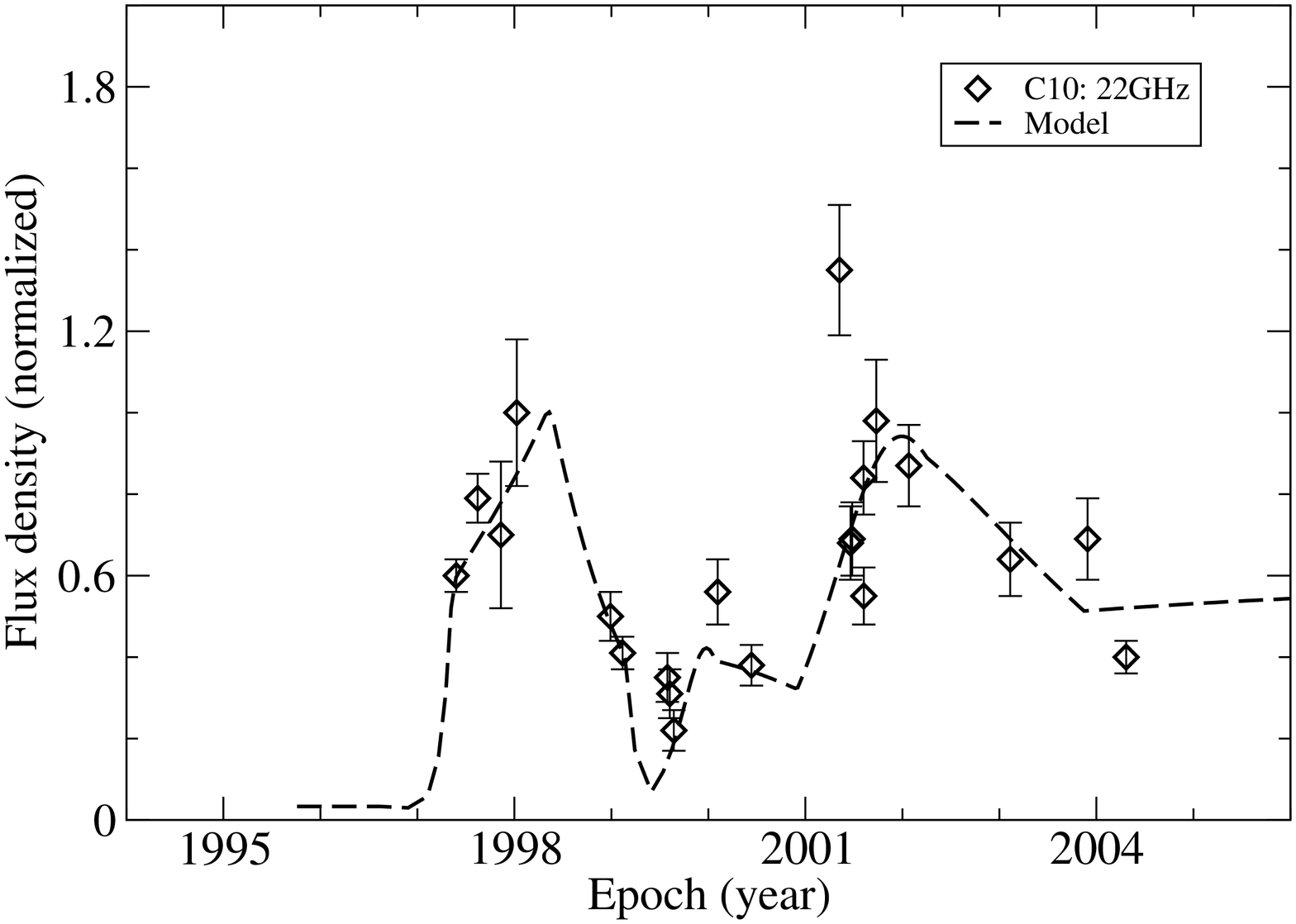}
   \caption{Knot C10: The 22\,GHz variations were extremely well fitted by 
   the Doppler boosting profile with an adopted spectral index $\alpha$=0.80
   for the  observed flux light-curve (left panel) and the normalized flux
   light-curve (right panel).}
   \end{figure*}
   \begin{figure*}
   \centering
   \includegraphics[width=5cm,angle=-90]{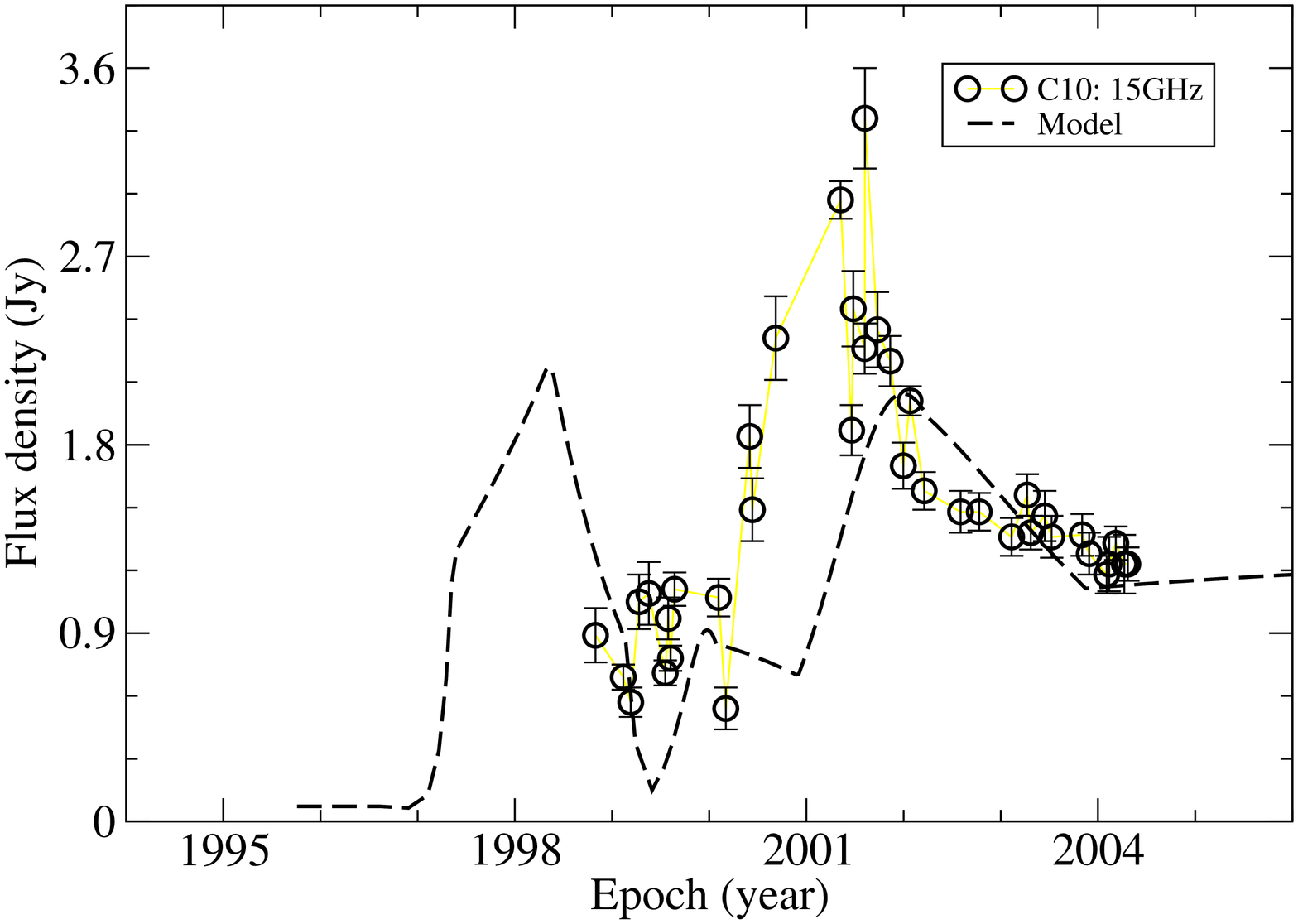}
   \includegraphics[width=5cm,angle=-90]{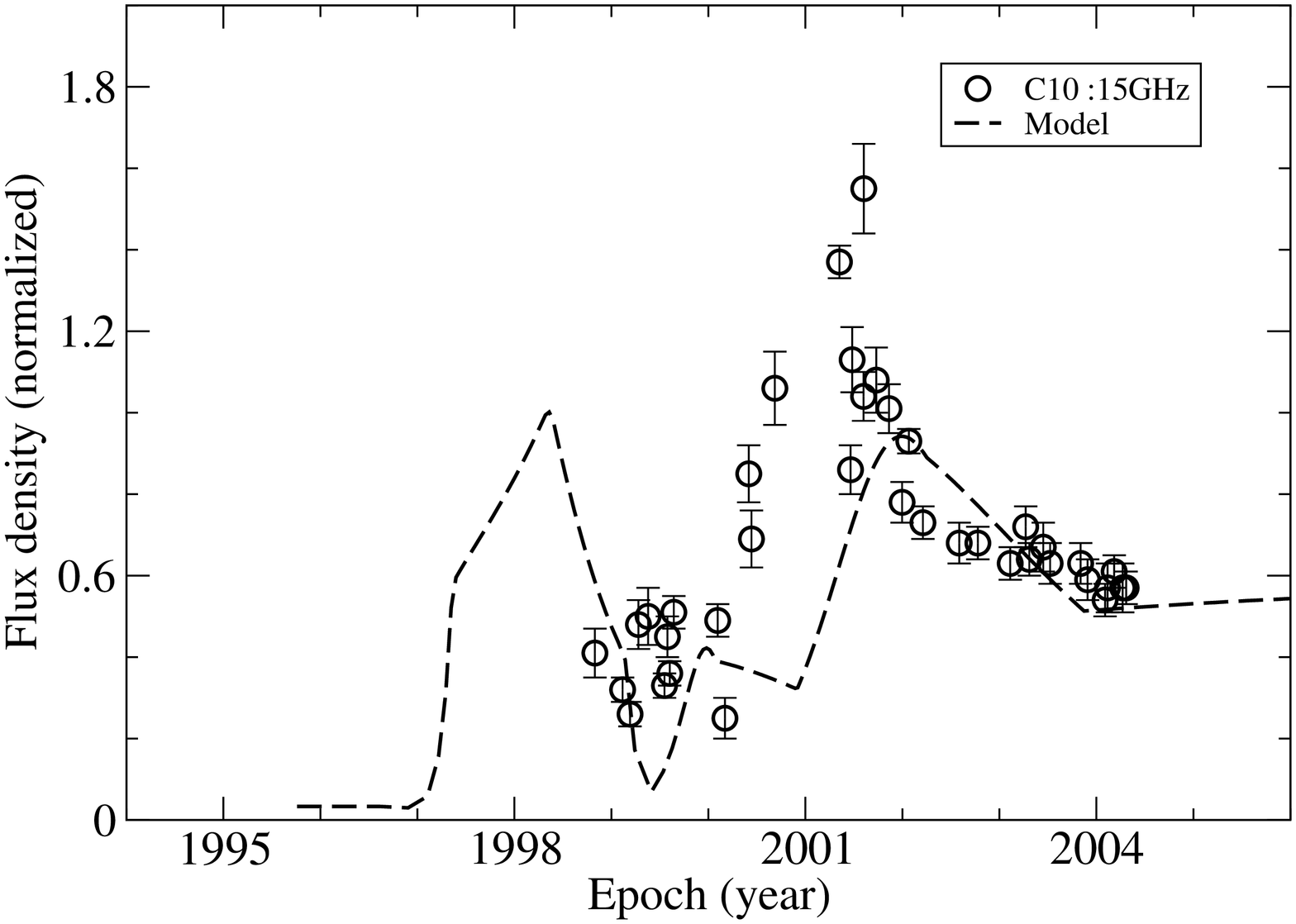}
   \includegraphics[width=5cm,angle=-90]{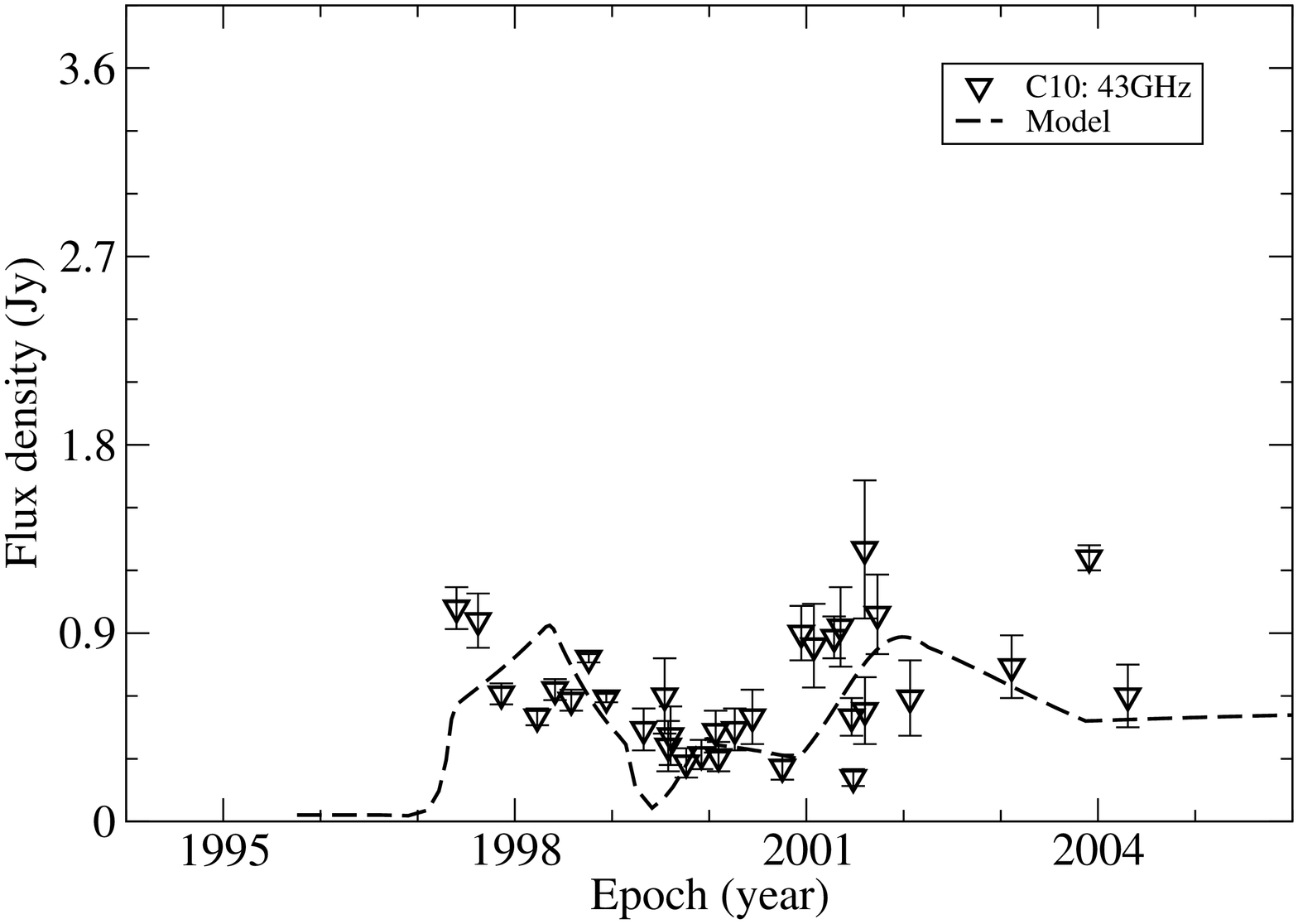}
   \includegraphics[width=5cm,angle=-90]{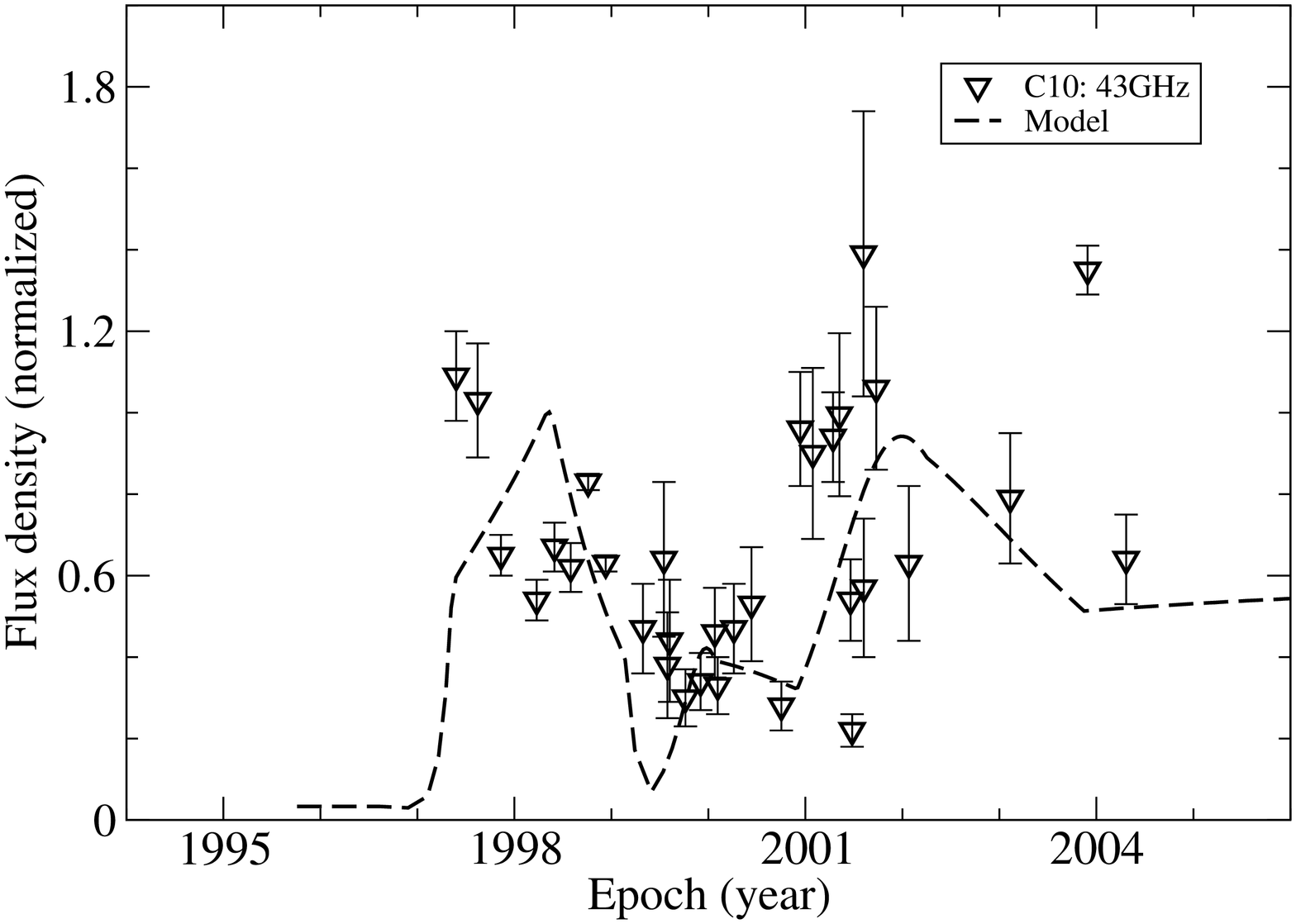}
   \caption{Knot C10: The 15GHz and 43GHz light-curves well fitted by the 
    Doppler-boosting profiles. Variations in its intrinsic flux density and
    spectral index at the two frequencies probably gave rise to the shorter
    time-scale flux fluctuations.} 
   \end{figure*} 
   \begin{figure*}
   \centering
   \includegraphics[width=4.5cm,angle=-90]{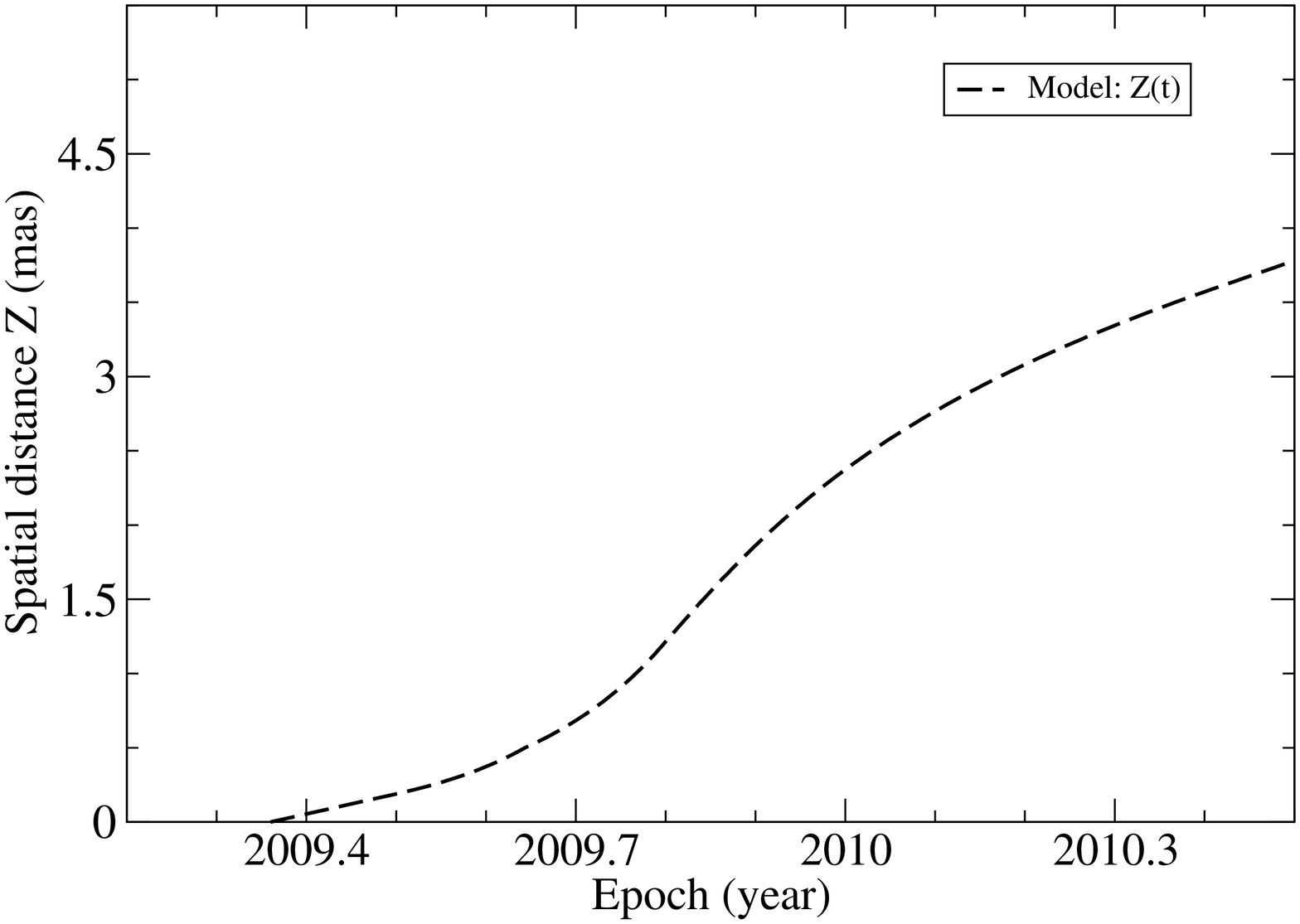}
   \includegraphics[width=4.5cm,angle=-90]{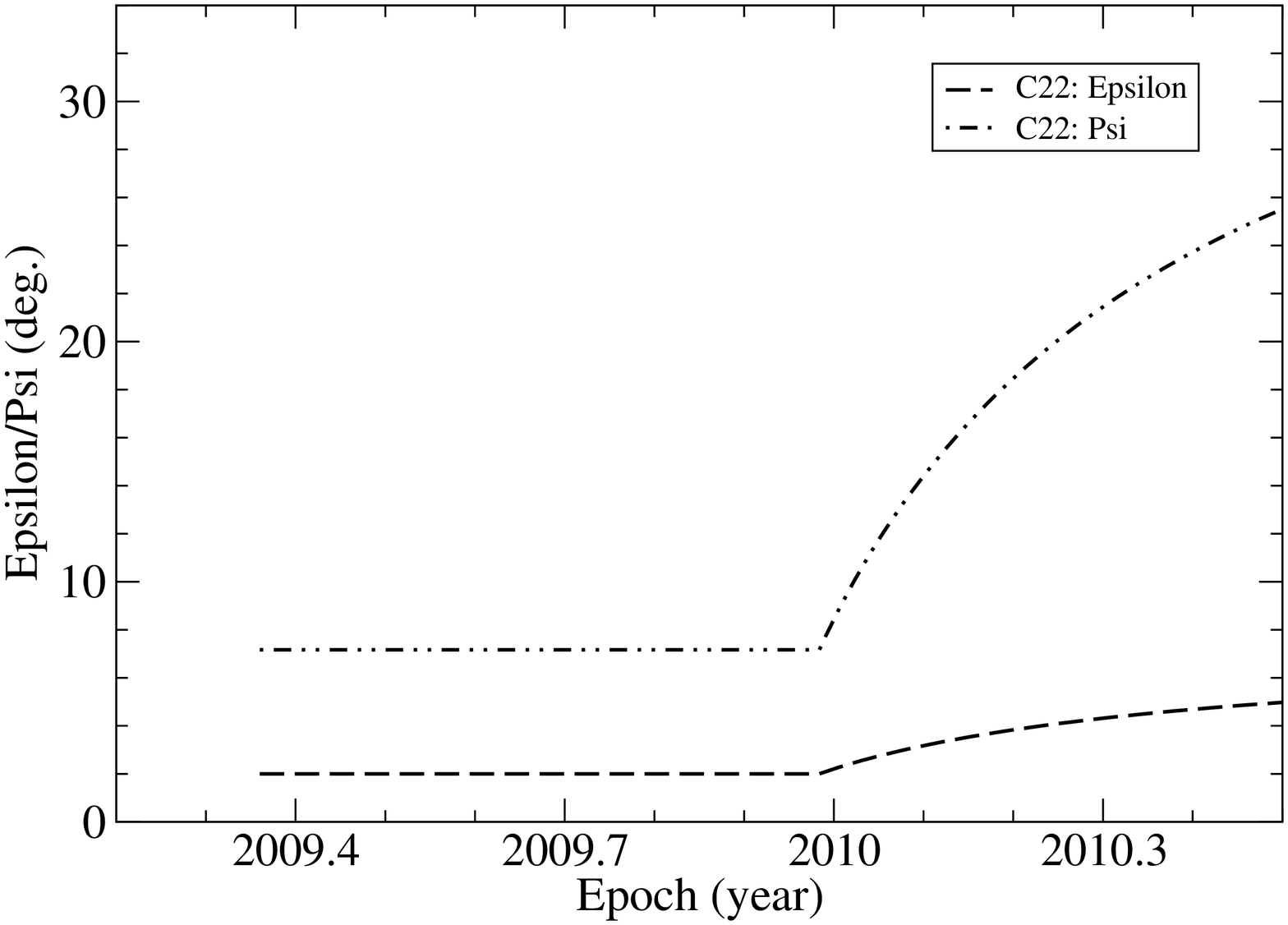}
   \caption{Knot C22: the traveled distance along the Z-axis (left panel)
    and the curves of $\epsilon$ and $\psi$ (right panel), describing the
    change in the modeled jet-axis direction. Before 2009.98 (Z$\leq$2.3\,mas)
    $\epsilon$=$2.0^{\circ}$ and $\psi$=$7.16^{\circ}$, knot C22 moved along
    the precessing common trajectory in its inner trajectory-section. After
    2009.98, $\epsilon$ and $\psi$ started to increase and knot C22 started
    to move along its own individual trajectory in its outer 
   trajectory-section.}
   \end{figure*}
   \begin{figure*}
   \centering
   \includegraphics[width=6cm,angle=-90]{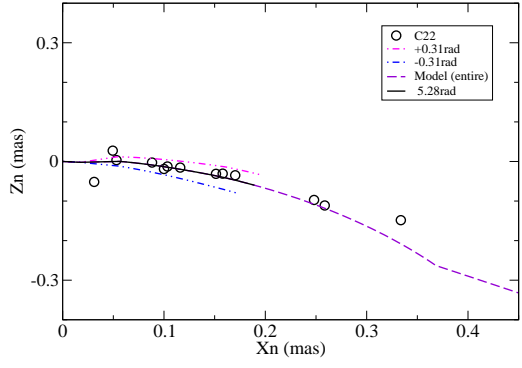}
   \caption{Knot C22: Model-fit to its inner trajectory-section by the 
    precessing common trajectory. Black curve represents the precessing common
   trajectory-section for the precession phase 5.28\,rad+8$\pi$, 
    the curves in magenta
    and blue represent the precessing common trajectory sections for
    precession phases 5.28+8$\pi$$\pm$0.31\,rad, respectively. the curve in violet
   represents the model fit to the whole trajectory.}
  \end{figure*}
   \begin{figure*}
   \centering
   \includegraphics[width=4.5cm,angle=-90]{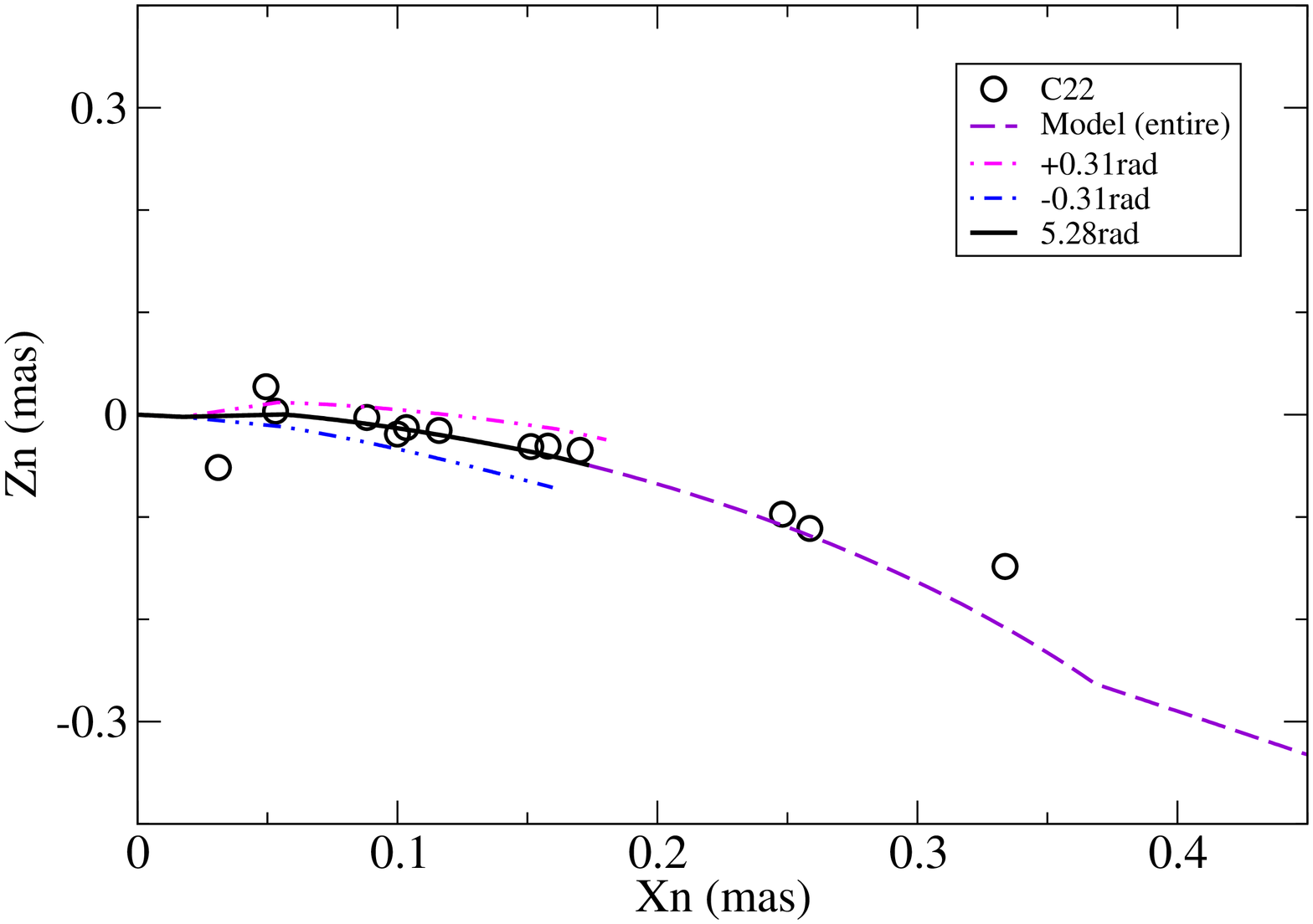}
   \includegraphics[width=4.5cm,angle=-90]{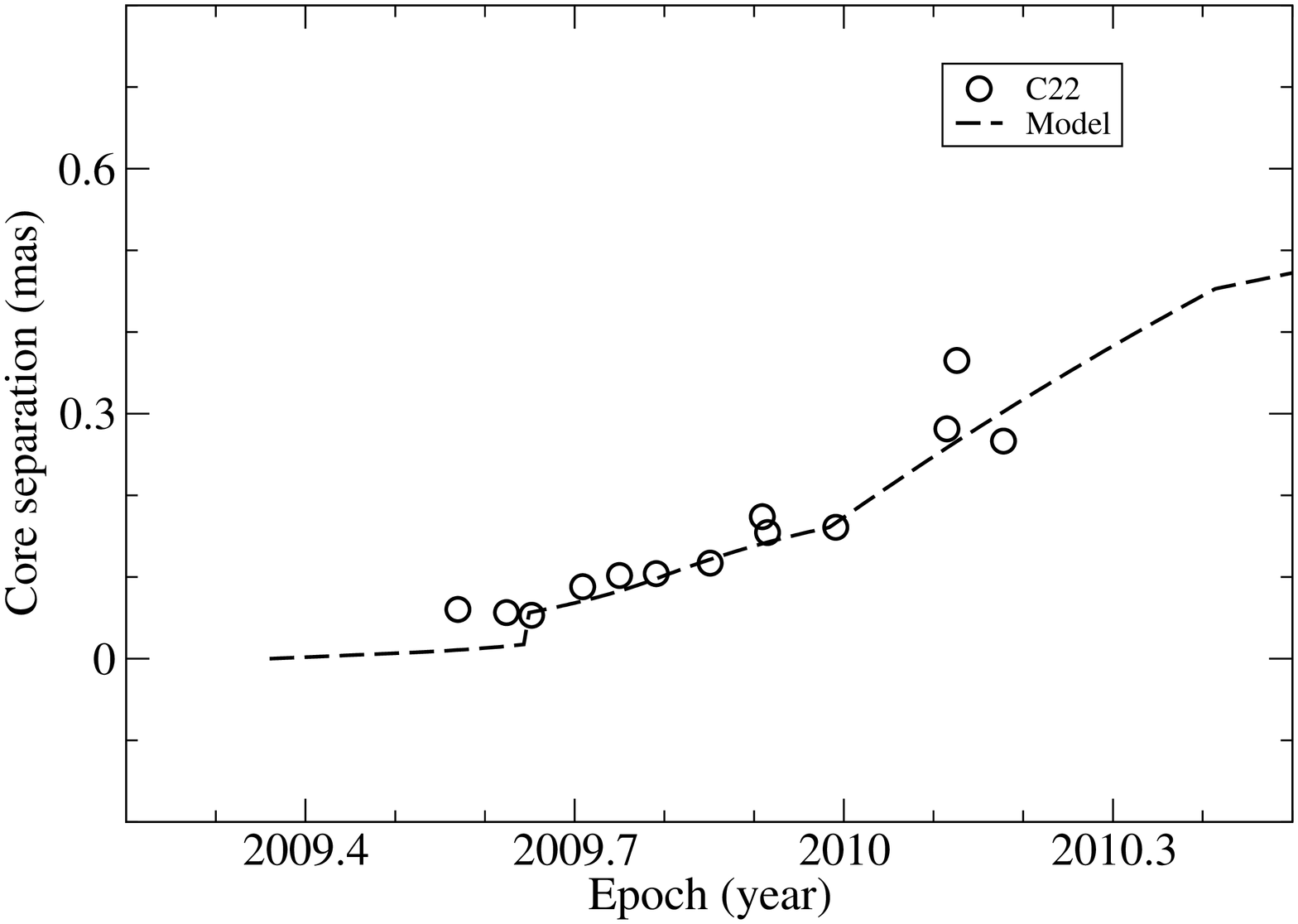}
   \includegraphics[width=4.5cm,angle=-90]{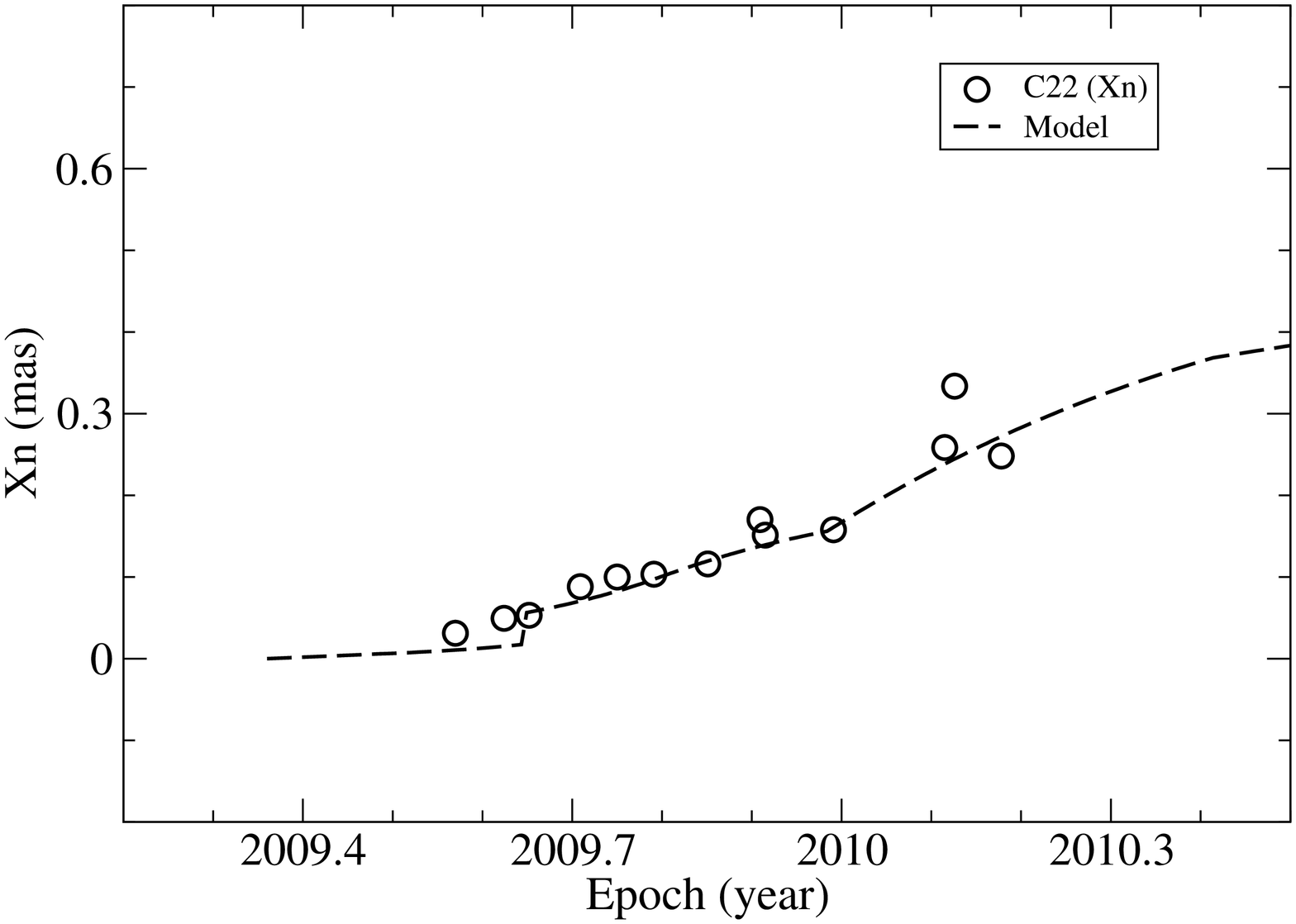}
   \includegraphics[width=4.5cm,angle=-90]{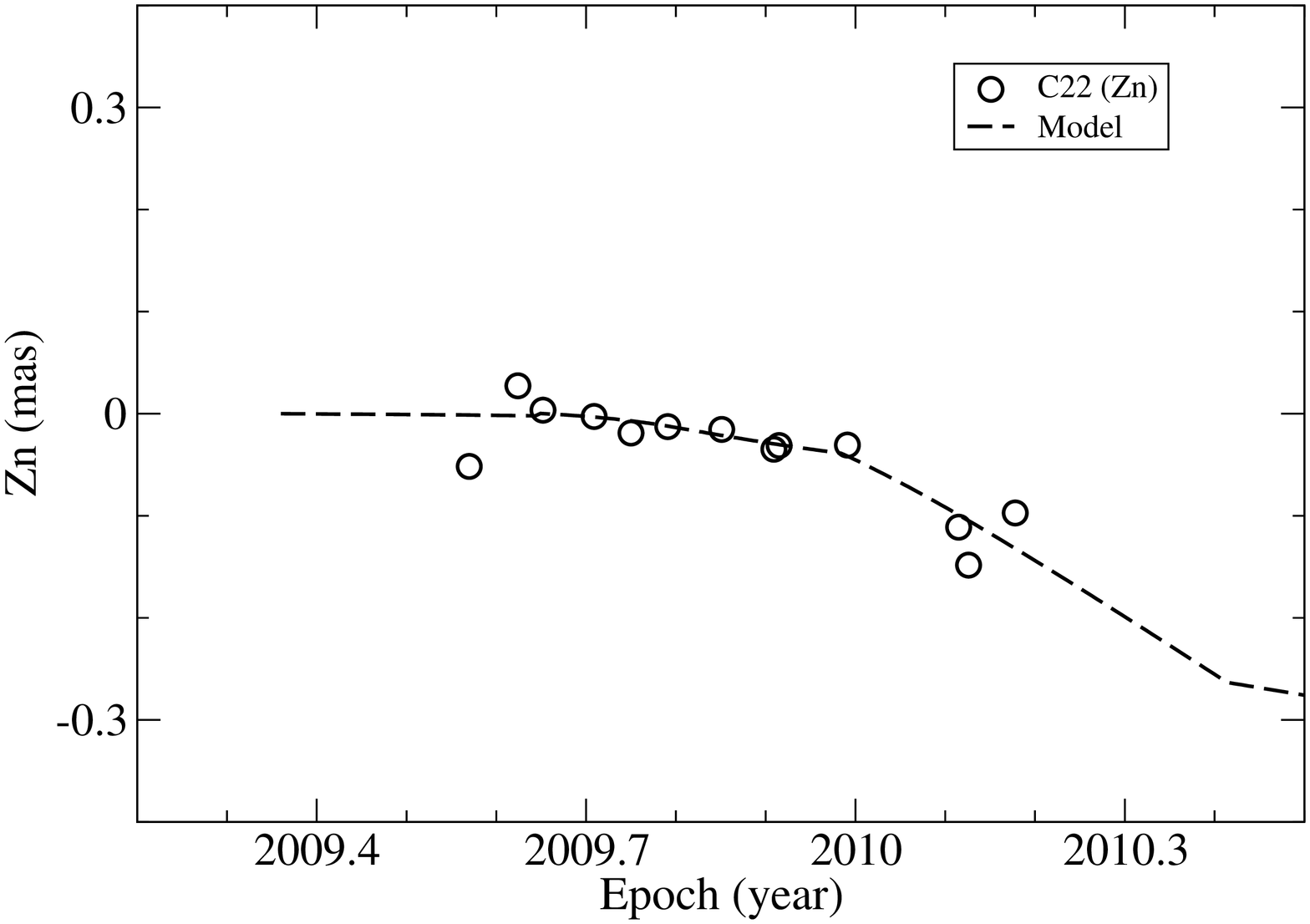}
   \includegraphics[width=4.5cm,angle=-90]{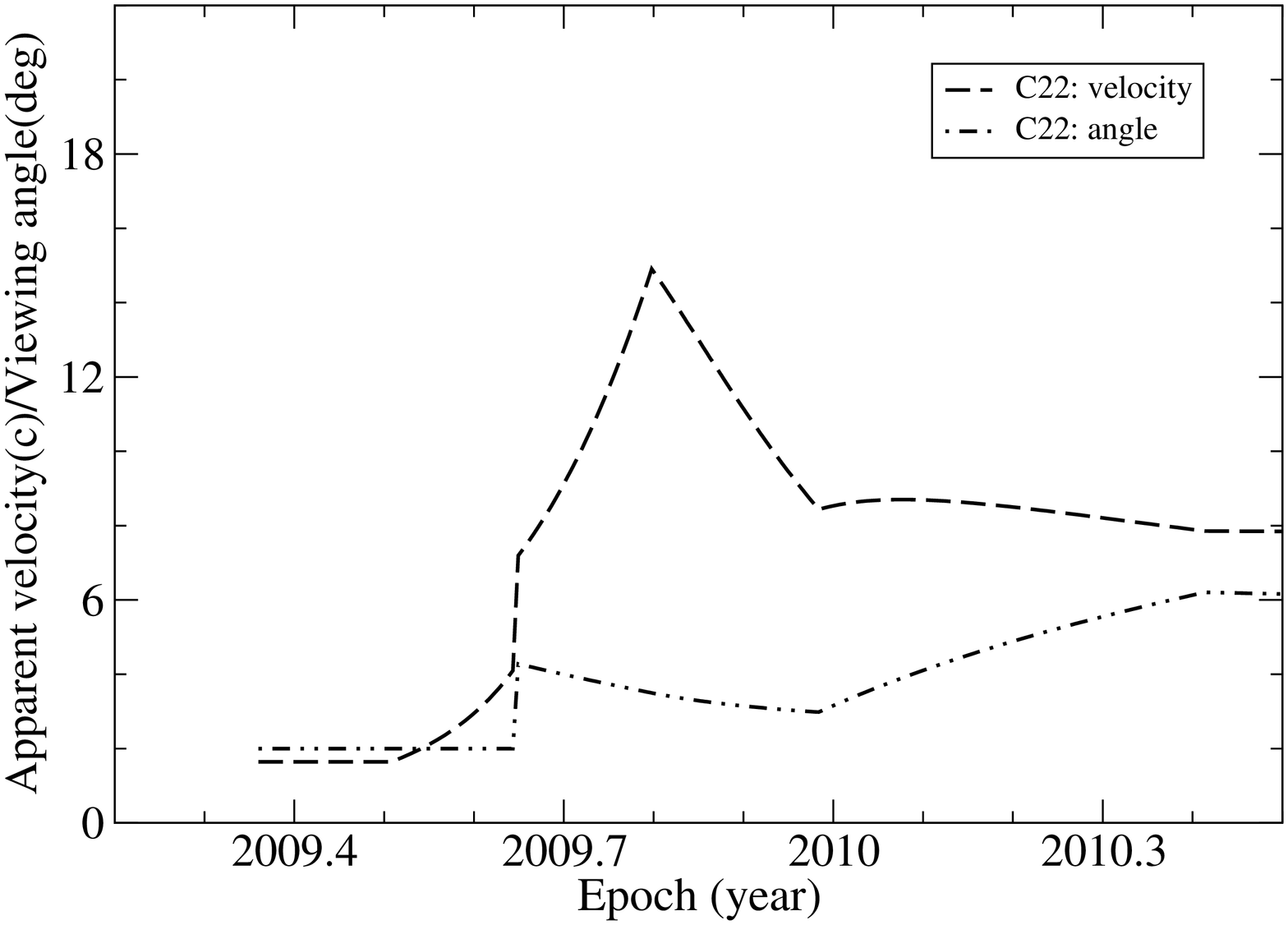}
   \includegraphics[width=4.5cm,angle=-90]{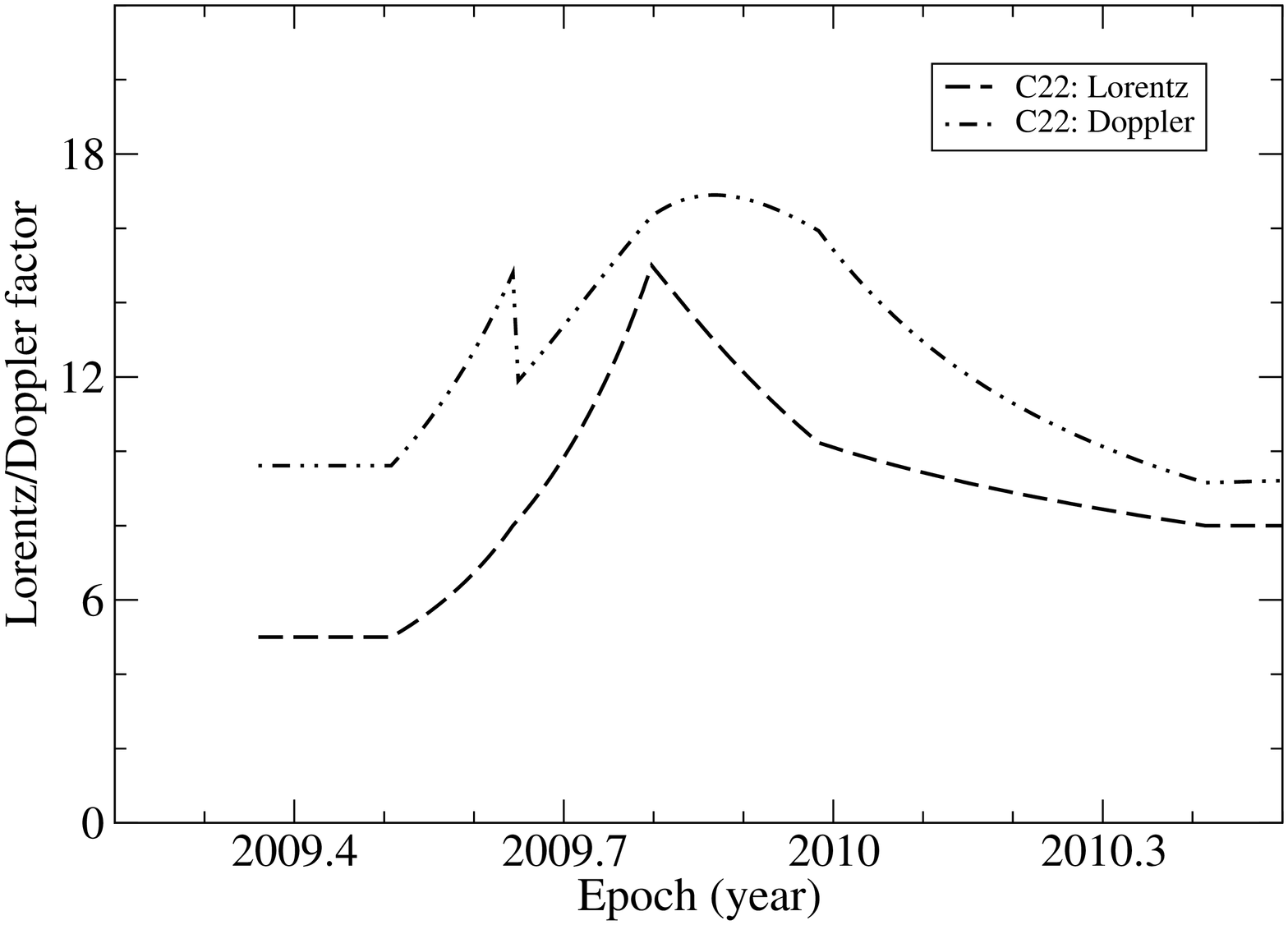}
   \caption{Knot C22: Model fits to its kinematic properties with its 
   precession  phase $\phi_0$=5.28\,rad+8$\pi$, $t_0$=2009.36, including the
    fits to the whole trajectory, core separation and  coordinate $X_n$ (upper
    three panels); coordinate $Z_n$, the derived 
    apparent-velocity/viewing-angle and bulk Lorentz-factor/Doppler-factor
    (bottom three panels).}
   \end{figure*}
    \begin{figure*}
    \centering
    \includegraphics[width=4.5cm,angle=-90]{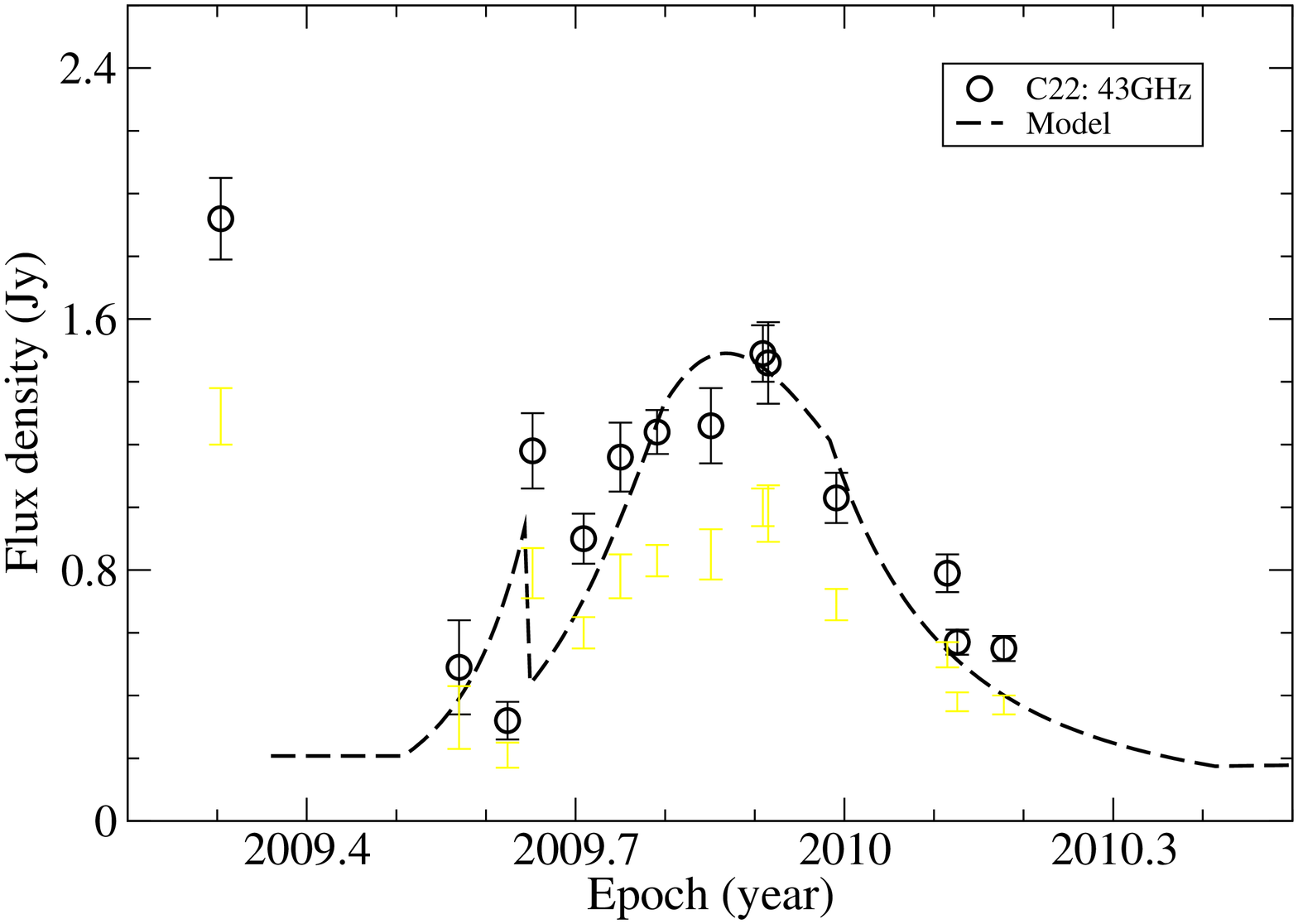}
    \includegraphics[width=4.5cm,angle=-90]{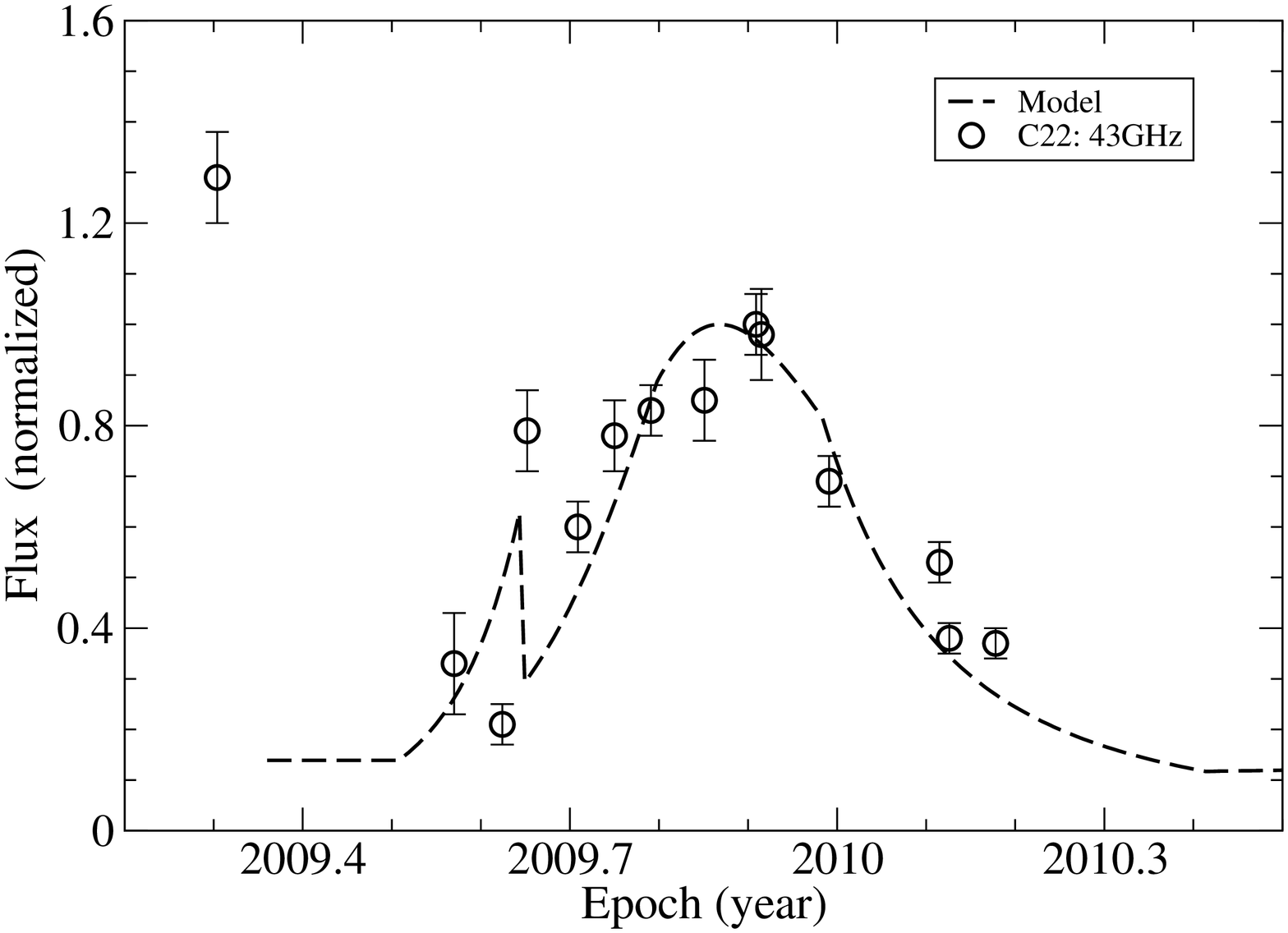}
    \caption{Knot 22: The observed 43\,GHz flux light-curve (left panel) 
    and the normalized flux light-curve (right panel) were very well fitted
    by the Doppler-boosting profiles.}
    \end{figure*}
   \section{Interpretation of kinematics and flux evolution for knot C22}
   As shown in the previous paper (Qian \cite{Qi22}), the model-fit to the
    kinematic behavior of knot C22 was quite important, because: (1) the 
   ejection time of knot C22 (2009.36) was approximately equal to that
    (1995.06) of knot C9 plus
    $\sim$two precession periods (14.6\,yr) and that (1980.80) of knot C5 
   plus $\sim$4 precession periods. Thus
   our precessing nozzle scenario could explain its kinematic behavior during
   a time-interval of four precession periods ($\sim$29.2 years);
   (2) the observed trajectories of knot C5, C9 and C22 quite clearly indicated
   the recurrence of the precessing common trajectory pattern, validating
    this helical pattern adopted in our precessing nozzle scenario; (3) the 
  three knots all revealed the intrinsic acceleration/deceleration in their 
   motion; (4) their Doppler factors derived from the model-simulation of their
   kinematics showed bump-structures, implying the existence of 
   Doppler-boosting effect.   \\
   Here in this paper we shall confirm that the flux variations observed in 
   knot C22 could be successfully interpreted in terms of its Doppler-boosting
    effect as for the flux evolution observed  in  knots C4, C5, C9 and C10.\\
    All these distinct features listed above may actually justify our
    precessing nozzle scenario for investigating the VLBI-phenomena in 
   QSO 3C345.
    \subsection{Model simulation of kinematics for knot C22}
   As shown in the previous paper (Qian \cite{Qi22}) the model-fitting of 
   the kinematics for knot C22 of jet-A in 3C345 was very important
    because the VLBI-observations of knot C22 
    extended the periodic behavior of jet-A to 4 precession periods relative 
   to the ejection of knot C4 (from $\sim$1979 to 2009, about 30 years).\\
    According to the precessing nozzle scenario the kinematics of C22 could be 
   model-fitted by using precession phase $\phi_0$(rad)=5.28+8$\pi$ and
    ejection epoch $t_0$=2009.36.\\
   In Figure 27 are shown its traveled distance Z(t) along the Z-axis (left
   panel) and the curves of functions $\epsilon(t)$ and $\psi(t)$. During
   the period 2009.36--2009.98 in its inner trajectory-section 
    (Z$\leq$2.3\,mas, $X_n{\leq}$0.156\,mas, spatial distance 
    $Z_c{\leq}$15.3\,pc), $\epsilon$=$2.0^{\circ}$ and $\psi$=$7.16^{\circ}$,
     and knot C22
    moved along the precessing common trajectory. After 2009.98 $\epsilon$ 
   and $\psi$ started to increase and
   knot C22 started to move along its own individual trajectory, which deviated
   from the precessing common trajectory in its outer trajectory-section.\\
   The model-fit to its inner trajectory-section in terms of the precessing 
   common trajectory is shown in Figure 28.\\
   The model-fitting results for its entire kinematics are shown in Figure 29.
    It can be seen that the entire trajectory, core separation, two 
   coordinates were very well fitted in terms of our precessing nozzle 
    scenario. \\
    The model-derived apparent velocity/viewing angle and the bulk Lorentz
    factor/Doppler factor are shown in the bottom/middle and bottom/right panels
    of Figure 29, respectively. \\
    Specifically, its bulk Lorentz factor was modeled as: for Z$\leq$0.20\,mas
    (2009.51) $\Gamma$=5.0; for Z=0.2--1.2\,mas(2009.51-2009.80) $\Gamma$
    increased form 5.0 to 15.0; for Z=1.2--2.3\,mas (2009.80-2009.98) 
    $\Gamma$ decreased from 15.0 t0 10.2; For Z=2.3--3.6\,mas (2009.98-2010.41)
    $\Gamma$ decreased from 10.2 to 8.0 and then kept constant beyond 
    Z=3.6\,mas (after 2010.41).\\
    As shown in Figure 29 (bottom/right panel) that the model-predicted 
    Doppler factor showed a bump-structure during the period
    $\sim$2009.5-2010.2 with its peak $\delta_{max}$=16.90 at 2009.87, which
    should induce Doppler-boosting effect in the flux evolution of knot C22
     and it really occurred as described below.
    \subsection{Doppler-boosting effect and flux evolution of knot C22}
     The close relation between the flux variations and the Doppler-boosting
    effect found in knot C22 provided one more valuable and successful test of
    our precessing nozzle scenario.\\
     It can be seen that both the observed 43\,GHz light-curve and its
     normalized light-curve were extraordinarily well fitted by the  
     Doppler-boosting profile, as shown in Figure 30:  a flaring event 
     during 2009.6--2010.2 with its peak of 
    $\sim$1.49Jy at $\sim$2009.91 was well coincident with the Doppler
     boosting profile $[{\delta}(t)/{\delta_{max}}]^{3.5}$.\,\,\footnote{Here we
    arbitrarily assumed a spectral index 0.50. The intrinsic flux density
    was then derived to be 5.72$\times{10^{-5}}$Jy.} That is, The
     flux-density evolution of knot C22 can be fully interpreted in terms of
     its Doppler boosting effect. Thus our precessing nozzle scenario gets 
    through one more valuable test.
    \section{Conclusive remarks}
    We can make some conclusions as follows.\\
   (1) We have applied our precessing jet-nozzle scenario previously proposed
    to interpret the VLBI-kinematics and flux evolution for five superluminal
    components (C4, C5, C9, C10 and C22) in QSO 3C345. It is shown that 
    the VLBI-measured kinematic
    properties of all these superluminal knots (including trajectory, core
    separation, coordinates, apparent velocity as functions of time) can be 
    well model-simulated and explained, adopting a precession period of 
    7.30$\pm$0.36\,yr and a precessing common trajectory pattern. The 
    model-simulation methods dealing with the model-fitting of multiple
    functions with multiple parameters were now validated.\\
   (2) The apparent trajectories observed for the five
     superluminal knots could be
    divided into two sections: inner section and outer section. In the inner 
    sections the superluminal components moved along the precessing common 
    trajectory (or common helical-pattern trajectory), while in the outer 
    sections they moved along their own individual trajectories which deviated
    from the common precessing trajectory pattern.\\
    (3) Through the model-simulation of their kinematic properties their bulk
    Lorentz factor and Doppler factor as functions of time were naturally 
    derived and  thus the relation between their Doppler boosting effect and 
     flux evolution could be investigated. It is shown that the knots' radio
     light-curves (radio flaring events) occurred during their motion in the
     inner regions could
     be successfully  explained in terms of their Doppler-boosting effect.
     The simultaneous determination of the multiple parameters as functions
     of time describing the kinematic and dynamic properties may be regarded
     as the  advantage of our precessing jet-nozzle scenario and 
     model-simulation  methods.\\
    (4) Using the anticipatively-determined Doppler factors, for the first time
    we were able to study the flux evolution of superluminal components in
    QSO 3C345.
     We found that the radio light-curves of the five components were fully
     coincident with the Doppler boosting profiles \footnote{Shorter 
    time-scale flux variations could be understood as due to the fluctuations
    in the intrinsic emission and spectral index of the knots, which could be
    produced by the interaction of the traveling relativistic shocks with
    the complex physical environments.}. Most importantly, for all
    the five superluminal components (C4, C5, C9, C10 and C22) the 
    radio flaring events produced by the Doppler-boosting effect occurred
    during their motion in the inner trajectory regions (i.e., the precessing
    common trajectory sections).\\
     Obviously, these results summarized above are significant for our
     precessing jet-nozzle scenario and  we can now make an overall review of 
     our scenario to provide a full-view for studying the VLBI-phenomena in
     QSO 3C345 as follows.
    \begin{itemize}
    \item The plasmoids and magnetized plasmas ejected by the precessing nozzle
     was assumed to form the entire jet of 3C345, thus VLBI-observations 
     could find many superluminal components moving along different 
    trajectories, forming a complex distribution of superluminal components 
   in the jet.
    \item In fact these observed trajectories were formed
    at different precession 
     phases by the precession of  a common helical trajectory pattern 
     associated with the nozzle-precession, although for different superluminal
     components they had outer trajectory-sections where they moved along
     their own individual trajectory patterns.
   \item The observed swing of the ejection position angle of superluminal 
     components and its quasi-periodicity could be understood in terms of the 
     nozzle-precession.
    \item The observed kinematics (including apparent trajectory, core distance,
     coordinates, apparent speed) can be well fitted consistently for all
     the superluminal components, if a double precessing-nozzle structure
     (or double-jet structure) was assumed. \footnote{Possible evidence
      of a double-jet structure has been discussed in Qian (\cite{Qi22}).
     The close relation between the flux evolution and Doppler-boosting effect
     for the superluminal components in jet-B of 3C345 (in addition to jet-A)
     will be  presented elsewhere, further justifying the double-nozzle
     structure in 3C345.} 
     \item The double-jet structure as an assumption was based on the 
     model-simulation results: jet-A and jet-B have different kinematic and
     dynamic  properties, including different jet direction, precessing common 
     trajectory  pattern and opening angle of precession cone,
      but having a similar precession period and direction of precession.
     This type of precession could originate from keplerian motion in the
     putative binary black hole in the nucleus of 3C345.
    \item The precession period 7.3$\pm$0.36\,yr for both the jet-A and jet-B
     seems model-simulated appropriately, since for both jets the kinematic
     behaviors have been well explained and model-fitted.
     \item Since the inner trajectory-sections observed for all the 
    superluminal components
     of both jet-A and jet-B in 3C345 are well model-fitted within an accuracy
     of $\pm$5\% of the precession period, the precessing common trajectory
     patterns derived for both jet-A and jet-B should be regarded to be 
     physically appropriate, although they are not unique. They should not
    deviate from the actual helical patterns too far.
     \item The modeled bulk Lorentz factor and Doppler factor derived for 
    the superluminal
     components by model-fits of their kinematics clearly revealed the 
     acceleration/deceleration in their motion. These anticipatively-derived
     Doppler factors provide valuable opportunities to study the relation
     between their flux-evolution and the Doppler-boosting effect: a key test
     for our precessing nozzle scenario.
     \item The flare events (flux variations) observed to  occur in their inner
      trajectory-sections for all the five superluminal components (C4, C5, C9,
     C10 and C22) could be fully interpreted in terms of their Doppler-boosting
     effects. The kinematics, dynamics and emission properties of all the five 
     superluminal components were  consistently well interpreted, 
    thus validating  the  whole scenario and 
    the model-simulation methods. 
   \footnote{The shorter time-scale flux variations 
   had no relation to the 
    Doppler-boosting effect induced by the intrinsic acceleration/deceleration
    of the knots 
    could be due to the variations in the intrinsic
    flux density and spectral index of the knots at different frequencies
    (Equation (22) in Sect.4).}  
   \end{itemize}
   
   \newpage
   \newpage
  \begin{appendix}
  \section{Doppler boosting effect and flux evolution of a superluminal
    component in blazar 3C279}
   3C279 is one of the most prominent blazars which have been intensively 
   observed and studied on VLBI-scales. The VLBI-kinematic behaviors observed 
   at 43\,GHz of its
   superluminal components have been analyzed in detail and successfully
   interpreted and model-simulated in terms of the precessing nozzle scenario
   with a double-jet structure
   for 31 components (Qian et al. \cite{Qi19a}).\\    
   Here as a supplement, we introduce the results of model-fitting of the 
   VLBI-kinematics for one superluminal component C13 in 3C279 and the 
    interpretation   of its flux evolution in terms of its Doppler-boosting 
   effect during its accelerated and decelerated motion. \\
    In the previous paper (Qian et al. \cite{Qi19a}) the superluminal 
   components in 3C279 was assumed to move along a common
   parabolic trajectory which precesses to give rise to the individual
   trajectories of the knots at their corresponding precession phases.
   In this framework, knot C13 had its ejection time $t_0$=1998.88 and
   corresponding precession phase $\omega$=3.65\,rad. Its
   bulk Lorentz factor $\Gamma(t)$ and Doppler factor $\delta(t)$ were
   derived as functions of time, showing its accelerated motion during the
    period 1999.2--1999.5. Its modeled Lorentz factor and
    Doppler factor increased. The anticipatively-derived Doppler factor 
    implied that knot C13 should have flux variations induced  by the
    Doppler boosting effect. However, in the previous model-simulation of the
    kinematics for knot C13 we did not take its deceleration after $\sim$1999.5
   into consideration. \\
   Here in this appendix, for analyzing the relation between its flux
   evolution and Doppler boosting effect during the period $\sim$1999.0-2001.0,
   we take both its accelerated and decelerated motion into consideration in
   order  to fit the 
   rising and decaying phases of its
   15\,GHz flux light curve. The 15GHz flux density data was adopted from
    Roland et al. (\cite{Ro13}), where the knot was designated as C5.\\
    \begin{figure}
    \centering
    \includegraphics[width=4.5cm,angle=-90]{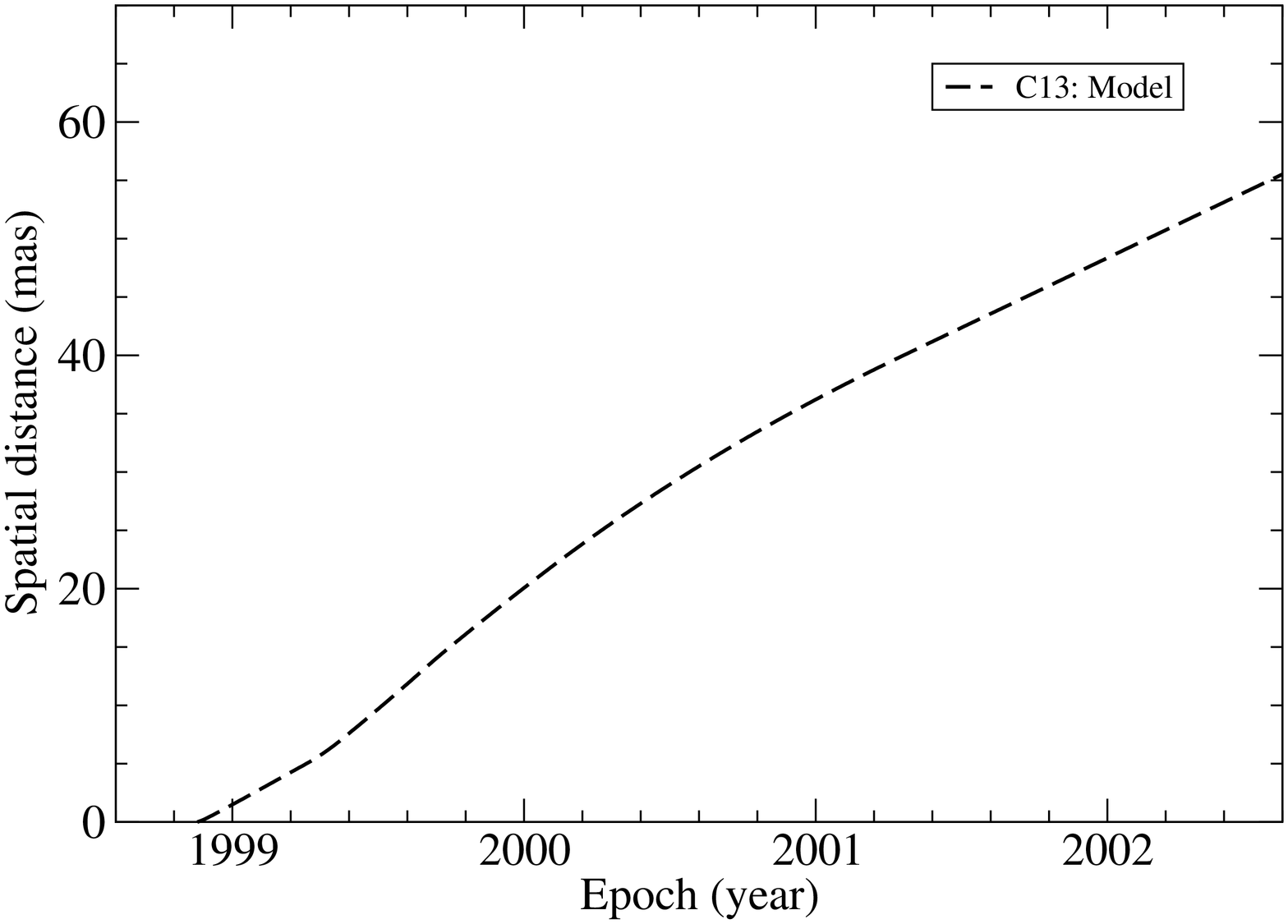}
    \includegraphics[width=4.5cm,angle=-90]{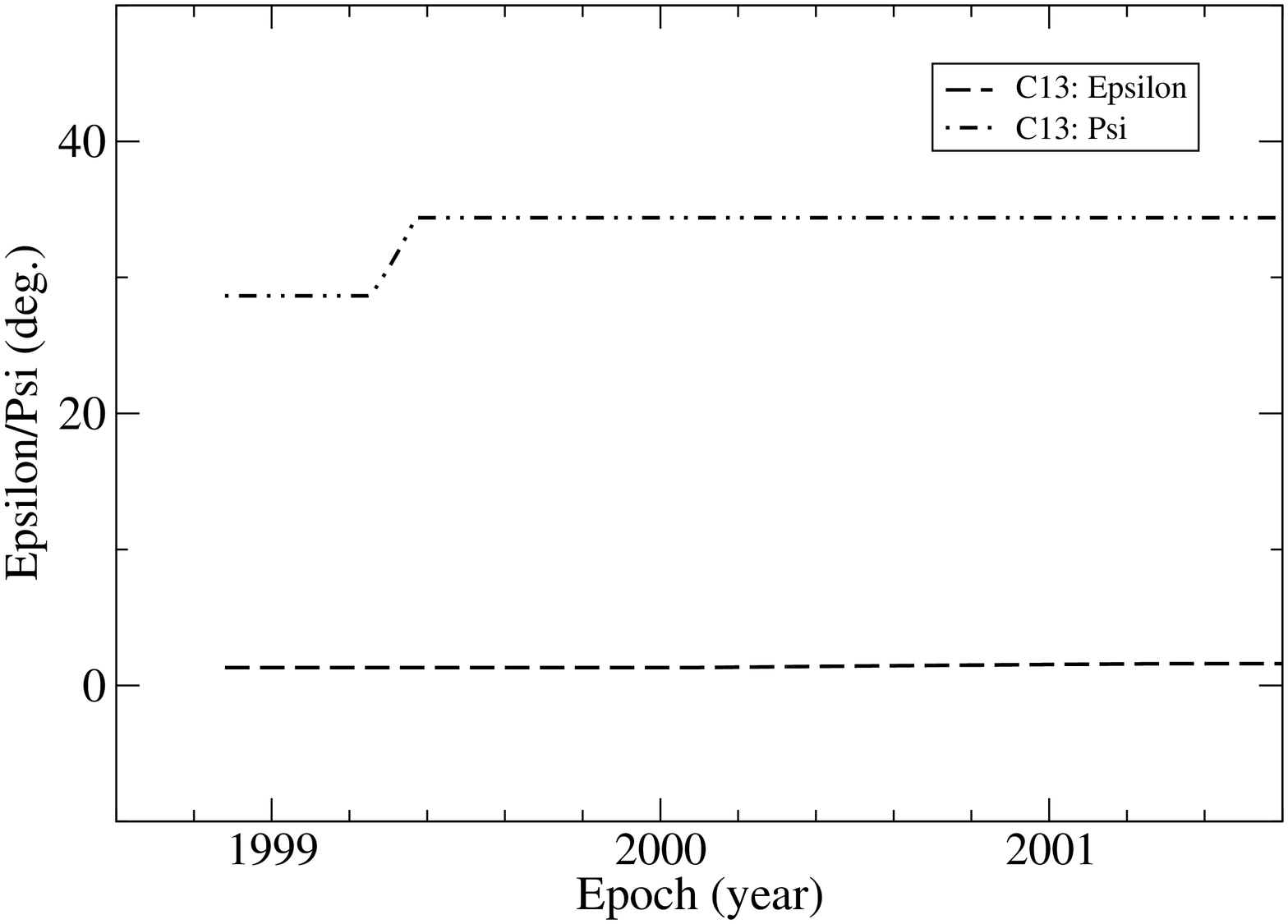}
    \includegraphics[width=4.5cm,angle=-90]{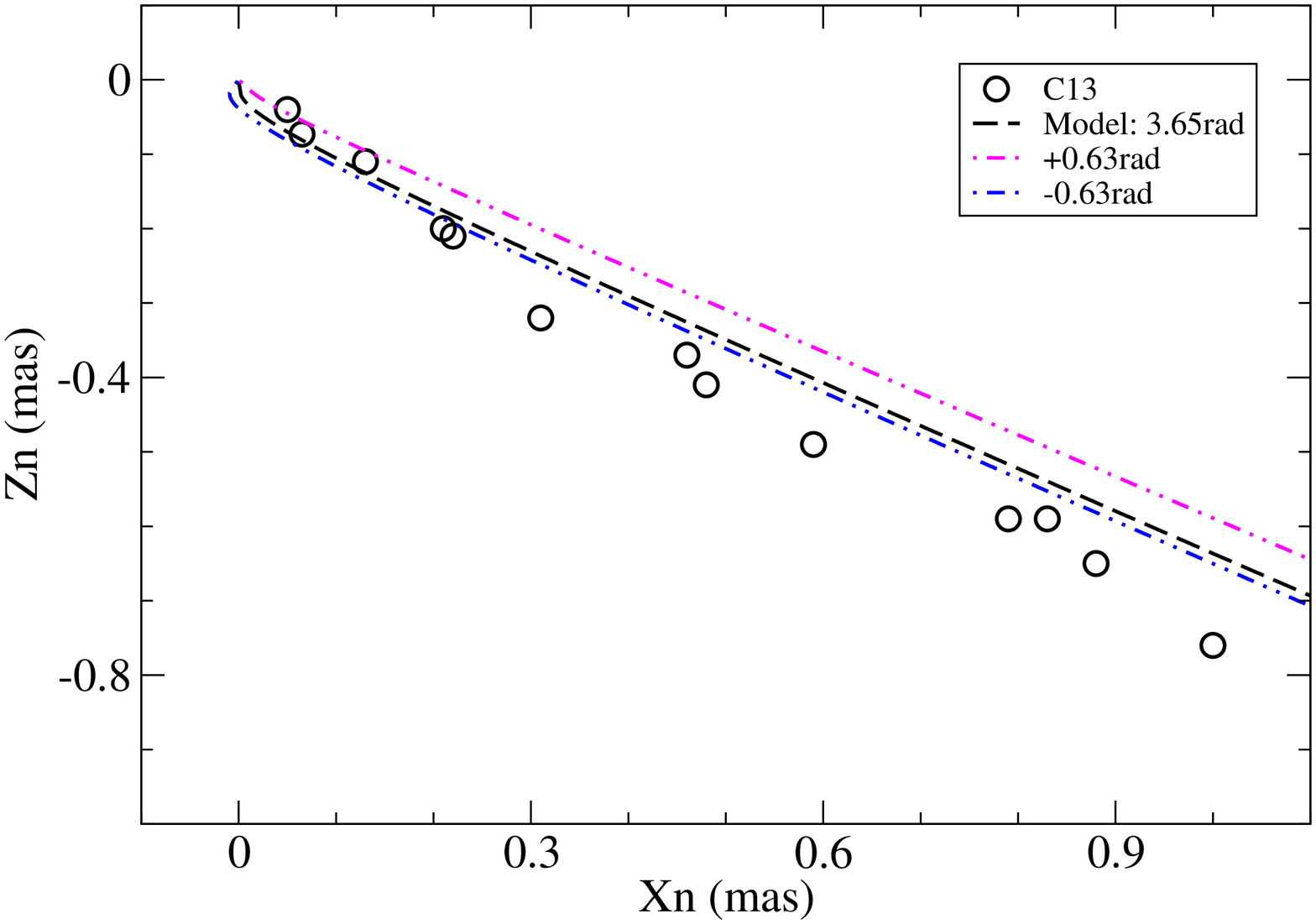}
    \caption{Knot C13 in 3C279. Top panel: the travelled distance Z(t) 
   along the Z-axis. The curves of $\epsilon(t)$ and $\Psi(t)$ (middle panel)
    shows that for Z$>$5.0\,mas (or after 1999.25) parameter $\psi$ started to
    increase and knot C13 started to move along its own individual trajectory.
    The model fit to its inner trajectory section ($X_n{\leq}$0.10\,mas, 
    $t{\leq}$1999.25) in terms of the  precessing common trajectory is shown
    in the bottom panel: the curves in black, magenta and blue represent the
    precessing common trajectories for the precession phases $\omega$=3.65\,rad
    and 3.65$\pm$0.63\,rad ($\pm$10\% of the precession period), respectively.} 
    \end{figure}
    \begin{figure*}
    \centering
    \includegraphics[width=4.6cm,angle=-90]{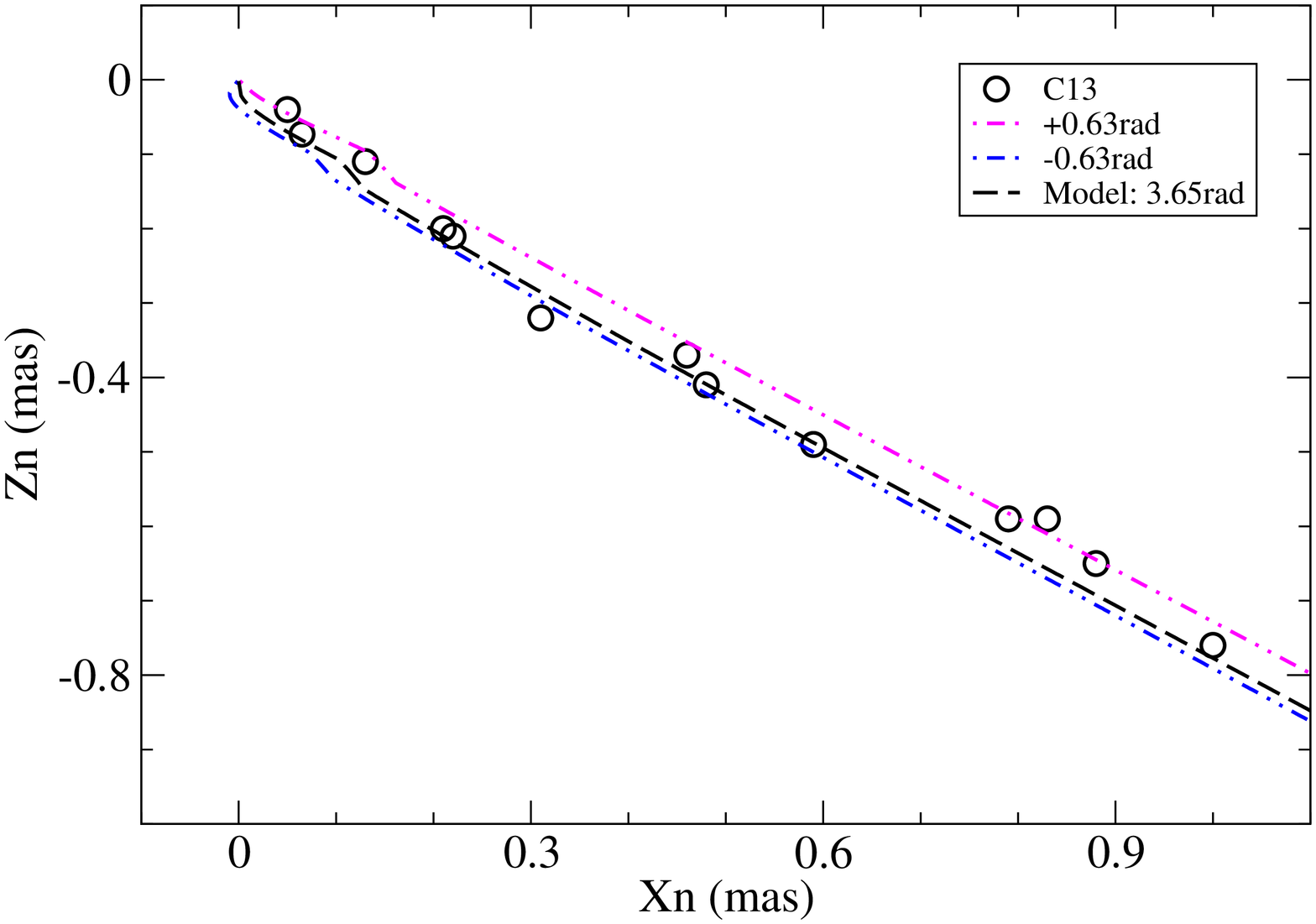}
    \includegraphics[width=4.6cm,angle=-90]{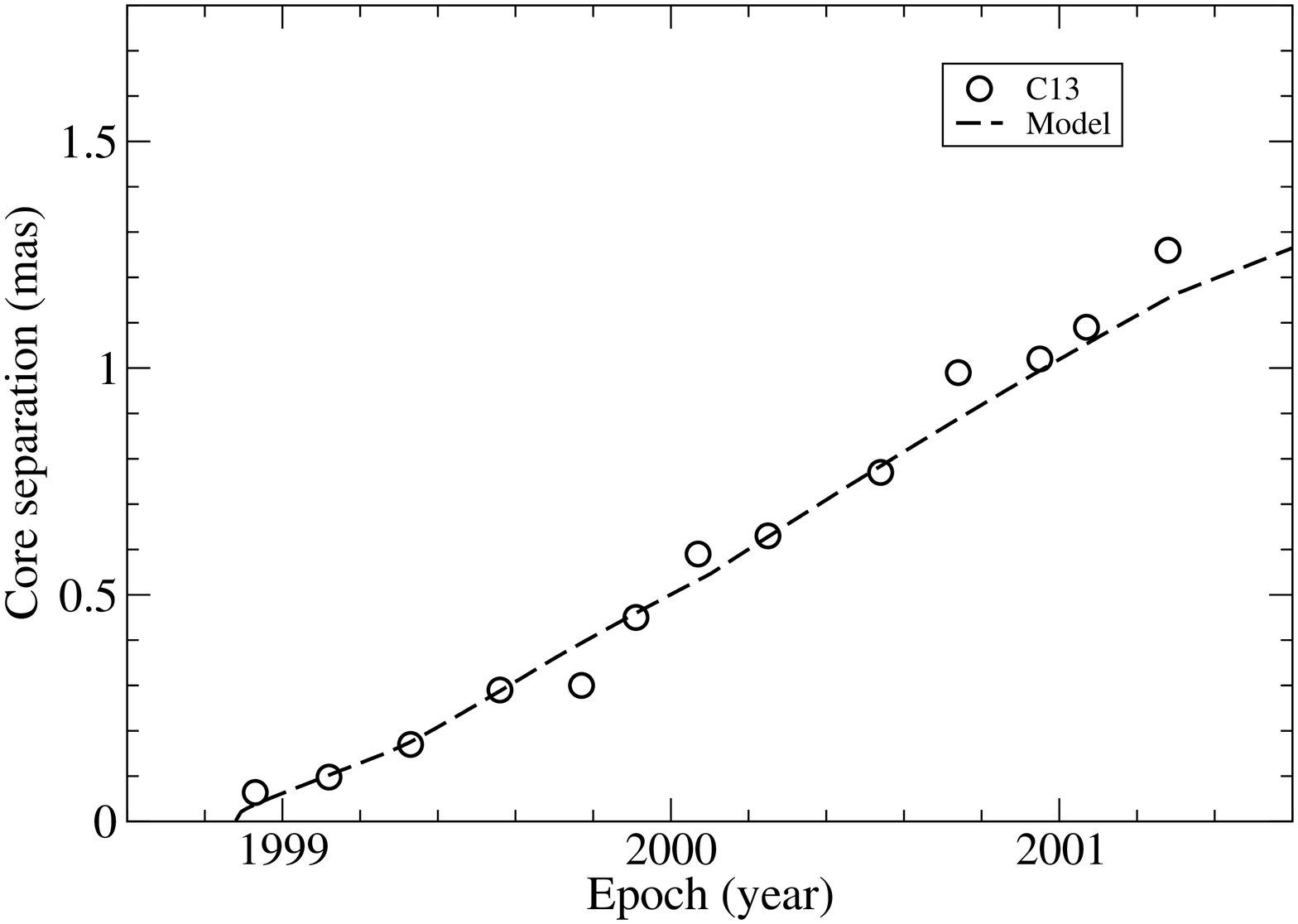}
    \includegraphics[width=4.6cm,angle=-90]{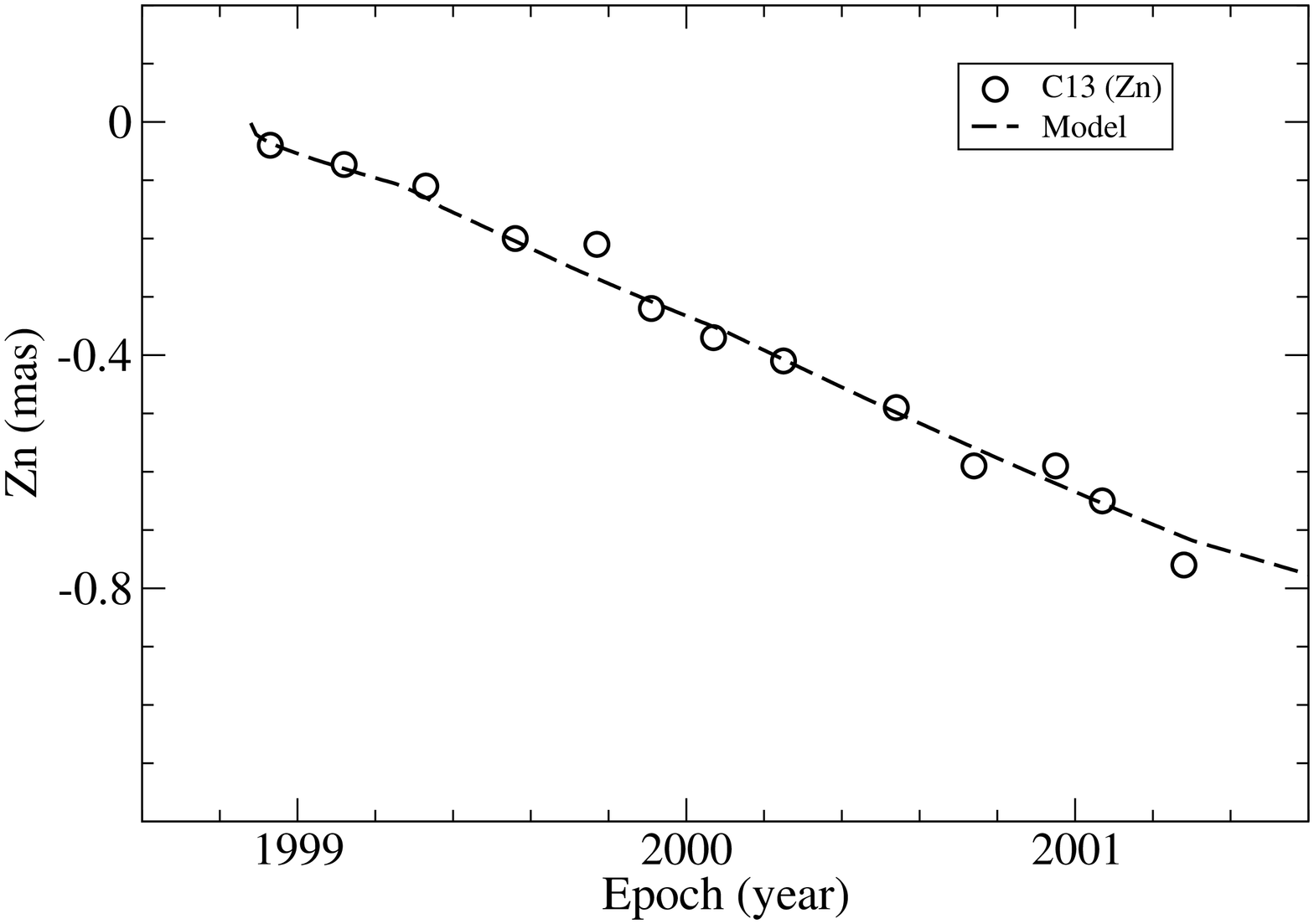}
    \includegraphics[width=4.6cm,angle=-90]{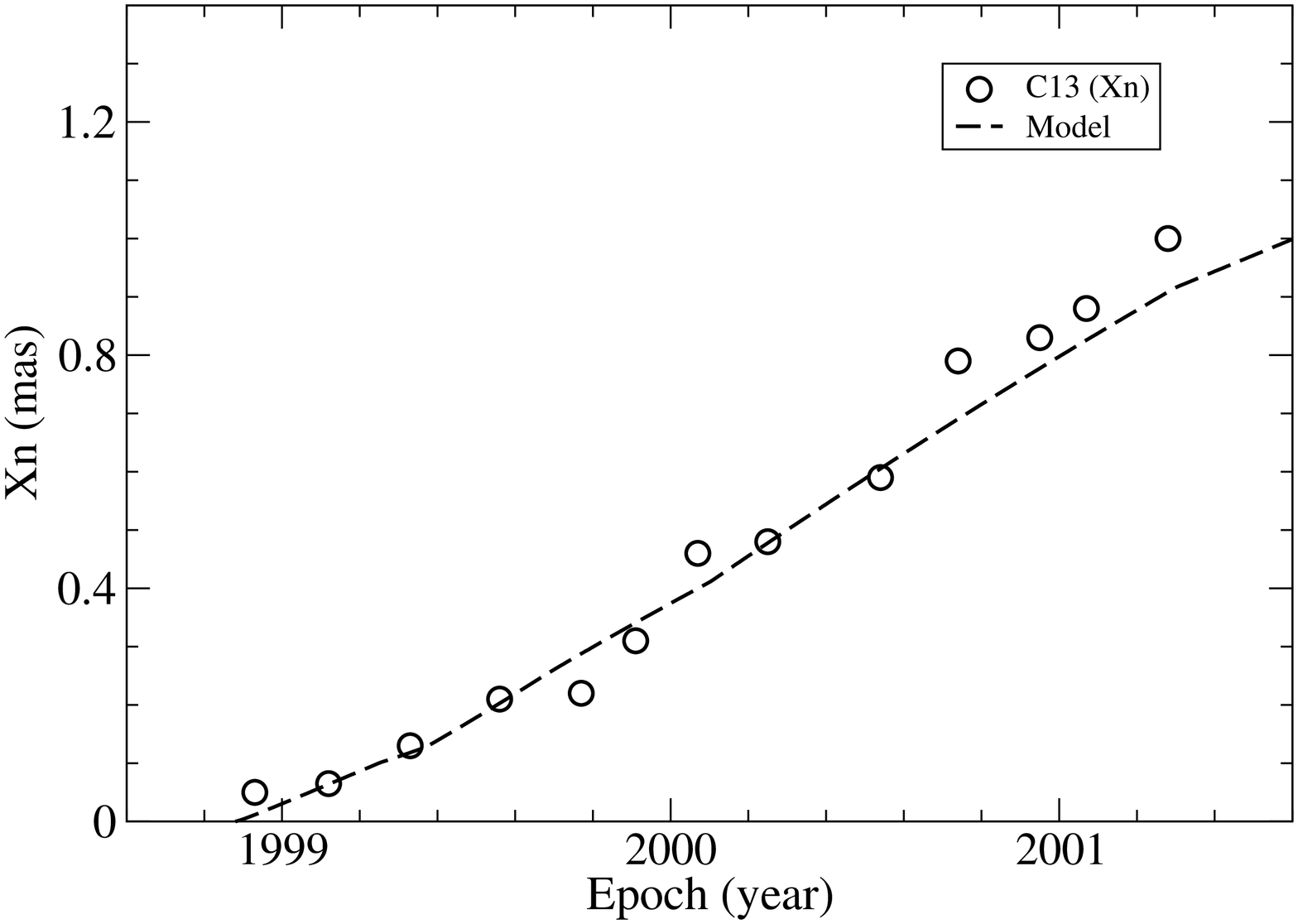}
    \includegraphics[width=4.6cm,angle=-90]{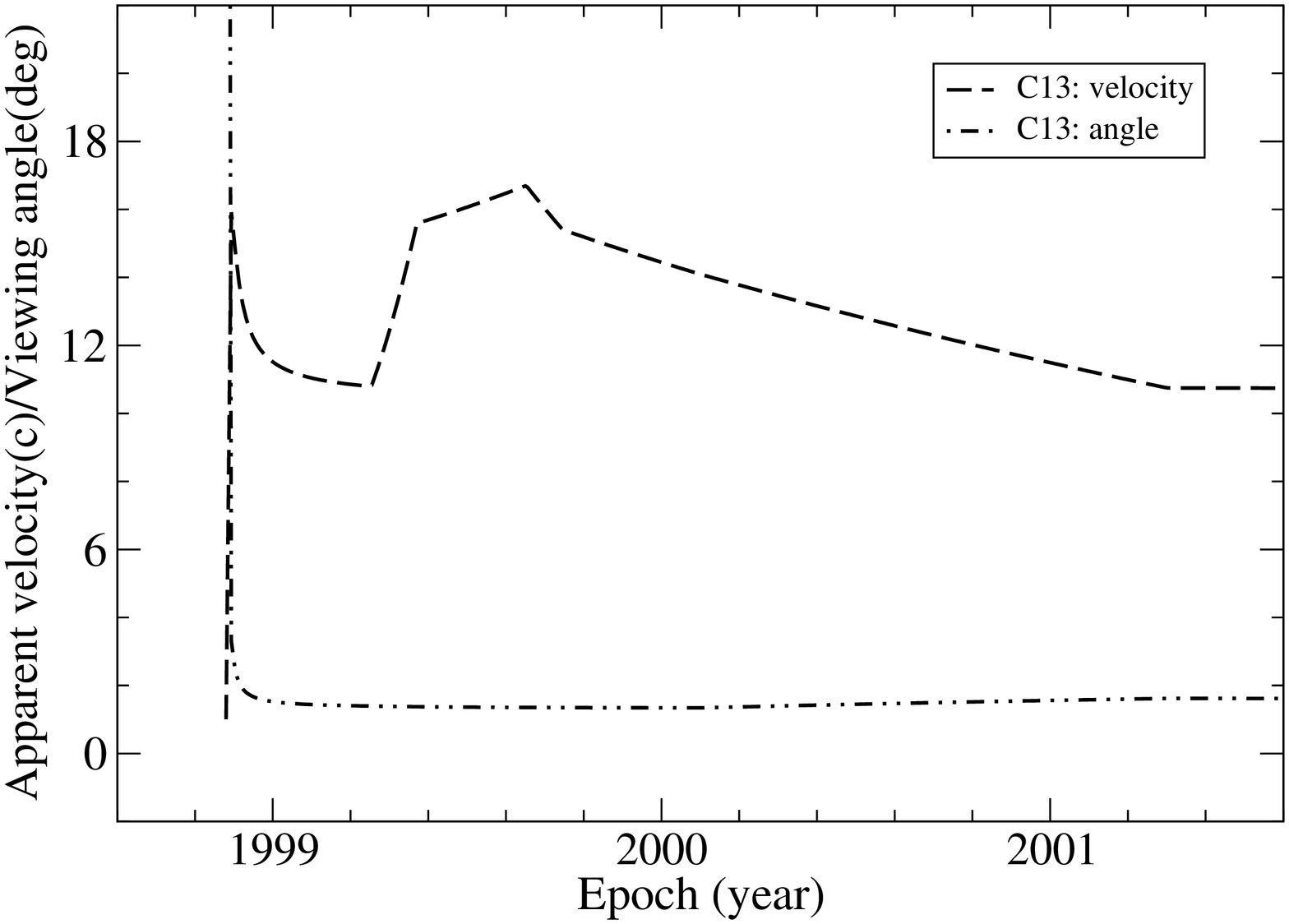}
    \includegraphics[width=4.6cm,angle=-90]{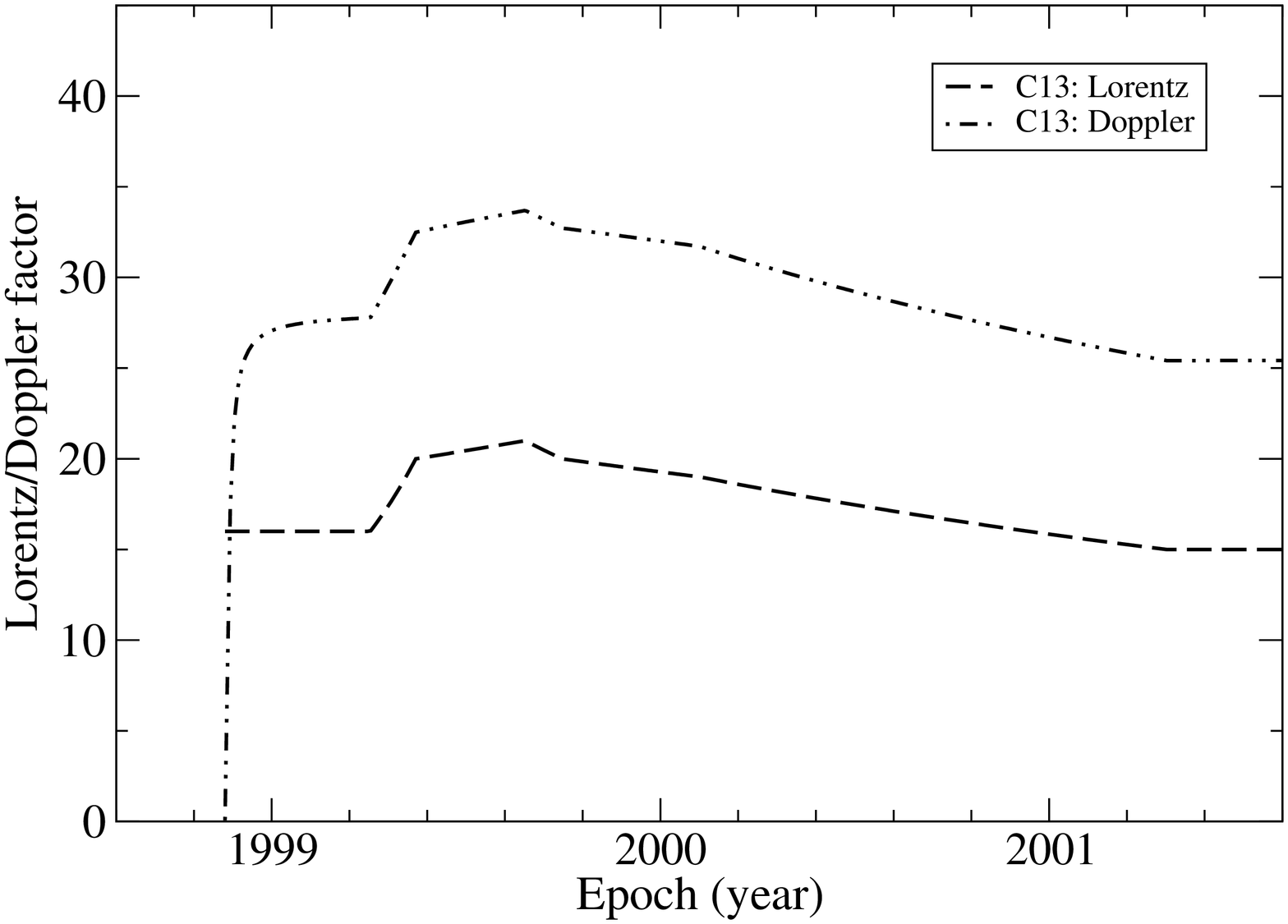}
    \caption{Knot C13 in 3C279: Model-fits to the whole kinematics,including
    both inner trajectory-section (within $r_n{\leq}$0.15\,mas, where
    knot C13 moved along the precessing common trajectory) and outer 
    trajectory section (beyond $r_n$=0.15\,mas, where knot C13 moved along
     its own individual trajectory): the whole trajectory, core separation 
   and  coordinate $X_n$ (top three panels); coordinate $Z_n$,
   apparent-velocity/viewing-angle and bulk Lorentz-factor/Doppler-factor
    (bottom three panels).}
    \end{figure*}
     In Figure A.1 are shown the travelled distance Z(t) along the Z-axis
    (top panel), the curves of parameters $\epsilon(t)$ and $\psi(t)$, and 
    the model-fit of the inner trajectory-section ($X_n{\leq}$0.1\,mas).\\
    The curves of $\epsilon$ and $\psi$ (middle panel of Fig.A.1) show that 
    $\epsilon$=$1.32^{\circ}$=constant, but parameter $\psi$ changed with time: 
    before 1999.25 (Z$\leq$5.02\,mas, $r_n{\leq}$0.15\,mas and 
    $X_n{\leq}$0.10\,mas,) 
    $\psi$=$28.70^{\circ}$, and after 1999.25 $\psi$ increased to 
    $34.38^{\circ}$ at 1999.37 (Z=7.0\,mas). Thus knot C13 moved along the 
    precessing common trajectory within $X_n{\leq}$0.10\,mas and its
    precessing common trajectory section extended to $\sim$33.4\,pc.
    Beyond $X_n$=0.10\,mas it moved along its own individual trajectory in its
    outer trajectory section.\\
    In the bottom panel of Fig.A.1 the model-fit to its inner trajectory
    section is shown,
    where knot C13 moved along the precessing common trajectory: the dashed 
    lines in black, magneta and blue represent the precessing common 
    trajectories corresponding to the precession phases $\omega$=3.65\,rad and
    3.65$\pm$0.63\,rad ($\pm$10\% of the preecssion period), respectively.\\
    The model-fits to its kinematics are shown in Figure A.2, including its
    whole trajectory $Z_n(X_n)$, core separation $r_n(t)$ and coordinate
    $X_n(t)$ (upper three panels), and coordinate $Z_n(t)$, the derived 
    apparent-speed $\beta_{app}(t)$/viewing-angle $\theta(t)$ 
    and the derived Lorentz-factor $\Gamma(t)$ and Doppler-factor $\delta(t)$
    are shown in the bottom three panels. Specifically, 
    its bulk Lorentz factor was 
    modeled as: for Z$\leq$5.0\,mas  $\Gamma$=16.0; for Z=5.0-13.0\,mas 
   $\Gamma$ increased from 16.0 t0 21.0 (maximum Lorentz factor at 1999.65);
    for Z=13.0-40.0\,mas $\Gamma$  decreased from 21.0 t0 15.0 and 
   then kept to be constant. Correspondingly, the Doppler factor had similar
    variations with a maximum value $\delta_{max}$=33.68 at Z=13.0\,mas
   (t=1999.65). During the period 1999.25-2001.0 the derived apparent speed 
   had a bump structure similar to that of Lorentz/Doppler factor, but the
   derived viewing angle changed in a very small range 
   [$1.39^{\circ}$-$1.34^{\circ}$-$1.56^{\circ}$].\\
   Obviously, the bump structure of its Doppler factor as a function of time 
   which was anticipatively-determined for knot C13 indicated the existence
   of Doppler boosting effect in its flux evolution.
    Using the derived  Doppler factor $\delta(t)$ and 
    arbitrarily assuming a spectral index $\alpha$=0.5, we performed the
    model-fits to the observed 15\,GHz light curves (one for the observed 
   flux density light-curve and  the other for the normalized flux light curve)
    in Figure A.3. It can be seen  that both the flux light-curves were very 
    well coincident with its Doppler boosting profile. Thus we see that the
    kinematic behavior and the emission  properties  observed  for
    superluminal component C13 in 3C279 could be consistently
    and successfully interpreted in terms of our precessing jet-nozzle
     scenario (Qian et al. \cite{Qi19a}). \\ 
    \begin{figure*}
    \centering
    \includegraphics[width=6cm,angle=-90]{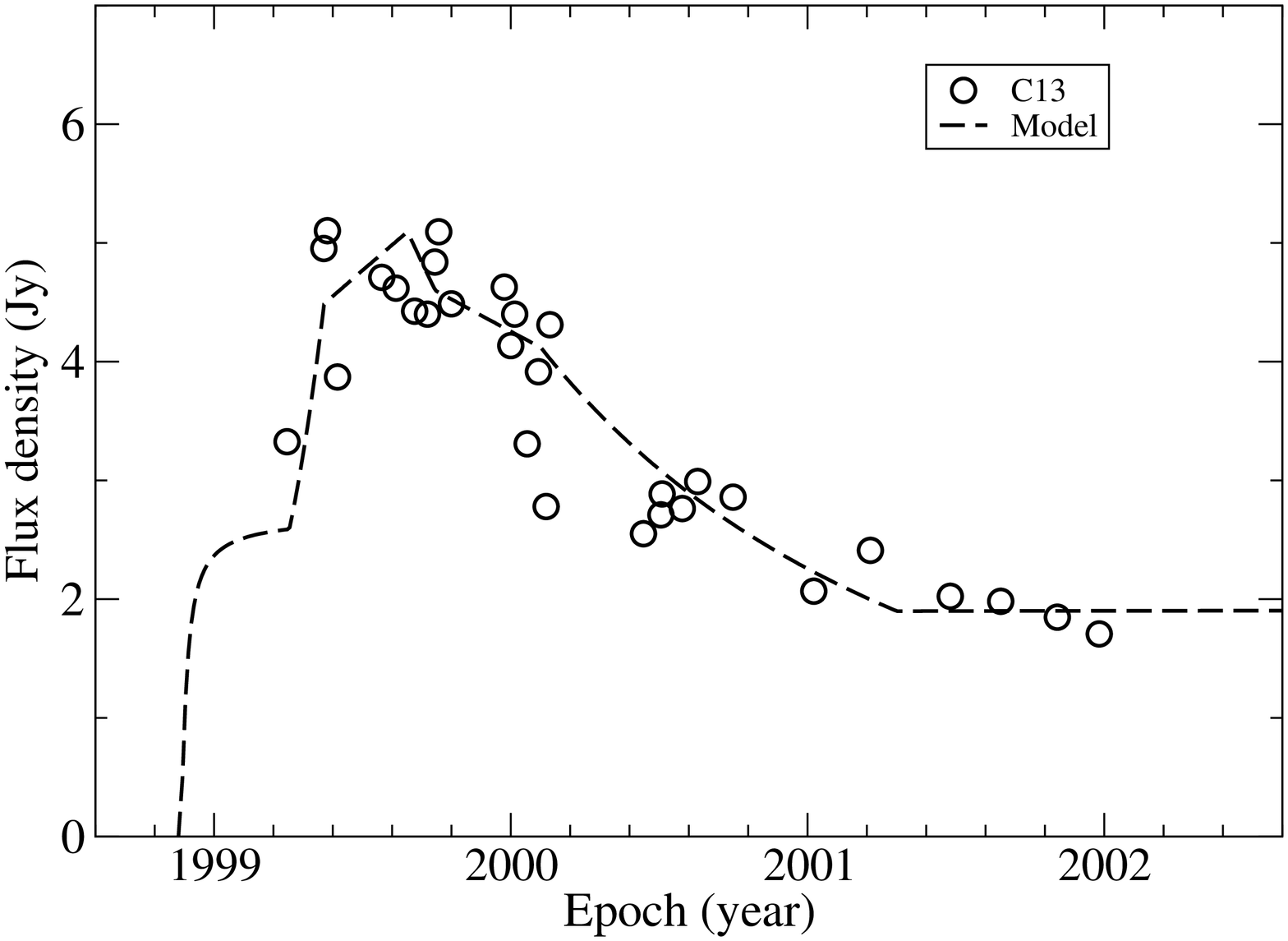}
    \includegraphics[width=6cm,angle=-90]{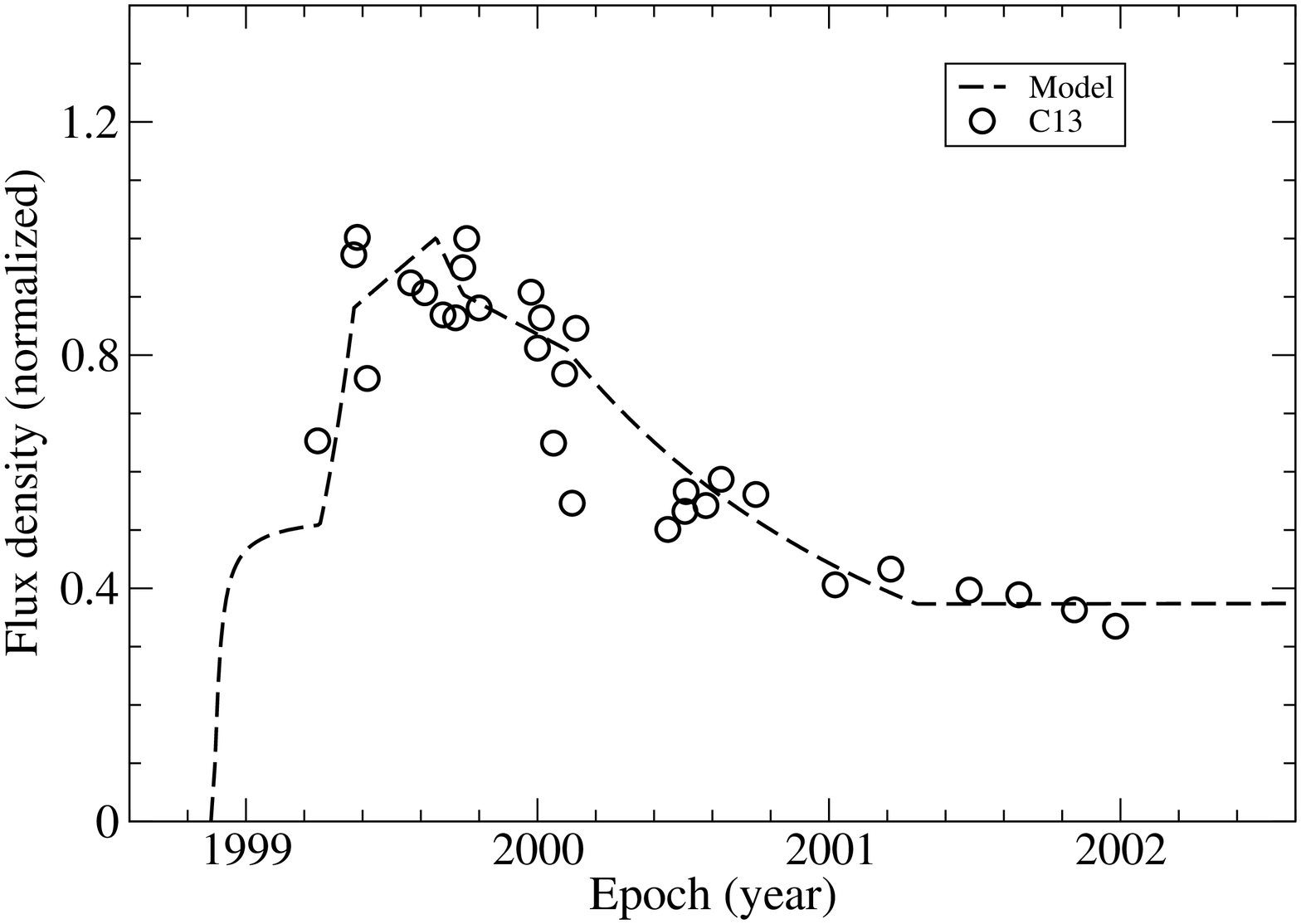}
    \caption{Knot C13 in 3C279: The observed 15\,GHz light curve (left panel)
   and the normalized flux light curve are well interpreted in terms of its
   Doppler boosting effect ($\propto{[\delta(t)/\delta_{max}]^{3.5}}$).}
    \end{figure*}
    We have made detailed studies of the close relation between 
    Doppler-boosting effect and flux evolution for the superluminal components
    in 3C345 and 3C279, and shown that their kinematic/dynamic behaviors and
    emission properties can  consistently and successfully interpreted in
    terms of our precessing nozzle scenario. Thus  the investigation of the
    relation between Doppler-boosting effect and flux evolution for the 
    superluminal components in other blazars or AGNs (e.g., 3C454.3, OJ287 and 
    B1308+326) would be important. We would like to try to do some work  
    in the near future.   
  \end{appendix}
 \end{document}